\documentclass[english,floatfix,twocolumn,showpacs, aps, prx, reprint, superscriptaddress, longbibliography]{revtex4-2}

\usepackage{amsmath}
\usepackage{amssymb}
\usepackage{physics}

\usepackage{hyperref}
\hypersetup{colorlinks=true, citecolor=blue, urlcolor=blue, linkcolor=blue, breaklinks=true}

\usepackage{graphicx}
\newcommand{\commentout}[1]{}
\usepackage{booktabs, footnote}

\newcommand{\NERSC}[0]{\affiliation{National Energy Research Scientific Computing Center, Lawrence Berkeley National Laboratory, Berkeley, CA 94720, USA}}
\newcommand{\UIUC}[0]{ \affiliation{Department of Physics and Institute for Condensed Matter Theory, University of Illinois at Urbana-Champaign, Urbana, IL 61801, USA}}
\newcommand{\UofT}[0]{\affiliation{Department of Physics, University of Toronto, Toronto, Ontario M5S 1A7, Canada}}
\newcommand{\QuEra}[0]{\affiliation{QuEra Computing Inc., 1284 Soldiers Field Road, Boston, MA, 02135, USA}}
\newcommand{\Princeton}[0]{\affiliation{Department of Physics, Princeton University, Princeton, NJ 08544, USA}}
\newcommand{\PCTS}[0]{\affiliation{Princeton Center for Theoretical Science, Princeton University, Princeton, NJ 08544, USA}}

\renewcommand{\vb}[1]{\boldsymbol{\mathbf{#1}}}

\begin{document}
%TC:ignore

\title{Quantum criticality and nonequilibrium dynamics on a Lieb lattice of Rydberg atoms}

\author{Mark~R.~Hirsbrunner$^{\dagger}$}
\thanks{These authors contributed equally.
\\
$^\dagger$ \href{mailto:mark@mhirsbrunner.com}{mark.hirsbrunner@utoronto.ca}; \\
$^\ddagger$ \href{mailto:mkornjaca@quera.com}{mkornjaca@quera.com}; \\
$^\S$ \href{mailto:rhine_samajdar@princeton.edu}{rhine\_samajdar@princeton.edu};\\
$^\bullet$ \href{mailto:plopes@quera.com}{plopes@quera.com}; \\
$^\circ$ \href{mailto:kklymko@lbl.gov}{kklymko@lbl.gov}
}
\NERSC
\UIUC
\UofT

\author{Milan~Kornja\v{c}a$^\ddagger$}
\thanks{These authors contributed equally.
\\
$^\dagger$ \href{mailto:mark@mhirsbrunner.com}{mark.hirsbrunner@utoronto.ca}; \\
$^\ddagger$ \href{mailto:mkornjaca@quera.com}{mkornjaca@quera.com}; \\
$^\S$ \href{mailto:rhine_samajdar@princeton.edu}{rhine\_samajdar@princeton.edu};\\
$^\bullet$ \href{mailto:plopes@quera.com}{plopes@quera.com}; \\
$^\circ$ \href{mailto:kklymko@lbl.gov}{kklymko@lbl.gov}
}
\QuEra

\author{Rhine~Samajdar$^\S$}
\Princeton
\PCTS

\author{Siva~Darbha}
\NERSC

\author{Majd~Hamdan}
\QuEra

\author{Jan~Balewski}
\author{Ermal~Rrapaj}
\NERSC

\author{Sheng-Tao~Wang}
\QuEra

\author{Daan~Camps}
\NERSC

\author{Fangli~Liu}

\author{Pedro~L.~S.~Lopes$^\bullet$}
\QuEra

\author{Katherine~Klymko$^\circ$}
\NERSC

\date{\today}
\linespread{1.0}
\setlength{\belowcaptionskip}{-10pt}

%%%%%%%%%%%%
% Abstract %
%%%%%%%%%%%%

\begin{abstract}
Neutral-atom quantum simulators offer a promising approach to the exploration of strongly interacting many-body systems, with applications spanning condensed matter, statistical mechanics, and high-energy physics. Through a combination of quantum experiments, numerical calculations, and analytical methods, we demonstrate a rich set of phenomena accessible on such quantum simulators by studying an array of Rydberg atoms placed on the Lieb lattice. First, we map out the ground states and phase diagram of the system, identifying a range of density-wave-ordered phases---including a collinear phase stabilized purely by quantum fluctuations---and find good agreement between theory and experiment. Allowing for local control of the detuning field thereafter, we discover a quantum analog of the classical liquid--vapor transition between two density-wave phases distinguished by sublattice occupation, and probe its underlying hysteretic dynamics. Furthermore, we study out-of-equilibrium quantum quenches and observe anomalously slow relaxation dynamics consistent with the kinetic constraints of an emergent string phase. These results highlight how geometric control offered by neutral-atom simulators can extend the frontiers of programmable quantum matter, enabling access to complex phases, metastability, and thermalization dynamics in many-body quantum systems.
\end{abstract}

\maketitle
%TC:endignore

%%%%%%%%%%%%%%%%
% Introduction %
%%%%%%%%%%%%%%%%

\section{Introduction}

A central organizing principle in physics is the notion of universality. For instance, in the context of continuous quantum phase transitions (QPTs), universality allows simple paradigmatic models to capture the long-wavelength phenomenology of a wide range of physical systems---from strongly correlated electrons to high-energy physics~\cite{wilson1972critical, sachdev2001quantum}---independent of their microscopic details. Rapid experimental progress over the last decade has established neutral-atom quantum simulators as a powerful computational platform for investigating quantum matter. These systems offer exceptional control over quantum states, functioning as both coherent analog simulators~\cite{Labuhn2016, Bernien2017} and gate-based quantum computers~\cite{Evered2023, Bluvstein2024}. This versatility has proven invaluable for the study of quantum phases and universality, including the experimental realization of highly entangled topological phases and strongly correlated phenomena predicted by theory~\cite{deLeseleuc2019, Samajdar2021, Verresen2021, Semeghini2021, Kornjaca2023}. Notably, these tools have also enabled quantum simulation of complex dynamical processes, leading to new \textit{discoveries} such as quantum many-body scars~\cite{Bernien2017, Bluvstein2021}. In the study of QPTs and criticality, the ability to probe coherent quantum dynamics has further enabled investigations of fundamentally nonequilibrium phenomena, such as the Kibble-Zurek mechanism~\cite{Keesling2019, Ebadi2021, Zhang2024} and beyond~\cite{samajdar2024quantum, Manovitz_2025}.

Despite this progress, many intriguing theoretical concepts
%many theoretically established concepts of critical phenomena 
remain unexplored with modern quantum simulation tools~\cite{sachdev2001quantum}. Some key open directions include (i) dynamics across first-order quantum phase transitions~\cite{darbha2024false,darbha2024longlived,Papic_nucleation,zhu2024probingfalsevacuumdecay,surace2024stringbreakingdynamicsquantumadiabatic, Luo2025}, (ii) paradigms of slow thermalization~\cite{PhysRevX.10.011055}, and (iii) multicriticality~\cite{RevModPhys.87.457,PhysRevLett.132.226502, wang2025tricriticalkibblezurekscalingrydberg}. These phenomena can have important implications across disciplines, for example, capturing peculiarities of the Higgs transition and other nucleation dynamics~\cite{PhysRevD.106.114507}. Yet these concepts can be computationally intensive to analyze, and have been previously inaccessible with tabletop experiments. This frontier therefore presents a unique opportunity to harness the capabilities of large-scale analog quantum simulators.

In this work, we access the broad range of complex phenomena described above by studying Rydberg atom arrays on the Lieb lattice, finding that the sublattice structure enables dramatically richer physics than that of the much-studied square lattice~\cite{Samajdar2020,Ebadi2021,Scholl2021}. Through a combination of numerical, analytical, and experimental methods, we first map out the phase diagram, finding a range of density-wave-ordered phases, including phases stabilized by quantum fluctuations. Importantly, we obtain good agreement between experiments on the QuEra Aquila platform~\cite{aquila2023quera} and density-matrix renormalization group (DMRG) simulations~\cite{White1992, itensor-r0.3}. We also systematically construct order parameters characterizing the different transitions in the Lieb-lattice phase diagram and analyze boundary-condition effects in the ground-state physics of 2D Rydberg atom arrays, addressing subtleties previously identified in the literature~\cite{Kalinowski2022, miles2023machine, Garnet_chan_rydberg}.

Thereafter, introducing a local detuning field, we experimentally access a quantum analog of the paradigmatic liquid--vapor phase diagram. This phase diagram consists of a first-order transition between two ordered phases that terminates in a quantum critical point. Beyond this point, the two phases are smoothly connected through a crossover driven by quantum fluctuations. Implementing adiabatic state-preparation protocols that traverse the phase transition, we further probe the underlying hysteretic dynamics.

Making use of Aquila's ability to probe real-time evolution, we extend our work to nonequilibrium dynamical behavior via quantum quench experiments. Compared to quenches into a trivial paramagnetic phase, we observe anomalously slow dynamics following quantum quenches into a regime hosting an emergent string phase with strong kinetic constraints. Such slow relaxation dynamics have been previously studied on the kagome lattice via imaginary-time algorithms~\cite{Samajdar2021,Yan2023}, but are challenging to simulate in real time. Taken together, our results highlight both the diversity of phenomena accessible on the Lieb lattice as well as the expansive scope of neutral-atom quantum simulators.

\section{Ground states and phase diagram}

The Aquila analog platform implements the dynamics of an array of Rydberg atoms described by the Hamiltonian~\cite{aquila2023quera}
\begin{equation}
    \begin{aligned}
        H &= \sum_i \frac{\Omega(t)}{2}\left(\ketbra{g_i}{r_i} + \text{H.c}\right) - \Delta(t) \sum_i \hat{n}_i \\
        &+ \sum_i \Delta_{\text{L}}(\mathbf{r}_i, t)\hat{n}_i + \sum_{i<j}V(|\mathbf{r}_i - \mathbf{r}_j|) \hat{n}_i\hat{n}_j \, ,
    \end{aligned}
    \label{eq:ham}
\end{equation}
where $i,j$ enumerate the atoms, $\ket{g_i}$ and $\ket{r_i}$ are the ground and Rydberg states of atom $i$, $\Omega(t)$ is the Rabi frequency, $\Delta(t)$ and $\Delta_{\text{L}}(\mathbf{r}_i, t)$ are the global and local detunings, $V(r) \equiv C_6/r^6$ is the repulsive van der Waals interaction between two atoms excited to the Rydberg state, and $\hat{n}_i=\ketbra{r_i}{r_i}$ is the Rydberg number operator. 

\begin{figure}
%TC:ignore
\centering
\includegraphics[width=\linewidth]{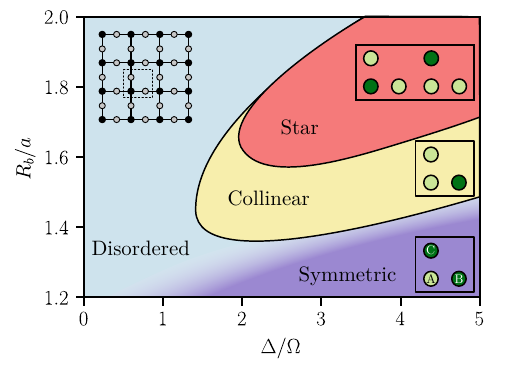}
\caption{\textbf{Numerical phase diagram.} A schematic ground-state phase diagram of the Lieb-lattice Rydberg atom array as a function of $\Delta/\Omega$ and $R_b/a$, with boundaries approximately matching those obtained from DMRG calculations performed on a cylindrical geometry~\cite{White1992, itensor-r0.3}. The disordered/symmetric, collinear, and star phases are depicted in light blue, purple, yellow, and red, respectively. Phase boundaries are marked by solid black lines, and the color gradient between the disordered and symmetric regimes indicates the smooth connection between the two. The top-left inset depicts the Lieb lattice, with A sites  as black circles and B and C sites as gray circles. The unit cell is given by the black dashed square. The insets in the symmetric, collinear, and star regimes each show one unit cell of the corresponding ordering pattern of Rydberg excitations, where light (dark) green corresponds to ground (Rydberg) states. The A, B, and C sublattices are explicitly labeled in the lower-right inset.}
%TC:endignore
\label{fig:Fig1_theory_PD}
\end{figure}

We arrange the Rydberg atoms on a Lieb lattice, depicted in the inset of Fig.~\ref{fig:Fig1_theory_PD}. It is a decorated version of the square lattice with one high-symmetry sublattice, denoted ``A'', and two low-symmetry sublattices denoted ``B'' and ``C''. The A sites form a square lattice, $\vb{R}^{\mathrm{A}}_{i,j} = (2i)a\hat{x} + (2j)a \hat{y}$; the B sublattice sites sit on the midpoints of the horizontal links of A, $\vb{R}^{\mathrm{B}}_{i,j} = (2i + 1)a \hat{x} + (2j)a \hat{y}$; and the C sublattice sites live on the midpoints of the vertical links of A, $\vb{R}^{\mathrm{C}}_{i,j}$\,$=$\,$(2i)a \hat{x} + (2j + 1)a \hat{y}$. The wallpaper group of the Lieb lattice is $p4mm$, consisting of translations along $\hat{x}$ ($T_x$) and $\hat{y}$ ($T_y$), fourfold rotations ($C_4$) and mirror reflections ($\sigma_v$) about the A sublattice sites, and twofold  rotations ($C_2$) around sites on the B and C sublattices.

We begin by setting $\Delta_{\text{L}}$\,$=$\,$0$ and describe the physics of the system in terms of the dimensionless ratio $\Delta/\Omega$, which can be thought of as a chemical potential for Rydberg excitations, and the Rydberg blockade radius, $R_b \equiv (C_6/\Omega)^{1/6}$~\cite{PhysRevLett.85.2208, PhysRevLett.87.037901}. In Fig.~\ref{fig:Fig1_theory_PD}, we plot a schematic ground-state phase diagram depicting the phase boundaries obtained via numerical DMRG simulations (see Appendix~\ref{sec:DMRG})~\cite{White1992, itensor-r0.3} as a function of $\Delta/\Omega$ and $R_b/a$. For small, positive values of $\Delta/\Omega$, the system is in a disordered phase that hosts a low density of Rydberg excitations. For larger values of $\Delta/\Omega$, we find that the system organizes into various density-wave-ordered phases. For each such phase, we plot a unit cell of the associated order in an inset in Fig.~\ref{fig:Fig1_theory_PD} and plot the typical ground-state Rydberg densities in Fig.~\ref{fig:FigE1}. Details of how the numerical phase boundaries are determined can be found in Appendix~\ref{sec:multicritical_point}.

For values of $R_b/a$ in the interval $[1, \sqrt{2}]$, the blockade prevents Rydberg excitations from occupying neighboring sites, so it is energetically most favorable to maximize the number of excitations on the B and C sublattices when the detuning is large and positive. This ordering of the excitations preserves all symmetries of the lattice, so this region, which we aptly label as symmetric in the phase diagram, is continuously connected to the disordered phase without an intervening phase transition. 

Above $R_b/a$\,$\sim$\,$\sqrt{2}$, we find a collinear phase in which the Rydberg excitations populate only either the B or the C sublattice. This phase breaks the rotational symmetry of the lattice and is characterized by a $\mathbb{Z}_2$ order parameter. Finite-size scaling analysis confirms that the boundary between the disordered and collinear phases is a second-order QPT (see Appendix~\ref{sec:multicritical_point}). Above the collinear phase, we observe another density-wave-ordered phase in which the Rydberg excitations form a lattice with primitive vectors $4a\hat{x} \pm 2a\hat{y}$ or $2a\hat{x} \pm 4a\hat{y}$, which we denote as the ``star" phase following the precedent for a similar phase on the square lattice~\cite{Samajdar2020}. The star phase further breaks a translation symmetry in addition to the rotation symmetry broken in the collinear phase. The transition from the disordered phase to the star phase is characterized by a multicomponent order parameter transforming under the symmetry group $D_4 \oplus \mathbb{Z}_2$, whereas the transition from the collinear phase to the star phase involves a single-component $\mathbb{Z}_2$ order parameter.
 In Appendix~\ref{sec:multicritical_point}, we provide a detailed symmetry analysis suggesting that the intersection of the disordered--collinear and collinear--star phase transitions can host a proximate tricritical point~\cite{Landau1937, Park2001, sachdev2002ground, shackleton_deconfined_2021, Kalinowski2022}.

Simply counting excitations in the unit cell and considering only up to second-nearest-neighbor interactions shows that the collinear and star patterns are degenerate in the classical limit ($R_b/a \gg \Delta \gg \Omega$). Taking longer-range interactions into account, however, the star phase becomes classically more stable. Thus, the formation of the collinear phase must be driven by quantum fluctuations stemming from the transverse field $\sim \Omega$. This can be understood in perturbation theory as a consequence of the collinear pattern having Rydberg excitations positioned at lower-coordinated sublattices (B or C). This allows for more quantum fluctuations on the empty sites compared to the star phase, whereas having excitations situated on the A sublattice incurs a greater cost for virtual density fluctuations. As a result, the collinear phase at lower $R_b/a$ is stabilized by quantum fluctuations while the $1/r^6$ tails of the long-range Rydberg interaction stabilize the star phase at higher $R_b/a$.

\begin{figure}[t!]
%TC:ignore
\centering
\includegraphics[width=\linewidth]{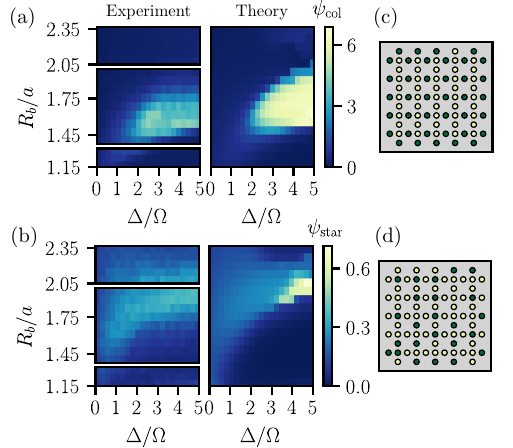}
\caption{\textbf{Experimental ground-state phase diagram.}
The order parameters for the (a) collinear and (b) star phases. The Lieb lattice studied here has $5\times5$ unit cells and boundaries terminating on the B and C sublattices. In each case, the left panel presents the order parameter as measured experimentally on Aquila after adiabatic state preparation, and the right shows numerical DMRG results on the same lattice geometry. The horizontal breaks in the left panels separate experimental runs, with the bottom two regions using $\Omega=2\pi \times 2.5$ MHz and the top region using $\Omega=2\pi \times 1.2$ MHz. Despite varying $\Omega$ to access the full range of $R_b/a$ on Aquila, the results are consistent between the regions, as seen from the continuity of the order parameters. We also present two experimental shots, each of which maximizes the order parameter in the respective phase: (c) collinear order parameter, maximized at $R_b/a=1.52$, $\Delta/\Omega=4.00$, and (d) star order parameter, maximized at $R_b/a=1.99$, $\Delta/\Omega=4.75$.
}
\label{fig:Fig2_exp_PD}
%TC:endignore
\end{figure}

\begin{figure*}
%TC:ignore
\centering
\includegraphics[width=\linewidth]{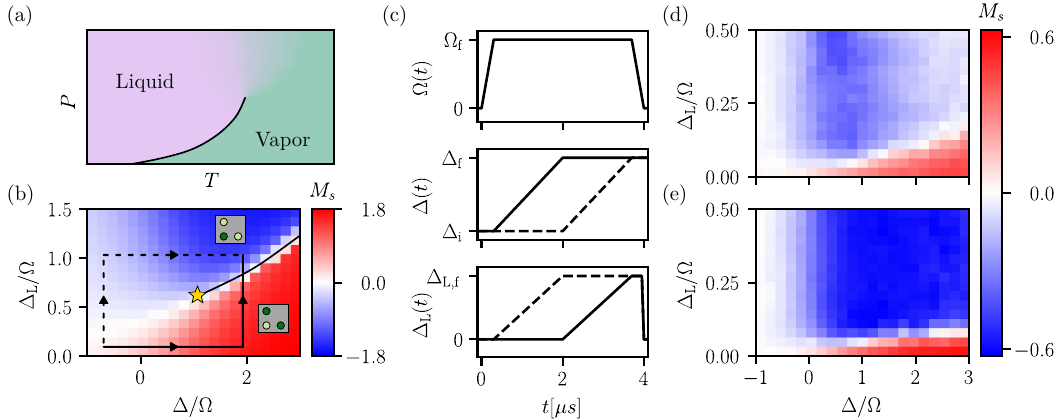}
\caption{\textbf{Experimental quantum liquid--vapor phase diagram.} (a) A schematic of the classical liquid--vapor phase transition in the pressure-temperature plane, wherein a first-order phase transition depicted by the black line terminates in a critical point. (b) The ground-state sublattice magnetization $M_s$ as a function of $\Delta$ and $\Delta_{\text{L}}$, obtained via DMRG with $R_b/a=1.2$. The star marks the approximate position of the critical point, and the line marks the first-order phase transition. The solid and dashed arrows indicate the paths traversed by the global- and local-first protocols. The insets show the density profiles in the corresponding A- and BC-symmetric phases. (c) The Rabi frequency (top), global detuning (middle), and local detuning (bottom) waveforms for the global- and local-first adiabatic state preparation protocols, depicted with solid and dashed lines, respectively. We note that each protocol begins at $\Delta=-2\Omega$ and $\Delta_{\text{L}}=0$, and not the value depicted schematically by the arrows in (b). (d,\,e) The sublattice magnetization measured after adiabatic state preparation with the (d) global-first and (e) local-first protocols. Both the numerics and experiments utilize the same lattice geometry as in Fig.~\ref{fig:Fig2_exp_PD}.
}
\label{fig:Fig3_liquid--vapor}
%TC:endignore
\end{figure*}

Guided by the above results, we experimentally  obtain the Lieb-lattice phase diagram by quasiadiabatically preparing the ground states (see Appendix~\ref{sec:adiabatic}) for each set of parameters $\Delta/\Omega$ and $R_b/a$, and measuring the order parameters for the collinear and star phases. The order parameters $\psi_{\text{col}}$ and $\psi_{\text{star}}$ are chosen so as to minimize interference between each other at high values of $R_b/a$, facilitating identification of the different phases (see Appendix~\ref{sec:OPs} for explicit definitions). We use a lattice with $5 \times 5$ unit cells and decorate the edges so that all boundary sites reside on the B or C sublattices. By choosing an odd number of unit cells along the $\hat{x}$ and $\hat{y}$ directions and terminating the lattice on B and C sites, we ensure compatibility with each of the density-wave-ordered phases. We supplement the experimental results with DMRG simulations performed on an identical lattice. 

In Fig.~\ref{fig:Fig2_exp_PD}, we plot the measured and calculated order parameters for the collinear and star phases (for a similar characterization of the symmetric ordering, refer to Fig.~\ref{fig:FigE1}). We find good qualitative agreement between the experiment and simulation, and are able to identify each of the density-wave-ordered phases. We also show in Fig.~\ref{fig:Fig2_exp_PD} two processed images of experimental shots, each of which maximizes either the collinear or the star order parameter. The collinear shot exhibits nearly perfect ordering across the entire lattice, and the star shot hosts a sizable domain following the expected ordering. It is important to note that the star phase is significantly more challenging to realize on this finite lattice, as the size of the unit cell of the ground state is doubled in one direction; consequently, the star order parameter remains lower in magnitude than the collinear order parameter on the lattice geometry used here, both in experiment and DMRG.

We take special care to evaluate the effect of boundary conditions on the experimental phase diagram, as they were found to be significant in a prior analysis of similar experiments on a square lattice~\cite{Kalinowski2022, Garnet_chan_rydberg}. It is energetically favorable for excitations to occupy boundary sites over bulk sites, as the former have fewer neighbors. This boundary pinning complicates the competition between the disordered and density-wave ordered phases, each of which occupies boundary sites to different extents. For example, the collinear order occupies half of the boundary sites of a BC-terminated lattice, while the star order occupies only one-eighth of them, which shifts the actual phase boundary deep into the star phase relative to that found on the cylindrical geometry. In Appendix~\ref{sec:ABC_experiment}, the results of repeating this experiment using a boundary termination containing A, B, and C sites are shown. There, the phase boundaries are similarly altered. Thus, although we find good agreement between the experiment and equivalent numerics, boundary effects preclude a quantitative study of criticality for the current experimentally accessible lattice sizes~\cite{Keesling2019, Ebadi2021}.

\section{Quantum liquid--vapor transition}
%3/4 page
The phases explored in the previous section all arise from a Hamiltonian that is identical for all sites of the lattice; we now relax this constraint by allowing a local variation in the laser detuning, $\Delta_{\text{L}}(\vb{r}_i,t)$, as described in Eq.~\eqref{eq:ham}. Considering the simplest case, we only vary the detuning within the unit cell such that $\Delta_{\text{L}}(\vb{r}_i, t) = 0$ on A sites and $\Delta_{\text{L}}(\vb{r}_i, t) = \Delta_{\text{L}}(t)$ on B and C sites, thus introducing a homogeneous energy penalty for the symmetric ordering. We explore the phase diagram in the $\Delta$-$\Delta_{\text{L}}$ plane, again through both adiabatic state preparation experiments and numerical DMRG computations. Utilizing the same geometry as before, a lattice with $5\times5$ unit cells decorated such that the boundary consists of only B and C sites, we choose $R_b/a=1.2$ so that sweeping the global detuning with $\Delta_{\text{L}}=0$ accesses the disordered and symmetric regimes. We characterize the resultant phases using the sublattice magnetization,
\begin{equation}
    M_s=\frac{1}{2N}\sum_{i=1}^{N}\expval{\hat{n}_{i,\text{B}} + \hat{n}_{i,\text{C}} - 2 \hat{n}_{i,\text{A}}} \, ,
\end{equation}
where $N$ is the number of unit cells. The sign of $M_s$ can distinguish between different realizations of the symmetric pattern preferring occupation of the high- and low-symmetry sublattices.

The state-preparation protocol is now more complex as the global and local detuning fields must both be slowly ramped up to explore the phase space. We choose to adiabatically turn on the global and local detuning fields sequentially, using half of the available simulation time for each. Two versions of this protocol are possible, in which either the global or local detuning is activated first~[Fig.~\ref{fig:Fig3_liquid--vapor}(c)].

In Fig.~\ref{fig:Fig3_liquid--vapor}(b), we plot the sublattice magnetization of the ground state as obtained numerically through DMRG using the same lattice geometry and parameters as the experiment, retaining nearest-neighbor van der Waals interactions only (see Appendix~\ref{sec:SM_liquid_vapor} for results on a larger cylinder geometry keeping the full $1/r^6$ interactions, which yield the same qualitative phase diagram and confirm the persistence of the terminal critical point). For small values of $\Delta_{\text{L}}$, we find that the sublattice magnetization is positive, indicating that the zero-anisotropy symmetric ordering extends to finite perturbations. For larger values of $\Delta_{\text{L}}$, the sublattice magnetization is instead negative, indicating that Rydberg excitations preferentially occupy the A sublattice. We refer to this phase as the A-symmetric phase, and to the positive-$M_s$ phase as the BC-symmetric phase. At low values of $\Delta$, the sublattice magnetization continuously changes from positive to negative upon increasing $\Delta_{\text{L}}$, indicating a nonsingular crossover between these two phases. We remark that boundary pinning likely favors the BC-symmetric phase (since the A-symmetric phase does not occupy any boundary sites with this termination), so the phase transition will occur at smaller values of $\Delta_{\text{L}}$ in the thermodynamic limit, as shown in Appendix~\ref{sec:SM_liquid_vapor}.

As the global detuning increases, we find that this crossover changes into a first-order transition between the A- and BC-symmetric phases. This is a quantum analog of a feature found in the paradigmatic liquid--vapor and Higgs--electroweak~\cite{Hindmarsh2021} phase diagrams depicted schematically in Fig.~\ref{fig:Fig3_liquid--vapor}(a): a first-order phase transition terminating at a critical point, beyond which the two phases are continuously connected. The sublattice magnetization, $M_s$, plays the role of the $\mathbb{Z}_2$ order parameter in this case. Precisely at the critical endpoint, the \textit{quantum} phase transition is second-order, even though neither of the surrounding (classically stabilized) symmetric phases break any symmetries of the Hamiltonian. Furthermore, the critical exponents at this quantum critical point are naturally identical to those of the classical liquid--vapor critical point, as the two transitions belong to the Ising universality class in $(2$\,$+$\,$1)$ and $(3$\,$+$\,$0)$ spacetime dimensions, respectively.

In Figs.~\ref{fig:Fig3_liquid--vapor}(d) and (e), we plot the sublattice magnetizations measured experimentally after attempting adiabatic state preparation using the global-first and local-first protocols, respectively. We note that the phase transition appears to occur at a lower value of $\Delta_{\text{L}}$ in the experimental data than predicted by the equilibrium DMRG ground state. We attribute this offset to a combination of two effects, both consistent with the data: (i) a possible miscalibration of the local detuning field on the device, which uniformly rescales the experimental $\Delta_{\text{L}}$ axis, and (ii) finite-time effects of the sequential adiabatic protocol, since neither protocol traces a path through the equilibrium $(\Delta, \Delta_{\text{L}})$ plane shown in Fig.~\ref{fig:Fig3_liquid--vapor}(b). Importantly, however, the position of the first-order line itself is reflected internally in the experimental data, independent of the absolute calibration of $\Delta_{\text{L}}$: in Fig.~\ref{fig:Fig3_liquid--vapor}(d), we see that the global-first protocol successfully prepares the BC-symmetric phase, appearing as the bright red triangular region in the bottom right.
There is a strip of negative $M_s$ above the BC-symmetric phase between $0\lesssim\Delta/\Omega \lesssim 1.5$, but the global-first protocol fails to prepare the A-symmetric phase for any larger value of $\Delta$. Indeed, adiabatic state preparation protocols are expected to fail when passing through a first-order transition~\cite{RevModPhys.90.015002}.

\begin{figure*}[tb]
%TC:ignore
\centering
\includegraphics[width=\linewidth]{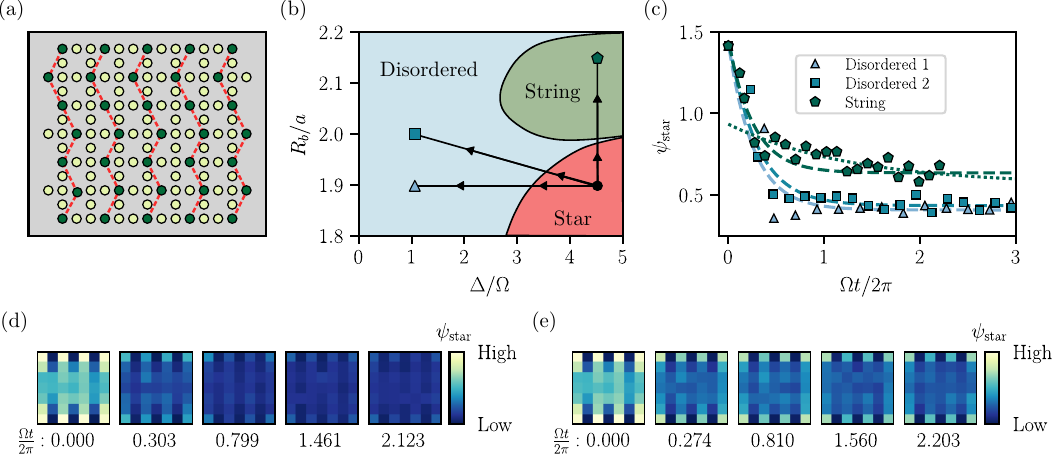}
\caption{\textbf{Experimental observation of kinetically constrained dynamics.}
(a) The lattice geometry employed for the quench experiment. The color depicts a representative configuration of ground-state (light green) and excited (dark green) atoms consistent with the kinetic constraints characteristic of the string phase. Red dashed lines serve as a visual guide to highlight how the excitations align into string-like patterns. 
(b) Schematic phase diagram, informed by DMRG, illustrating the predicted disordered, star, and string phases. The black circle denotes the parameters corresponding to the initial star-ordered state prepared in the experiment, while the arrows indicate the three quench protocols.
(c) Star order parameter plotted as a function of post-quench evolution time, scaled by the post-quench value of $\Omega$. All three quenches display an initial transient decay; however, only the quench into the string phase exhibits a second, anomalously slow decay at late times. In contrast, the quenches into the disordered phase are followed by a rapid equilibration of the star order parameter to a constant value. The blue dashed lines represent exponential fits for the two star-to-disordered quenches. For the quench into the string phase, the green dashed and dotted lines correspond to exponential fits for the transient decay ($\Omega t/2\pi < 0.6$) and long-time decay ($\Omega t/2\pi > 0.6$), respectively. Notably, the long-time decay in this case is approximately five times slower than that observed for the star-to-disordered quenches. 
(d, e) Spatially resolved star order parameter for (d) the high-$R_b/a$ disordered quench, corresponding to the blue square in panel (b), and (e) the string quench. The labels below each plot indicate the respective post-quench measurement times.
}
\label{fig:Fig4_glassy}
%TC:endignore
\end{figure*}

Conversely, in Fig.~\ref{fig:Fig3_liquid--vapor}(e), we observe that the local-first protocol can prepare the entire A-symmetric phase, which appears as the large blue region, but it fails to prepare the top of the BC-symmetric phase above $\Delta_{\text{L}}/\Omega\sim0.07$: the top of the red region in the bottom-right of Fig.~\ref{fig:Fig3_liquid--vapor}(d) has disappeared. To clarify the success and failure of these protocols in different regions, we plot in Fig.~\ref{fig:FigE3} the average Rydberg densities obtained after each protocol in three regions of the phase diagram: (1) at low $\Delta$ and $\Delta_{\text{L}}$, away from the critical point; (2) close to the first-order transition deep in the A-symmetric phase; and (3) close to the first-order transition deep in the BC-symmetric phase. The final state strongly depends on the path taken through the phase diagram, and the characteristic hysteretic behavior confirms both the first-order phase transition and the presence of the terminal critical point predicted by the numerics. Our experiments thus demonstrate how Rydberg atom arrays provide a tunable tabletop platform for the study of metastability and other phenomena specific to first-order transitions~\cite{darbha2024false, Papic_nucleation}.

\section{Slow quantum relaxation dynamics}

In addition to the equilibrium quantum phases and phase transitions studied thus far, a novel opportunity afforded by quantum simulation lies in understanding the collective quantum dynamics of nonequilibrium many-body systems. An important generic problem in this regard is the question of how a quantum system
approaches equilibrium---or relaxes---following an external perturbation.

Recently, large-scale quantum Monte Carlo (QMC) studies on a kagome lattice of neutral atoms provided evidence suggestive of anomalously slow relaxation dynamics~\cite{Yan2023} and a rugged energy landscape~\cite{hibatallah2024}, reminiscent of classical glasses. The origin of these slow dynamics can be understood from the underlying kinetic constraints~\cite{ritort2003glassy,PhysRevLett.121.040603} in the system.
Such dynamical constraints naturally arise in the so-called string phase of the kagome-lattice Rydberg array~\cite{Samajdar2021}; here, the (ordered) ground state is composed of a superposition of exponentially many configurations in which Rydberg excitations are arranged in extended ``strings'' that traverse the lattice. Due to the close-packed nature of the strings, the strong blockade
forbids several processes that rearrange Rydberg excitations, resulting in a drastic reduction of the number of allowed configurations that the dynamics can explore~\cite{angelone2016superglass,feldmeier2019emergent}.
Motivated by the fact that the kagome lattice can be converted into the Lieb lattice by a shear deformation, and since the latter also hosts a similar string phase, we attempt to uncover signatures of such slow dynamics in real-time evolution, as opposed to the indirect imaginary-time dynamics accessible to QMC.

To do so, we return to the homogeneous case with $\Delta_{\text{L}}=0$ and experimentally study the dynamics of the system after a controlled quantum quench~\cite{mitra2018quantum} in which an ordered phase is prepared and then driven into the putative kinetically constrained phase. We plot a schematic phase diagram in Fig.~\ref{fig:Fig4_glassy}(b) showing the disordered, star, and string phases, along with three quench protocols. The location of the string phase is determined via DMRG (see Appendix~\ref{sec:multicritical_point} for details). Each protocol begins with the adiabatic preparation of the star phase in the same deterministic symmetry-broken state, followed by a sudden quench of $\Delta$ (and possibly $\Omega$), terminating inside either the disordered phase or the string phase. The lattice, depicted in Fig.~\ref{fig:Fig4_glassy}(a), was chosen to optimize the preparation of the star phase in order to compensate for the reduced time allowed for state preparation. The dimensions and the boundary terminations of the lattice are engineered such that it is commensurate with a particular orientation of the star order and that the maximum possible number of boundary sites are occupied, exploiting boundary pinning to maximize the order parameter. We consider two quenches into the disordered phase at different values of $\Omega$ (and consequently, different $R_b/a$) to ensure that any differences in the dynamical response between the disordered and string quenches do not arise simply due to the change in $\Omega$. The system is allowed to evolve for a time $t$ after the quench before the star order parameter is measured. Details of the quench protocols are provided in Appendix~\ref{sec:adiabatic}.

In Fig.~\ref{fig:Fig4_glassy}(c), we plot the star order parameter after each quench as a function of the evolution time scaled by the post-quench Rabi frequency, $\Omega$, including the measured pre-quench value of the star order parameter at $t=0$. All three quenches display a sharp transient decay, which we fit to an exponential (shown with dashed lines). The order parameters in the case of the star-to-disordered quenches are static after this transient decay, quickly equilibrating to some finite thermal expectation value. In stark contrast, the quench into the numerically predicted string phase demonstrates a second long-time exponential decay at a much slower rate. The strikingly different dynamics between the disordered and string quenches are also highlighted in Figs.~\ref{fig:Fig4_glassy}(d, e), which plot snapshots of a spatially resolved star order parameter (defined in Appendix~\ref{sec:OPs}) as a function of time.

This anomalous long-time decay is consistent with the kinetic constraints expected to be operative in the string phase. To make this connection more quantitative, we identify the leading dynamical process responsible for breaking down star order in the constrained subspace and estimate its perturbative timescale. In the relevant blockade regime, $R_b/a \in (2, \sqrt{5})$, the blockade forbids occupation of any pair of sites separated by $\sqrt{2}\,a$ or less, so that ground states are organized into parallel strings of Rydberg excitations spaced by $3a$ (see Appendix~\ref{sec:multicritical_point} for the explicit string order parameter $\psi^{}_{\text{string}}$ and its DMRG characterization on the cylinder). Within this manifold of string configurations, an off-resonant Rabi-driven transition on any single atom generically either creates a blockade violation or removes an excitation from a string, incurring an energy cost of order $\Delta$. The leading resonant rearrangement is instead a correlated process involving several atoms, a ring-exchange-like move that bends a string by displacing one node by a single unit cell while preserving both the blockade constraints and the total Rydberg density. This appears at $k$th order in $\Omega/\Delta$ in degenerate perturbation theory, where $k\geq 3$ is set by the minimum number of intermediate blockade-violating states that must be virtually traversed. The associated rate scales as $\Gamma_{\text{string}}\sim\Omega(\Omega/\Delta)^{2(k-1)}$, parametrically smaller than the bare $\Omega$ rate that governs equilibration into the disordered phase.

With the post-quench parameters of the string quench [Table~\ref{table:T1_quench}, $\Delta/\Omega=4.5$], even a $k=3$ process gives a slowdown of $(\Omega/\Delta)^4\approx 2\times10^{-3}$ in the bulk; on the modest experimental lattice, however, dynamical processes seeded at the boundary contribute significantly, and the observed factor of $\sim\!5$ between the long- and short-time decay rates in Fig.~\ref{fig:Fig4_glassy}(c) is broadly compatible with such a low-order constrained process. We emphasize that, while a precise extraction of $k$ would require a finer scan of $\Delta/\Omega$ on a larger lattice, the qualitative dichotomy between the rapid star-to-disordered relaxation and the protracted star-to-string relaxation provides direct real-time evidence---complementary to the imaginary-time QMC results of Refs.~\cite{Samajdar2021,Yan2023}---that the kinetic constraints inherent to the string phase imprint themselves on the post-quench dynamics. The constrained dynamics therefore manifest as an anomalously slow decay of the order parameter after the quench, which is consistent with our observations here upon measuring $\psi^{}_{\text{star}}$ in repeated experiments with increasing delays between the quench and the measurement.

\section{Discussion and outlook}
In this work, we establish the Lieb lattice as a robust, experimentally accessible setting for the quantum simulation of strongly correlated phenomena, advancing the science of Rydberg atom arrays in three concrete ways. First, we demonstrate that the decorated sublattice structure of the Lieb lattice stabilizes a quantum-fluctuation-driven collinear phase that has \textit{no classical analog}; the perturbative analysis of Eqs.~\eqref{eq:cldelECS}--\eqref{eq:delECS} traces this phase to the asymmetric cost of virtual density fluctuations on the high-coordination A sublattice versus the low-coordination B/C sublattices, and the experimental phase diagram of Fig.~\ref{fig:Fig2_exp_PD} confirms its existence at low $R_b/a$ on hardware. The proximity of this collinear phase to the classically degenerate star phase furthermore generates a multicomponent $D_4\oplus\mathbb{Z}_2$ order parameter and at least one tricritical point, opening a direct experimental route to quantum multicriticality~\cite{RevModPhys.87.457,PhysRevLett.132.226502, wang2025tricriticalkibblezurekscalingrydberg} that has historically been difficult to realize in solid-state systems~\cite{Friedemann2018, Wang2021}.

Second, by introducing a sublattice-resolved local detuning, we realize a quantum analog of the classical liquid--vapor phase diagram in which a first-order transition between two density-wave phases terminates at a quantum critical endpoint~\cite{Hindmarsh2021,PhysRevD.106.114507}. The hysteretic response observed in Figs.~\ref{fig:Fig3_liquid--vapor}(d,\,e) under inequivalent state-preparation paths, reinforced by both DMRG on a larger cylinder and an analytical product-state model in Appendix~\ref{sec:SM_liquid_vapor}, confirms the first-order character of the transition and demonstrates direct quantum-simulator access to metastability and path-dependent dynamics in a many-body setting. This positions the Lieb-lattice array as a tunable platform for testing classical and quantum nucleation theories, including false-vacuum decay~\cite{darbha2024false, darbha2024longlived, surace2024stringbreakingdynamicsquantumadiabatic, Papic_nucleation, Luo2025} and the quantum analog of hysteresis loops familiar from classical first-order transitions.

Third, our quench experiments provide direct, real-time evidence that the kinetic constraints of an emergent string phase produce parametrically slow relaxation, complementing the imaginary-time QMC studies of related kagome systems~\cite{Samajdar2021, Yan2023, hibatallah2024} that can only access these constraints indirectly. The factor-of-five separation between the post-quench transient and the long-time decay rate, taken together with the perturbative analysis above, makes the Lieb lattice a concrete platform for studying frustration-driven slow dynamics, a route toward quantum glassiness, and, more broadly, thermalization and ergodicity-breaking in strongly correlated quantum systems~\cite{Papic2021, Moudgalya2022, ritort2003glassy, PhysRevLett.121.040603}.

These physical findings are enabled by, and in turn motivate the further development of, three experimental capabilities that we leverage throughout this work: (i) lattice decoration, as employed in our strategic modification of lattice connectivity from the square to the Lieb; (ii) single-site addressability of local detuning fields, which makes the liquid--vapor experiment of Fig.~\ref{fig:Fig3_liquid--vapor} possible; and (iii) boundary-condition engineering, i.e., the controlled variation of system terminations~\cite{Kalinowski2022}. The ability to precisely manipulate boundary conditions in fact proves essential for analyzing state-preparation protocols and comparing experimental findings with theoretical predictions, as shown in Fig.~\ref{fig:Fig2_exp_PD} and Fig.~\ref{fig:FigS4_abc_OP}; these protocols also establish a systematic route to studying surface criticality in future investigations~\cite{Kalinowski2022}. The scalability of current Rydberg platforms suggests that increasing the number of atoms by an order of magnitude is well within reach~\cite{PhysRevLett.130.180601, Pause:24, Manetsch2024}, which would enable quantitative finite-size scaling at the disordered--collinear and liquid--vapor critical points and a finer dissection of the perturbative structure of the constrained string-phase dynamics. Combined with quantum quenches and adiabatic traversals through first-order transitions~\cite{PhysRevResearch.6.013271,lukin2024}, our work points toward a unified experimental program for probing exotic physics beyond conventional symmetry breaking---highly frustrated magnetism~\cite{Semeghini2021,Samajdar2021, Verresen2021,bintz2024,tian2025}, multicriticality, glassy quantum matter, and the classical computational complexity of simulating nonequilibrium quantum systems~\cite{Daley2022Practical, Kashyap2025}---on a single tabletop architecture.

\begin{acknowledgments}
We thank David Huse and Chris Laumann for useful discussions, Johannes Blaschke for assistance with numerics, and Pavel Dolgirev for careful review of the manuscript. This research was supported by the U.S. Department of Energy (DOE) under Contract No. DE-AC02-05CH11231, through the National Energy Research Scientific Computing Center (NERSC), an Office of Science User Facility located at Lawrence Berkeley National Laboratory. R.S. is supported by the Princeton Quantum Initiative Fellowship. This work was performed in part at the Aspen Center for Physics, which is supported by a grant from the Simons Foundation (1161654, Troyer).
\end{acknowledgments}
\newpage

\appendix
% \section*{Methods}
% \label{sec:methods}

\begin{figure*}[ht]
\centering
\includegraphics[width=\linewidth]{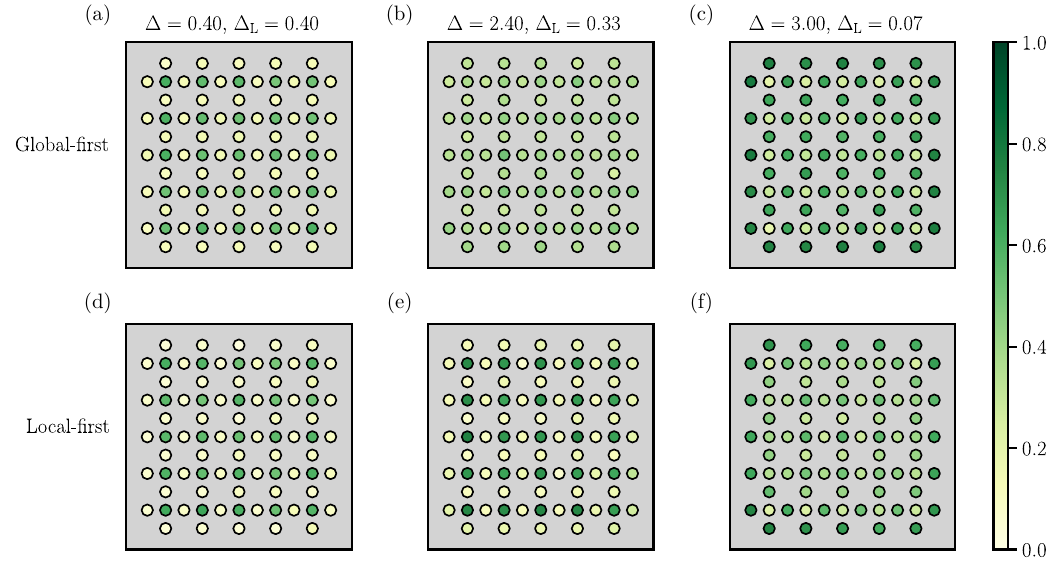}
\caption{\textbf{Experimental Rydberg excitation densities after global- and local-first adiabatic state preparation.} Average Rydberg densities obtained from adiabatic state preparation using the global-first (a--c) and local-first (d--f) protocols. Preparing the A-symmetric phase at $\Delta=0.40$, $\Delta_{\text{L}}=0.40$ succeeds using both protocols, as seen by comparing (a) and (d). Attempting to prepare the A-symmetric phase at $\Delta=2.40$, $\Delta_{\text{L}}=0.33$ with the (b) global-first protocol fails because this entails crossing the phase transition. The (e) local-first protocol avoids the phase transition and succeeds. The reverse is true for preparing the BC-symmetric phase at $\Delta=3.00$, $\Delta_{\text{L}}=0.07$: the (c) global-first protocol successfully prepares the BC-symmetric phase, while the (f) local-first protocol scrambles the phase.\\
}
\label{fig:FigE3}
\end{figure*}

\section{Adiabatic state preparation}
\label{sec:adiabatic}
The general protocol employed for adiabatic state preparation on Aquila is as follows. First, the detuning is initialized to a large negative value, preparing the system in the all-$\rvert g \rangle$ state. Next, the Rabi frequency is rapidly increased from zero to the desired value. The detuning is then slowly linearly ramped up to its end value. Finally, the Rabi drive is quickly ramped back to zero, after which the state is measured.

For preparing the ground states in Fig.~\ref{fig:Fig2_exp_PD} and Fig.~\ref{fig:FigE3}, we use Rabi drives with $\Omega=2\pi\times 2.5$ MHz and $2\pi\times 1.2$ MHz, and a ramp time of $0.3\,\mu \text{s}$. The detuning was initialized to $\Delta=-2\Omega$ and ramped up over the remaining $3.4\,\mu \text{s}$ of the total $4\,\mu \text{s}$ evolution time available on Aquila. We collected 500 shots for each data point, discarding any shots for which fewer than 98\% of the atoms were loaded properly. This typically resulted in discarding 5--20\% of the shots. The remaining shots were used to estimate the values of observables in the computational basis, which is sufficient to estimate the order parameters of all phases with sub-$5\%$ precision. 

The global- and local-first state-preparation protocols described in Fig.~\ref{fig:Fig3_liquid--vapor} also used $\Omega=2\pi\times 2.5 $ MHz, ramping up and down over $0.3\,\mu \text{s}$. The global detuning was again initialized to $-2\Omega$, and the local detuning was initialized to $0$. The global and local detunings were ramped up to their final values sequentially, with the order reversed between the two protocols, each over $1.7\,\mu \text{s}$.

\begin{figure*}[ht]
\centering
\includegraphics[width=\linewidth]{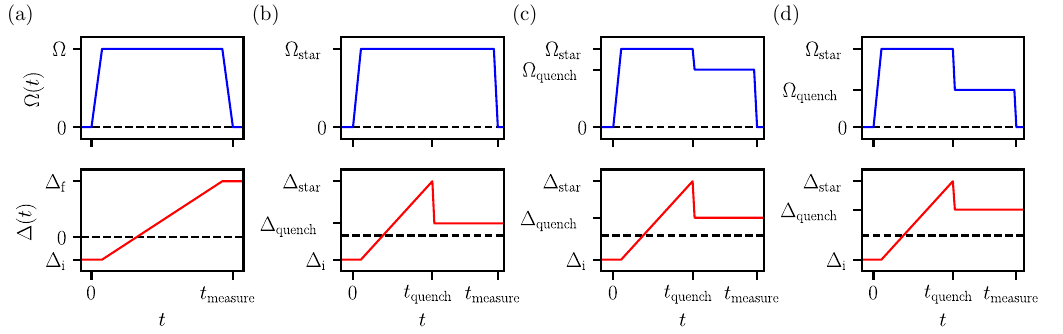}
\caption{\textbf{Experimental waveforms.}
Representative schematics of the waveforms for $\Omega(t)$ and $\Delta(t)$ used to implement (a) the adiabatic state preparation protocol, and the quenches from the star phase to the (b) low-$R_b/a$ disordered phase, (c) the high-$R_b/a$ disordered phase, and (d) the string phase.
}
\label{fig:FigE2}
\end{figure*}

The protocol for the quench experiments described in Fig.~\ref{fig:Fig4_glassy} required steps beyond the simple linear ramps. For all quenches, the preparation of the initial star phase was identical. The Rabi drive was ramped up to $\Omega=2\pi\times 2.5$ MHz over $0.2\,\mu \text{s}$ and the detuning was ramped up from $\Delta=-2\Omega$ to $4.5\Omega$ over $1.8\,\mu \text{s}$. The systems were quenched over $0.05\,\mu \text{s}$ and then allowed to evolve for a variable amount of time. Finally, $\Omega$ was ramped down over $0.1$, $0.075$, and $0.05\,\mu \text{s}$ for the low-$R_b/a$ disorder, high-$R_b/a$ disorder, and string quenches, respectively. The post-quench values of $\Omega$, $R_b/a$, and $\Delta$ for each quench are reported in Table~\ref{table:T1_quench}. Schematic depictions of the waveforms for the adiabatic state preparation protocol and quench protocols are provided in Fig.~\ref{fig:FigE2}.

\setlength{\tabcolsep}{12.5pt}
\begin{table}[h]
\begin{tabular}{c|cccc}
\hline
Quench & $\Omega$ ($2\pi\times$ MHz) & $R_b/a$ & $\Delta/\Omega$ \\
\hline
\hline
Disorder (1) & 2.50           & 1.90     & 1.00                                                \\
Disorder (2) & 1.84           & 2.00     & 2.00                                                \\
String       & 1.19           & 2.15    & 4.50  \\
\hline
\end{tabular}
\caption{The post-quench values of $\Omega$, $R_b/a$, and $\Delta$ for each quench.}
\label{table:T1_quench}
\end{table}

\section{Many-body order parameters}
\label{sec:OPs}
Here, we define the order parameters used to identify the symmetric, collinear, and star phases. All these definitions proceed from the symmetry analysis showcased in Appendix~\ref{sec:multicritical_point}. For analyzing experimental results, however, we co-design the order parameters to provide the best contrast between proximate phases. The collinear and star phases exhibit multiple degenerate symmetry-breaking ground states. As such, we employ order parameters constructed from two-point correlation functions, which can distinguish between phases while accounting for such degeneracies (see Appendix~\ref{sec:multicritical_point}). All the order parameters that we consider are normalized such that their value is 0 in the disordered phase and $N$ in the perfectly ordered classical state, where $N$ is the number of unit cells. To minimize boundary effects, we do not include excitations on sites decorating the boundary when computing these order parameters.

The symmetric phase, as the name suggests, does not break any lattice symmetries and thus, strictly speaking, does not have an order parameter (in the absence of a local detuning field). Nonetheless, we define a metric that we refer to as a ``symmetric order parameter'' as the average two-point correlation function of the sublattice magnetization,
\begin{equation}
    \psi^{}_{\text{sym}} \equiv \frac{1}{4N}\sum_{i, j}\expval{\hat{M}^{}_i\hat{M}^{}_j},
\end{equation}
where the local sublattice magnetization operator for site $i$ is defined as $\hat{M}_i\equiv\hat{n}_{i,\text{B}} + \hat{n}_{i,\text{C}} - 2 \hat{n}_{i,\text{A}}$. Despite not being a true order parameter, $\psi^{}_{\text{sym}}$ indeed quantifies the degree to which the state observed corresponds to the ordering pattern of the relevant classical symmetric product state.

The collinear phase does not host excitations on the A sublattice and is characterized by an unequal occupation of the B and C sublattices. Therefore, we define the BC-sublattice magnetization as $\hat{M}^{\text{BC}}_i=\hat{n}^{}_{i,\text{B}} - \hat{n}^{}_{i,\text{C}}$, and take the average of its correlation function as the order parameter for the collinear phase,
\begin{equation}
    \psi^{}_{\text{col}} \equiv \frac{1}{N}\sum_{i, j}\expval{\hat{M}^{\text{BC}}_i\hat{M}^{\text{BC}}_j}.
\end{equation}

Similarly, the star phase is also characterized by selective occupation of either the B or C sublattice, but also hosts Rydberg excitations on some A sites; the expectation value of $\hat{M}^{\text{BC}}_i$ does not vanish in the star phase and can exhibit a finite value of $\psi^{}_{\text{col}}$. To avoid further interference between the order parameters, we define the star order parameter using only the density on the A sublattice. Depending on the orientation of the ordering, the A sublattice excitations in the star phase oscillate in space with wavevector $\vb{k}=(\pi/2a, 0)$ or $\vb{k}=(0, \pi/2a).$ This is reflected as peaks in the density-density correlation function,
\begin{equation}
    G^{\text{A},\text{A}}(\vb{k}) \equiv \sum_{i,j=1}^N\expval{\hat{n}^{}_{i,\mathrm{A}}\hat{n}^{}_{j,\mathrm{A}}}e^{-i\vb{k}\cdot\big(\vb{R}^{\mathrm{A}}_i-\vb{R}^{\mathrm{A}}_j\big)}.
\end{equation}
Accordingly, we define the star order parameter as
\begin{equation}
    \psi^{}_{\text{star}} \equiv \frac{1}{N}\left[G^{\text{A},\text{A}}(\pi/2a,0) + G^{\text{A},\text{A}}(0, \pi/2a)\right].
\end{equation}
Because each of these order parameters is defined as a correlation function of operators defined over unit cells, we include only sites contained within whole unit cells when evaluating them. Accordingly, we exclude sites on the left and bottom edges of the BC-terminated lattice, as well as on the right and top edges of the ABC-terminated lattice. We similarly exclude sites added to the cylindrical geometry to make the boundary terminations symmetric.

We further define a local version of the star order parameter as
\begin{equation}
    \psi^{}_{\text{star},i} \equiv \frac{1}{N}\left[G^{\text{A},\text{A}}_i(\pi/2a,0) + G^{\text{A},\text{A}}_i(0, \pi/2a)\right],
\end{equation}
for which we use a modified density-density correlation function
\begin{equation}
    G^{\text{A},\text{A}}_i(\vb{k}) \equiv \sum_{j=1}^N\expval{\hat{n}^{}_{i,\mathrm{A}}\hat{n}^{}_{j,\mathrm{A}}}e^{-i\vb{k}\cdot(\vb{R}_i-\vb{R}_j)}.
\end{equation}
Unlike for the other order parameters, we do include the boundary-decorating A sites when computing the local star order parameter, as it is difficult to visually resolve otherwise. 

\begin{figure*}[tb]
\centering
\includegraphics[width=\linewidth]{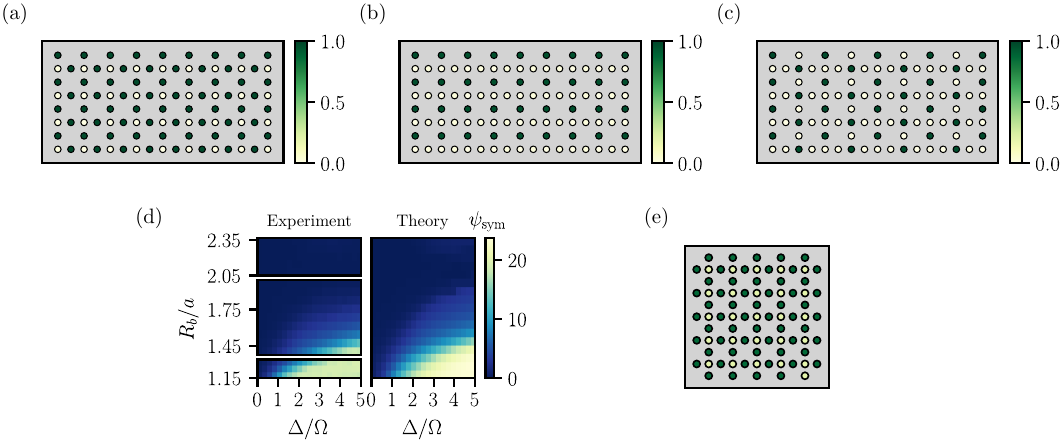}
\caption{\textbf{Density of Rydberg excitations from DMRG.}
The ground-state Rydberg densities computed using DMRG on a cylindrical geometry deep in the (a) BC-symmetric, (b) collinear, and (c) star phases. This data was obtained from the same calculations as Fig.~\ref{fig:Fig1_theory_PD} of the main text. The symmetric order parameter is plotted in (d) and was obtained from the same experiment and simulation as the collinear and star order parameters presented in Fig.~\ref{fig:Fig2_exp_PD}. The shot maximizing the symmetric order parameter at $R_b/a=1.37$, $\Delta/\Omega=4.75$ is plotted in (e).
}
\label{fig:FigE1}
\end{figure*}

\section{Numerical DMRG simulations}
\label{sec:DMRG}
All the ground-state phase diagrams presented in this work were obtained numerically via the density matrix renormalization group (DMRG) algorithm~\cite{dmrg_1, dmrg_2, dmrg_3} as implemented in the \textsc{ITensor} library~\cite{itensor, itensor-r0.3}. We use a singular value cutoff of $10^{-10}$ and consider the ground state to be converged when the energy and entanglement entropy vary by less than $10^{-6}$ from the previous bond-dimension sweep, and the smallest truncation error is less than $10^{-8}$. We begin each calculation with a small bond dimension, $\chi\sim10$, and ramp it up over approximately 100 sweeps to a maximum value of $\chi=1600$. To avoid being trapped in local minima, we add a small amount of noise to the density matrix after each sweep and slowly ramp down the noise to zero over the first $\sim50$ sweeps. Regardless of convergence criteria, we always perform at least 60 sweeps.

The ground-state phase diagram in Fig.~\ref{fig:Fig1_theory_PD} was obtained using a cylindrical geometry with $8\times4$ unit cells, with this aspect ratio chosen so as to minimize finite-size corrections in $1/L_y$~\cite{PhysRevLett.99.127004}. We decorated one end of the cylinder with an additional row of A and C sublattice sites such that the boundaries were symmetric and compatible with the symmetric, collinear, and star order parameters. We retained van der Waals interactions up to a distance of $4a$ in the Hamiltonian, which faithfully captured the possible density-wave-ordered phases over the full range of $R_b/a$ studied. The phase boundaries were determined qualitatively from maxima in the half-cut bipartite entanglement entropy, $S=-\text{Tr}(\rho_r\ln\rho_r)$, where $\rho_r$ is the reduced density matrix, and the circumferential cut was taken at the center of the cylinder. For the calculations in Fig.~\ref{fig:Fig2_exp_PD} and Fig.~\ref{fig:Fig3_liquid--vapor}(b), the DMRG numerics were performed on the same geometry as the experiment, a $5\times5$ lattice with open boundaries decorated to terminate on B and C sites. The calculations in Fig.~\ref{fig:FigS4_abc_OP} were performed on a $4\times4$ lattice with open boundaries formed by A, B, and C sites, again matching the experiment.

\onecolumngrid

\section{Analysis of the Lieb-lattice phase diagram}
\label{sec:multicritical_point}

\subsection{Stability of the collinear phase}
The existence of the collinear region in the phase diagram is driven by quantum fluctuations, akin to the striated phase on the square lattice~\cite{Ebadi2021}. Indeed, while both the star and collinear phases satisfy the first- and second-neighbor blockade constraints, the interaction tails prefer the star phase classically. The classical energy difference between the star and collinear phases with up to fifth-neighbor interactions included is
\begin{equation}\label{eq:cldelECS}
    \delta E^{}_{\text{col} - \text{star, cl}}\approx\frac{1}{2}V(2a)-V(\sqrt{5}a)+V(2\sqrt{2}a) > 0.
\end{equation}

With quantum fluctuations driven by the transverse-field term in the Hamiltonian, both phases lower their energy by allowing finite occupation of classically ``empty'' sites and reducing the occupation of classically ``filled'' sites. However, the collinear phase gains more energy from such quantum fluctuations as empty B/C-sublattice sites have a lower cost of fluctuations than equivalent empty sites in the star phase. We can approximately capture this stabilization of the collinear phase in second-order perturbation theory, where the energy difference becomes
\begin{equation}
\label{eq:qdelECS}
    \delta E^{(2)}_{\text{col} - \text{star, q}}\approx -\left(\frac{\Omega}{2}\right)^2\left(\frac{1}{4V(\sqrt{2}a)-\Delta} - \frac{1}{V(a)+2V(\sqrt{2}a)-\Delta} \right) < 0.
\end{equation}

Putting the two energy contributions together, we get an approximate star-collinear energy difference, up to second order in the Rabi frequency, of
\begin{equation}\label{eq:delECS}
    \delta \tilde{E}^{(2)}_{\text{col} - \text{star}}\approx \frac{113}{64000}\tilde{R}_b^6-\frac{3}{16}\tilde{R}_b^{-6}\left(\frac{1}{2}-\frac{\tilde{\Delta}}{\tilde{R}_b^6}\right)^{-1}\left(\frac{5}{4}-\frac{\tilde{\Delta}}{\tilde{R}_b^6}\right)^{-1},
\end{equation}
where we have normalized $\delta \tilde{E}=\delta E/\Omega$, $\tilde{\Delta}=\Delta/\Omega$, $\tilde{R}_b=R_b/a$, and $\Omega=1$. It is evident that for large $\tilde{R}_b$, the classical interaction tails, scaling as $\tilde{R}_b^6$, eventually stabilize the star phase. In contrast, quantum fluctuations stabilize the collinear phase at lower $\tilde{R}_b$. A more detailed approximation of the star--collinear phase boundary can be found by considering a product-state \textit{Ansatz}, similar to the analysis of quantum liquid--vapor criticality in Appendix~\ref{sec:SM_liquid_vapor}. However, the high von Neumann entanglement entropy in the vicinity of the point where the phases meet highlights the importance of a fully quantum treatment beyond second-order perturbation theory and product-state approximations.

\subsection{Numerical determination of phase boundaries}
\begin{figure}[t]
\centering
\includegraphics[width=\linewidth]{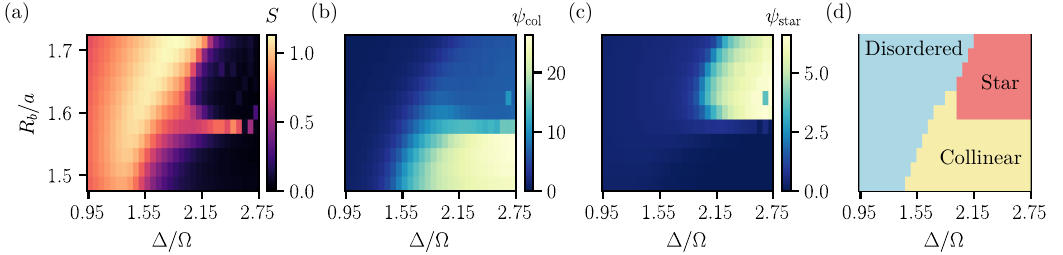}
\caption{\textbf{Numerical phase diagram.}
The (a) entanglement entropy, (b) collinear order parameter, and (c) star order parameter, as obtained from DMRG on a cylindrical geometry, zoomed in around the point where the three phases intersect. The phase diagram in (d) is constructed by coloring the area red if the star order parameter is greater than 25\% of its maximum value in the field of view, yellow if the same is true of the collinear order parameter, and blue otherwise.
}
\label{fig:FigS1_collinear_wedge}
\end{figure}

Here, we provide a more detailed look into the region of the phase diagram depicted in Fig.~\ref{fig:Fig1_theory_PD} where the disordered, collinear, and star regions meet using density-matrix renormalization group (DMRG) calculations. In Fig.~\ref{fig:FigS1_collinear_wedge}, we plot the entanglement entropy, the collinear order parameter, and the star order parameter, along with the  numerically obtained phase boundaries. Upon examining these three quantities, it is clear that the collinear phase wraps around the left boundary of the star phase. The star and collinear order parameters in this wedge-shaped region evolve smoothly, with the underlying ground states exhibiting compatible density-wave profiles between the resulting collinear and star orderings. The heuristic phase diagram, obtained by coloring each region according to whether the appropriate order parameter is greater than 25\% of its maximum value, also demonstrates the presence of a wedge of the collinear phase between the disordered and star phases. As such, the disordered--collinear and collinear--star phase boundaries must intersect, above which point they merge into the disordered--star boundary. In the following discussion, we argue that the region surrounding this intersection likely hosts a tricritical point.

\subsection{Symmetry analysis of the Lieb-lattice phases}
\label{sec:symmetry}

\begin{figure}
\centering
\includegraphics[width=1.0\textwidth]{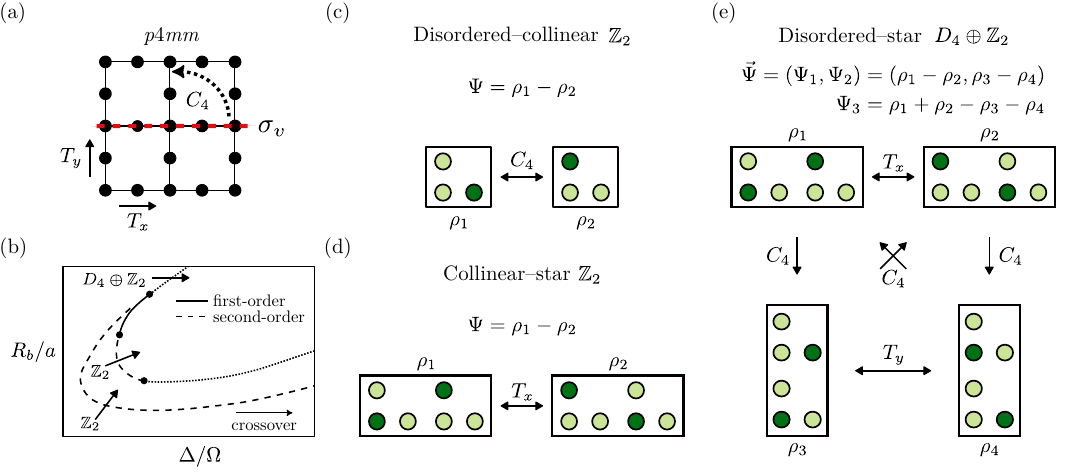}
\caption{\textbf{Symmetry analysis and quantum phase transitions.} (a) Generators of the wallpaper group of the Lieb lattice, $p4mm$. (b) Phase diagram of the Lieb lattice with allowed scenarios for the nature of the transitions, as determined from  symmetry analyses and numerics. At least one and at most three tricritical points (black dots) occur in the phase diagram, depending on the exact scenario realized. (c--e) 
Real-space symmetry analysis for three candidate second-order phase transitions. For each case, we identify the degenerate configurations of the low-symmetry phase that are symmetry-allowed descendants of a common high-symmetry parent phase. The symmetry operations of the high-symmetry phase are represented by the generators of the wallpaper group $p4mm$, or more generally by a subgroup thereof. By examining the transformation properties of the configuration densities under these symmetry operations, we construct linear combinations that transform according to the irreducible representations (irreps) of the high-symmetry group. These symmetry-adapted linear combinations define the appropriate order parameters for each transition.
}
\label{fig:FigS2_symmetries}
\end{figure}

We now proceed to classify the order parameters of the phases found on the Lieb lattice by employing a real-space symmetry analysis. The starting point for such an analysis is a collection of density profiles $\rho_{\lambda}\left(\boldsymbol{r}\right)=\left\{\langle n_{\boldsymbol{r}}\rangle \right\}_{\lambda}$, where $\lambda$ labels all the linearly independent density-wave-order combinations relevant to a given phase. While there exist an infinite number of possible linearly independent density orderings, for the phases appearing on a typical lattice, it suffices to consider only a finite number of such profiles, defined by the relatively small unit cells of the different ordered phases. The second ingredient in constructing the order parameter is the symmetry group of the lattice in question. The Lieb lattice has the wallpaper group $p4mm$, which is generated by the four symmetry operations illustrated in Fig.~\ref{fig:FigS2_symmetries}(a): translation along $\hat{x}$ $(T_x)$, translation along $\hat{y}$ $(T_y)$, fourfold rotation $(C_4)$, and mirror reflection about the $x$-axis $(\sigma_v)$. These operations are specified by their action on the lattice sites:
\begin{align}
    T^{}_x \vb{R}_{i,j}^{\mathrm{A,B,C}} = \vb{R}_{i+1,j}^{\mathrm{A,B,C}}, \qquad T^{}_y \vb{R}_{i,j}^{\mathrm{A,B,C}} = \vb{R}_{i,j+1}^{\mathrm{A,B,C}}, \qquad
\end{align}
\begin{align}    
    C^{}_4 \vb{R}_{i,j}^{\mathrm{A}} = \vb{R}_{-j,i}^{\mathrm{A}}
    , \qquad
    C^{}_4 \vb{R}_{i,j}^{\mathrm{B}} = \vb{R}_{-j,i}^{\mathrm{C}}
    , \qquad
    C^{}_4 \vb{R}_{i,j}^{\mathrm{C}} = \vb{R}_{-j-1,i}^{\mathrm{B}}
    , \qquad
\end{align}
\begin{align}    
    \sigma^{}_v \vb{R}_{i,j}^{\mathrm{A}} = \vb{R}_{i,-j}^{\mathrm{A}}
    , \qquad
    \sigma^{}_v \vb{R}_{i,j}^{\mathrm{B}} = \vb{R}_{i,-j}^{\mathrm{B}}
    , \qquad
    \sigma^{}_v \vb{R}_{i,j}^{\mathrm{C}} = \vb{R}_{i,-j-1}^{\mathrm{C}}.
     \qquad
\end{align}
Thus, starting with any profile corresponding to one density ordering for the ground state of a given phase, we can  generate all others in the same phase by acting with the symmetry generators ($G$) according to:
\begin{equation}
\hat{G}\rho^{}_{\lambda}\left(\boldsymbol{r}\right) = \rho^{}_{\lambda}\left(G\boldsymbol{r}\right) = \rho^{}_{\lambda'}\left(\boldsymbol{r}\right).
\end{equation}
This process constructs the permutation representation of the symmetry group on the space of density profiles, $\hat{G}$, together with the complete basis of possible linearly independent density-wave orderings. In order to obtain the set of order parameters that define a specific phase in the relevant density-profile basis, $\Psi_m=c_\lambda \rho_{\lambda}$, we decompose the representation $\hat{G}$ into irreducible representations of the symmetry group, with the symmetry-adapted basis directly specifying the order parameters:
\begin{equation}
    \hat{G} = \oplus^{}_{m} \hat{G}^{(m)}, \qquad \hat{G}\Psi^{}_m=\hat{G}^{(m)}\Psi^{}_m, \qquad \hat{G}^{(m)} \neq E.
\end{equation}
The last expression describes the condition for the $\Psi_m$ to define a symmetry-broken phase: that it does not transform as the identity representation ($E$) of the symmetry group. This real-space symmetry analysis is completely generic for all density-wave-ordered phases on an arbitrary lattice and provides a generalization of the more familiar momentum-space construction to lattices with decorated unit cells~\cite{Balents2005, Samajdar2020}. \smallskip

Proceeding along these lines, the symmetry analyses for the three main phase transitions observed on the Lieb lattice are showcased in Fig.~\ref{fig:FigS2_symmetries}(b), and we discuss each in turn below. The order parameters depicted in Fig.~\ref{fig:FigS2_symmetries} are the basis for the ones utilized in the main text to analyze the experimental data. It is straightforward to check that the relevant density profiles for the symmetric phase transform as the identity over the symmetry group of the lattice. As such, the disordered and symmetric phases can only be separated by a first-order transition or a crossover, and the numerical phase diagram in Fig.~\ref{fig:Fig3_liquid--vapor} conclusively shows a smooth crossover in the absence of local detuning fields. Allowing for local detuning can further drive a first-order transition between two different symmetric density profiles~\cite{PhysRevResearch.6.013271}, as we discuss in Appendix~\ref{sec:SM_liquid_vapor}.
\smallskip

\begin{figure}
\centering
\includegraphics[width=0.95\linewidth]{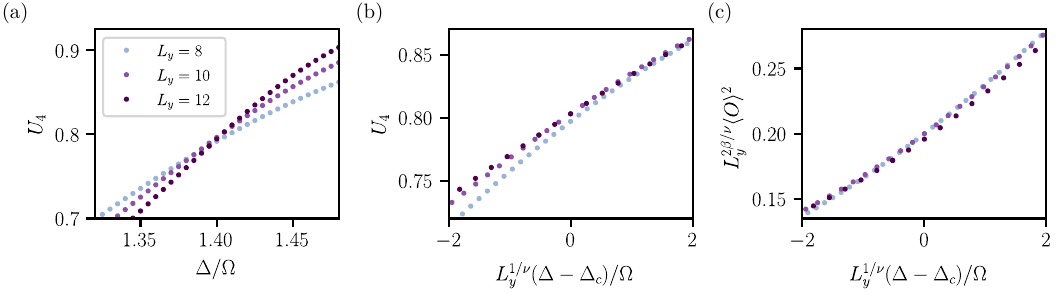}
\caption{\textbf{Finite-size scaling analysis of the disordered--collinear phase transition.} (a) The Binder cumulant plotted as a function of $\Delta/\Omega$ for a range of system sizes. The intersection of the lines for different system sizes informs the location of the quantum critical point, which we approximate to be $\Delta_c = 1.405 \Omega$. Rescaling the (b) Binder cumulant, and (c) square of the collinear order parameter shows excellent data collapse.
}
\label{fig:FigS3_fss}
\end{figure}

To study the disordered--collinear transition, we consider a $\mathbb{Z}_2$ order parameter that characterizes the imbalance of excitations between the B and C sublattices. The associated quantum phase transition is described by a $(2+1)$D $\mathbb{Z}_2$-symmetric Landau-Ginzburg-Wilson (LGW) field theory~\cite{Landau1937}, in which the effective action includes all relevant terms up to quartic order in the order parameter. Going beyond a mean-field approximation and incorporating fluctuations via a renormalization group (RG) analysis reveals the existence of a nontrivial interacting fixed point---the celebrated Wilson-Fisher fixed point~\cite{wilson1972critical}---which governs the infrared behavior of the theory. The presence of this non-Gaussian fixed point in the space of RG flows implies that if the transition were to be continuous, it belongs to the $(2+1)$D Ising universality class. However,  a first-order transition cannot be ruled out based on symmetry considerations alone; e.g., the fixed point might be inaccessible from certain regions in parameter space. Therefore, we additionally carry out a finite-size scaling analysis, detailed below, finding that the disordered--collinear transition is indeed second-order. To facilitate this, we define the collinear order parameter operator $\hat{O}$ as
\begin{equation}
    \hat{O}=\frac{1}{3N}\sum_i \left(\rho^{}_{i,\mathrm{B}}-\rho^{}_{i,\mathrm{C}}\right),
\end{equation}
where $N$ is the total number of unit cells, and consider the Binder cumulant~\cite{Binder1981}
\begin{equation}
    U^{}_4=\frac{3}{2}-\frac{\langle\hat{O}^4\rangle}{2\langle\hat{O}^2\rangle^2}.
\end{equation}
In Fig.~\ref{fig:FigS3_fss}(a), we plot the Binder cumulant computed using DMRG on cylinders of increasing transverse size. The Binder cumulant curves for different transverse sizes are observed to intersect at a point, which is a hallmark of a second-order transition~\cite{Binder1981}. We directly probe the predictions of the $\mathbb{Z}_2$ critical theory by testing the data collapse using a value of the quantum critical point estimated from this crossing, $\Delta_c$, and the known $(2+1)$D Ising critical exponents ($z=1$, $\nu \approx 0.63$, $\beta\approx 0.33$)~\cite{Hasenbusch1999}. The finite-size scaling \textit{Ans\"atze} for the Binder cumulant and the order parameter take the form \cite{Goldenfeld1992}:
\begin{align}
    \langle \hat{O} \rangle^2 L_y^{\frac{2\beta}{\nu}}=f_1\left( \frac{\Delta-\Delta_c}{\Omega}L_y^{\frac{1}{\nu}}\right),\qquad 
    U^{}_4 =f_2\left( \frac{\Delta-\Delta_c}{\Omega}L_y^{\frac{1}{\nu}}\right),
\end{align}
where $f_2$ is some universal scaling function.
We observe high-quality data collapse of the $\mathbb{Z}_2$ order parameter using the unbiased independent estimate of $\Delta_c$, as presented in Figs.~\ref{fig:FigS3_fss}(b,\,c), confirming the predicted universality class and order of the transition.
\smallskip

As noted earlier in Appendix~\ref{sec:multicritical_point}, the star phase can be accessed both directly from the disordered phase and indirectly from the collinear phase. Our symmetry analysis for the collinear--star transition starts from the reduced wallpaper group of the collinear phase, $pmmm$ (generated by $T^{}_x, T^{}_y, \sigma^{}_v, C_4^2$), and results in a $\mathbb{Z}_2$ order parameter capturing the additional symmetry breaking (on top of the symmetries already broken by the collinear order) upon entering the star phase. The direct disordered--star transition is described by the orbit of four density profiles that defines a composite order parameter transforming as $D_4\oplus \mathbb{Z}_2$. The $\mathbb{Z}_2$ $\Psi_3$ component of the order parameter is equivalent to the collinear order, while the $D_4$ components represent additional translational symmetry breaking in the star phase. The presence of such a multicomponent order parameter directly follows from the decorated unit cell structure.
\smallskip

Both the collinear--star and disordered--star transitions can be either first- or second-order. In the former case, the $\mathbb{Z}_2$ criticality could turn first-order, as described above, depending on the coefficients in the LGW action. For the disordered--star transition, an RG analysis of the  $D_4\oplus \mathbb{Z}_2$ LGW theory provides the possibility of, among others, a stable XY fixed point, although a first-order transition is allowed as well in a part of the $D_4$ theory's parameter space~\cite{Pelissetto2002}. Conclusive numerical evidence regarding the nature of the phase transitions into the star phase remains currently inaccessible, primarily due to the enlarged unit cell associated with this phase. In general, large system sizes are required to reliably probe the quantum critical behavior of the Lieb lattice. For comparison, on the square lattice, current quantum hardware can access critical properties only in the $\mathbb{Z}_2$-ordered phase, which features the smallest unit cell, and only up to system sizes of $16 \times 16$ atoms~\cite{Ebadi2021}. Among the ordered phases of the Lieb lattice, only the collinear phase possesses a similarly compact unit cell; however, even this case is inaccessible in the relevant parameter regime due to geometric constraints imposed by existing hardware architectures~\cite{aquila2023quera}. Accessing the star phase is even more challenging, as its larger unit cell demands significantly larger system sizes that lie beyond current experimental capabilities.
\smallskip

However, the wedge of the collinear phase between the disordered and star phases described in Appendix~\ref{sec:multicritical_point} and Fig.~\ref{fig:FigS2_symmetries} supports the scenario of  a second-order phase transition between compatible collinear and star orderings. If both the disordered--collinear and collinear--star phase transitions are second-order, a first-order line, and consequently, a tricritical point, must exist in the vicinity of their intersection~\cite{Landau1937}. This scenario is further supported by the effective LGW theory governing this transition. We construct this theory by forming invariant polynomials from direct products of the order parameters, producing the following Landau functional:
\begin{equation}\label{eq:landauf_d4z2}  \mathcal{L}^{}_F
=r^{}_{12}\left(\Psi_1^2+\Psi_2^2\right)
+r^{}_3\Psi_3^2
+ \gamma^{}_{123}\Psi^{}_3(\Psi_1^2 - \Psi_2^2)
+u^{}_{12}\left(\Psi_1^4+\Psi_2^4\right)
+ v^{}_{12}\Psi_1^2\Psi_2^2
+ u^{}_3\Psi_3^4
+w^{}_{123}\left(\Psi_1^2+\Psi_2^2\right)\Psi_3^2.
\end{equation}
The third-order invariant $\Psi^{}_3(\Psi_1^2 - \Psi_2^2)$ necessarily drives the transition first-order  close to the tricritical point~\cite{Park2001, sachdev2002ground, Silva2018, kornjaca2020, Kalinowski2022}.
\smallskip

It is possible that the disordered--star transition becomes second-order away from the intersection at larger values of $R_b/a$, requiring the existence of an additional tricritical point. Furthermore, the collinear--star phase transition numerically  appears to be first-order at larger detunings away from the wedge (although we cannot determine this with certainty), which would entail  a third tricritical point. The schematic phase diagram presented in Fig.~\ref{fig:FigS2_symmetries}(c) depicts each of these scenarios. Finally, given the absence of conclusive numerical evidence, we note that it is possible for the entire boundary of the star phase to be first-order, in which case there is no tricritical point.
\smallskip

The existence of at least one tricritical point in the phase diagram would open the door to experimental studies of quantum multicriticality using neutral atoms. Such tricritical behavior has conventionally been difficult to realize in quantum materials and models~\cite{Friedemann2018, Wang2021}, yet it offers access to exotic and potentially unexplored universality classes. More broadly, the interplay of multicomponent order parameters and the emergence of possible quantum multicritical points highlight the versatility of the neutral-atom Lieb-lattice platform as a powerful setting for investigating complex quantum critical phenomena.

\subsection{Experimental phase diagram with alternative boundary terminations}
\label{sec:ABC_experiment}

\begin{figure}[tb]
\centering
\includegraphics[width=\linewidth]{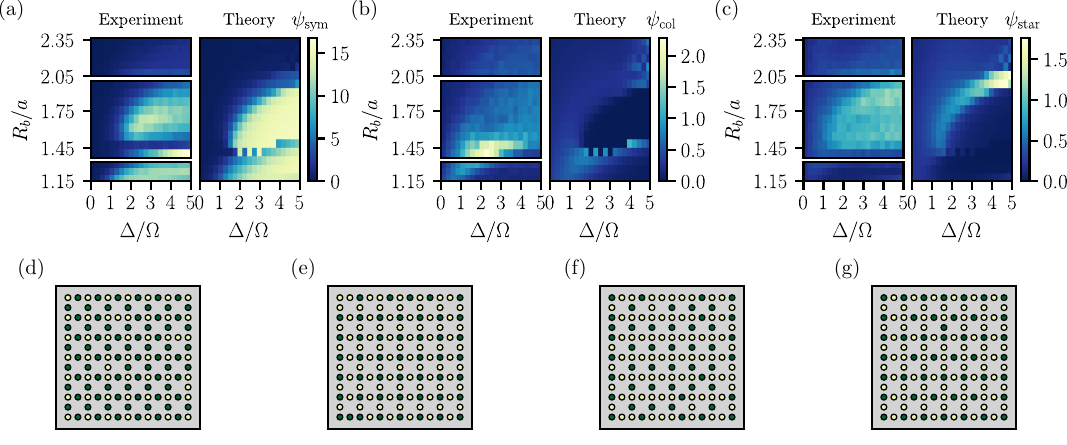}
\caption{\textbf{Experimental phase diagram with A, B, and C boundary sites.} The (a) symmetric, (b) collinear, and (c) star order parameters. The left panel of each depicts the order parameter as measured experimentally on Aquila and the right shows numerical DMRG results. The bottommost experimental panel corresponds to a $4\times4$ lattice, the others use a $6\times6$ lattice, and the DMRG calculations were performed on a $4\times4$ lattice. The bottom two panels of each experimental result are for $\Omega=2\pi\times 2.5$\,MHz, while the top panel is with $\Omega=2\pi\times 1.2$\,MHz. Each experiment used 500 shots, and all shots with greater than 98\% of the atoms loaded properly are included in the calculation of the order parameters. The initial value of $\Delta$ was $-2\Omega$, and the ramp time for the Rabi drive was 0.3 $\mu$s. The bottom row shows experimental shots that maximize each order parameter: (d) symmetric order parameter, maximized at $R_b/a=1.37$, $\Delta/\Omega=5.00$; (f) collinear order parameter, maximized at $R_b/a=1.37$, $\Delta/\Omega=2.25$; and (g) star order parameter, maximized at $R_b/a=1.82$, $\Delta/\Omega=4.25$. Figure~(e) shows an additional shot that maximizes the symmetric order parameter in the higher-$R_b/a$ region where the A sublattice is preferred, measured at $R_b/a=1.61$ and $\Delta/\Omega=3.00$.
}
\label{fig:FigS4_abc_OP}
\end{figure}

In Fig.~\ref{fig:FigS4_abc_OP}, we plot the experimental phase diagram together with supporting DMRG simulations for an alternative choice of boundary conditions, in which the boundary forms a smooth line consisting of A, B, and C sublattice sites. Although the precise phase boundaries are dramatically affected by this change of boundary conditions, we still observe the symmetric, collinear, and star phases. Note that the system size used in the DMRG simulations is smaller than that of the experiment for all but the smallest values of $R_b$, which significantly affects the agreement between theory and experiment.

The main difference from the BC-boundary phase diagram reported in the main text is that the A-symmetric phase emerges above the usual BC-symmetric phase, leaving only a sliver of the collinear phase between them. In the classical limit, the collinear and A-symmetric phases are degenerate; however, with this boundary termination, the A-symmetric phase has a lower energy because it occupies corner sites, which have the fewest neighbors. The collinear phase is stabilized in the thermodynamic limit by quantum fluctuations and therefore emerges here only for small values of $\Delta/\Omega$, where quantum fluctuations are strongest. That only a narrow sliver of the collinear phase remains in this geometry underscores the significance of this effect and provides experimental evidence thereof.

Similar to the BC-boundary case, the bottom of the star phase is pushed to large values of $R_b/a$ by boundary pinning effects. The small remaining region of the star phase predicted by DMRG is difficult to discern experimentally because the A-symmetric phase also exhibits a finite value of the star order parameter, likely as a result of the nonzero energy density introduced by the quasiadiabatic state-preparation protocol~\cite{samajdar2024quantum}. However, closer inspection of the data reveals that the symmetric order parameter weakens while the star order parameter grows as $R_b$ increases, consistent with our expectations.

Overall, despite the strong boundary dependence observed in the phase diagram, including the emergence of the boundary-seeded A-symmetric phase~\cite{Garnet_chan_rydberg}, we successfully detect all of the three main phases under both types of boundary conditions. The close agreement between classical and quantum computation confirms that our approach effectively captures the ground-state physics of the Lieb lattice. At the same time, the pronounced boundary effects underscore the need for larger system sizes in order to accurately resolve bulk properties and, in particular, to access the quantum critical behavior associated with the phase transitions.

\subsection{Numerical identification of the string phase}
\label{sec:SM_string_pd}
\begin{figure}
\centering
\includegraphics[]{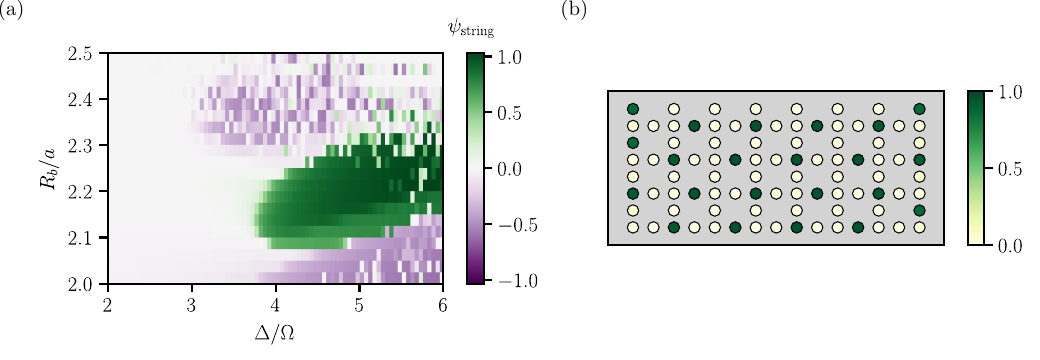}
\caption{\textbf{Numerical string phase diagram.} (a) The string order parameter $\psi_{\text{string}}$ obtained via DMRG on a cylindrical geometry with $7 \times 4$ unit cells as a function of $\Delta/\Omega$ and $R_b/a$, retaining interactions up to a distance of $\sqrt{13}\,a.$ A lobe of the string-ordered phase appears in the regime $2<R_b/a<\sqrt{5}$ and $\Delta/\Omega > 4,$ indicated by the large, positive value of the string order parameter. (b) The ground-state Rydberg density obtained deep within the string-ordered phase, demonstrating strings of excitations extending in the periodic (vertical) direction, separated by a distance of $3a$ in the open (horizontal) direction.
\label{fig:FigS_string}
}
\end{figure}

The quantum relaxation dynamics experiment considers quenches into a parameter regime characterized by kinetic constraints. Here, we show numerically that such constraints give rise to a distinct string phase in finite systems with commensurate boundary conditions. In the string region, Rydberg excitations organize into extended strings spanning the lattice, and there exists an exponentially large set (scaling with the system’s circumference) of classically degenerate string configurations, equivalent to the string patterns described on the kagome lattice~\cite{Samajdar2021}. The classical degeneracy arises from the ability of excitation strings to ``turn'' left or right at each node without changing the classical energy, provided that adjacent strings remain parallel. In a finite system, boundary conditions lift this degeneracy and select a string-ordered phase with a specific orientation, as depicted in Fig.~\ref{fig:FigS_string}(b). Although different string shapes are allowed, the strings are constrained to remain parallel and separated by a distance of $3a$. We characterize this behavior with the following order parameter:
\begin{equation}\label{eq:stringOP}
    \psi_{\text{string}}^{y}
    =
    \frac{1}{\mathcal{N}}
    \text{Re}
    \sum_{x,y}
    \sum_{\alpha = A,B}
    \left[
    \expval{n_{(x,y), \alpha}\,n_{(x+3,y), \alpha}}
    +
    \omega \expval{n_{(x,y), \alpha}\,n_{(x+2,y), \alpha}}
    +
    \omega^2 \expval{n_{(x,y), \alpha}\,n_{(x+1,y), \alpha}}
    \right],
\end{equation}
where $\omega = e^{2\pi i/3}$ and $\mathcal{N} = 2N_y\sum_{x} \Theta(x+3 \le N_x)$. This order parameter detects vertical string orientations, while horizontal string orientations are captured by the analogous quantity $\psi_{\text{string}}^{x}$. Although other Lieb-lattice phases can exhibit finite values of this string order parameter, the phase factors of $\omega$ ensure that it acquires a positive value uniquely in the string phase.

The numerical identification of the string phase is performed using DMRG calculations on the cylindrical geometry with $7 \times 4$ unit cells described in Appendix~\ref{sec:DMRG}, retaining interactions up to a distance of $\sqrt{13}\,a$. The boundaries of the string phase detected in the $7 \times 4$ system using the string order parameter are shown in Fig.~\ref{fig:FigS_string}. We find that the string phase appears in the range $2 < R_b/a < \sqrt{5}$ for large values of the detuning. These phase boundaries serve as a guide for the quench protocols discussed in the main text.

\section{Quantum liquid--vapor criticality}
\label{sec:SM_liquid_vapor}

\begin{figure}
\centering
\includegraphics{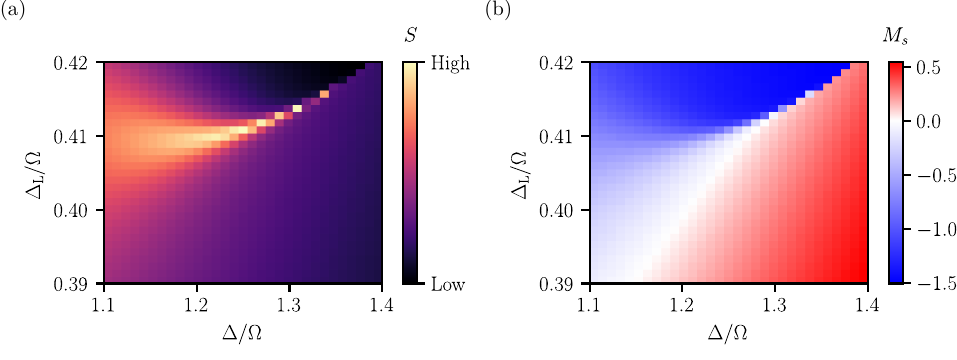}
\caption{\textbf{Numerical quantum liquid--vapor phase diagram on a large cylinder.} The (a) entanglement entropy and (b) staggered magnetization as a function of the global and local detunings, calculated via DMRG on a $9\times5$ cylinder with $R_b/a=1.2$, retaining nearest-neighbor interactions only. We observe a clear first-order transition (from the staggered magnetization) that terminates at a critical point surrounded by a smooth crossover. We exclude boundary sites from the calculation of the staggered magnetization to minimize boundary effects.
}
\label{fig:FigS5_pbc_local_detuning}
\end{figure}

In this section, we further investigate the quantum liquid--vapor transition~\cite{Landau1996-tu} introduced in the main text. Our preliminary evidence for this transition was based on quantum simulator experiments and DMRG calculations performed on (small) finite systems with open boundary conditions. To reinforce these findings, we now present additional DMRG calculations on a larger system with a cylindrical geometry and construct a minimal theoretical model that supports the existence of a terminal critical point in the thermodynamic limit. As shown in Fig.~\ref{fig:FigS5_pbc_local_detuning}, the DMRG results capture the quantum liquid--vapor critical point, the location of which remains stable against further increases in bond dimension and system size. Notably, we observe a large von Neumann entanglement entropy near this critical point, in contrast to the two adjacent symmetric phases, which are well-approximated by classical descriptions. Finite-size scaling calculations for quantitatively extracting the associated $(2+1)$D $\mathbb{Z}_2$ critical exponents are impeded by the need to scan and scale across a two-dimensional parameter space, and are thus left for future work.
\smallskip

We can understand the critical properties of the liquid-vapor transition from a simple product-state model. As the Rydberg Hamiltonian is real, the following \textit{Ansatz} can capture any product ground state that does not enlarge the unit cell:
\begin{equation}\label{eq:ansatz}
\ket{\Psi}=\otimes_{i}\left[\left(\sin{\alpha}\ket{g_{i,\mathrm{A}}}+\cos{\alpha}\ket{r_{i,\mathrm{A}}}\right)\otimes \left(\sin{\beta}\ket{g_{i,\mathrm{B}}}+\cos{\beta}\ket{r_{i,\mathrm{B}}}\right) \otimes\left(\sin{\gamma}\ket{g_{i,\mathrm{C}}}+\cos{\gamma}\ket{r_{i,\mathrm{C}}}\right)\right].
\end{equation}
Such a description will necessarily fail to reproduce the quantitative DMRG data, and, of course, fails to incorporate the quantum critical correlations. However, the \textit{Ansatz} does account for the lowest-order effect of quantum fluctuations. Our restriction to \textit{Ans\" atze} that do not enlarge the unit cell is motivated by the DMRG results.  
\smallskip

\begin{figure}[h]
\centering
\includegraphics{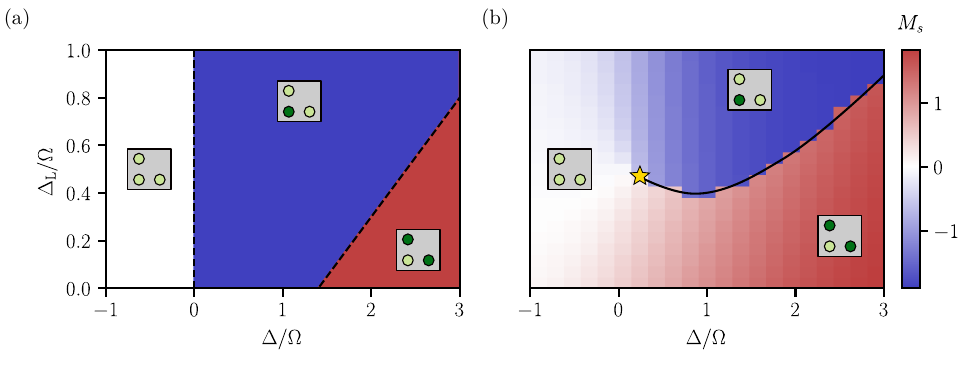}
\caption{\textbf{Effective model for liquid--vapor criticality.} Phase diagrams of the Lieb lattice with local and global detunings and $R_b/a=1.2$ for the product-state model. (a) The classical phase diagram at $\Omega=0$ shows three phases for $\Delta_{\text{L}}>0$ separated by first-order transitions (collinear not shown). (b) Introducing quantum fluctuations with $\Omega=1$ for $\Delta_{\text{L}}>0$ gives rise to liquid--vapor criticality; we find good qualitative agreement with DMRG results. The insets in each phase depict the ordering of Rydberg excitations within the unit cell, and the star in (b) indicates the approximate location of the critical point.
}
\label{fig:Fig_S6_effective_model}
\end{figure}

 In the thermodynamic limit, keeping at most third-neighbor interactions, the energy density of this \textit{Ansatz} is given by
\begin{equation}
    \label{eq:ansatz_energy}
    \begin{aligned}
        E\left[\{\alpha, \beta, \gamma\}, \Delta, \Delta_{\text{L}}, \Omega \right] &= \frac{\Omega}{2}\left(\sin{2\alpha} + \sin{2\beta} + \sin{2\gamma}\right) -\Delta\left(\cos^2{\alpha}+\cos^2{\beta}+\cos^2{\gamma}\right) \\
        &+\Delta_{\text{L}}\left(\cos^2{\beta}+\cos^2{\gamma}\right)+2V(a)\cos^2{\alpha}\left(\cos^2{\beta}+\cos^2{\gamma}\right) \\
        &+4V(\sqrt{2}a) \cos^2{\beta} \cos^2{\gamma}+2V(2a) \left(\cos^4{\alpha}+\cos^4{\beta}+\cos^4{\gamma}\right).
    \end{aligned}
\end{equation}
We employ simulated annealing in order to determine the ground states~\cite{Kirkpatrick1983}. The classical limit corresponds to $\Omega=0$, for which the phase boundaries can be mapped out analytically as
\begin{equation}\label{eq:cl_boundaries}
    \Delta=0, \qquad \Delta_{\text{L}}=\Delta, \qquad \Delta_{\text{L}}=\Delta+4V(\sqrt{2}a), \qquad \Delta_{\text{L}}=\frac{\Delta}{2}-2V(\sqrt{2}a)+V(2a),
\end{equation}
separating the four phases arising for positive and negative local detuning: the disordered phase, the collinear phase (at negative detunings only), the A-symmetric phase, and the BC-symmetric phase. The phase diagram for $\Omega=0$ and $\Delta_{\text{L}}>0$ is shown in Fig.~\ref{fig:Fig_S6_effective_model}(a). All the phase boundaries here represent first-order transitions.

Next, setting $\Omega=1$, we consider the effect of quantum fluctuations in Fig.~\ref{fig:Fig_S6_effective_model}(b). The previously sharp first-order line between disordered and symmetric phases now becomes a crossover, as seen in the Lieb lattice without local detuning. The first-order transition between the two symmetric phases persists only in the presence of nonzero local detuning and ends at a critical point. A comparison with the $\Omega = 0$ case reveals that even the lowest-order inclusion of quantum fluctuations is sufficient to account for the qualitative structure of the phase diagram, including the emergence of this terminal critical point.

\smallskip

There are, however, essential features of the quantum liquid--vapor transition that are not captured by the product-state model. By construction, our \textit{Ansatz} yields zero entanglement entropy throughout the entire phase diagram, in stark contrast to the DMRG results shown in Fig.~\ref{fig:FigS5_pbc_local_detuning}(a), especially around the critical point where quantum fluctuations lead to highly entangled states. As expected, the critical exponents derived from the product-state model correspond to those of Landau mean-field theory for a $\mathbb{Z}_2$ order parameter (e.g., $\nu = 1$, $\beta = 1/2$), which differ significantly from the Wilson-Fisher  $(2+1)$D $\mathbb{Z}_2$ values. Reliably extracting the correct critical behavior, including non-mean-field critical exponents, would require larger-scale experiments and DMRG simulations.

\twocolumngrid

\textbf{}

\begin{thebibliography}{78}%
\makeatletter
\providecommand \@ifxundefined [1]{%
 \@ifx{#1\undefined}
}%
\providecommand \@ifnum [1]{%
 \ifnum #1\expandafter \@firstoftwo
 \else \expandafter \@secondoftwo
 \fi
}%
\providecommand \@ifx [1]{%
 \ifx #1\expandafter \@firstoftwo
 \else \expandafter \@secondoftwo
 \fi
}%
\providecommand \natexlab [1]{#1}%
\providecommand \enquote  [1]{``#1''}%
\providecommand \bibnamefont  [1]{#1}%
\providecommand \bibfnamefont [1]{#1}%
\providecommand \citenamefont [1]{#1}%
\providecommand \href@noop [0]{\@secondoftwo}%
\providecommand \href [0]{\begingroup \@sanitize@url \@href}%
\providecommand \@href[1]{\@@startlink{#1}\@@href}%
\providecommand \@@href[1]{\endgroup#1\@@endlink}%
\providecommand \@sanitize@url [0]{\catcode `\\12\catcode `\$12\catcode
  `\&12\catcode `\#12\catcode `\^12\catcode `\_12\catcode `\%12\relax}%
\providecommand \@@startlink[1]{}%
\providecommand \@@endlink[0]{}%
\providecommand \url  [0]{\begingroup\@sanitize@url \@url }%
\providecommand \@url [1]{\endgroup\@href {#1}{\urlprefix }}%
\providecommand \urlprefix  [0]{URL }%
\providecommand \Eprint [0]{\href }%
\providecommand \doibase [0]{https://doi.org/}%
\providecommand \selectlanguage [0]{\@gobble}%
\providecommand \bibinfo  [0]{\@secondoftwo}%
\providecommand \bibfield  [0]{\@secondoftwo}%
\providecommand \translation [1]{[#1]}%
\providecommand \BibitemOpen [0]{}%
\providecommand \bibitemStop [0]{}%
\providecommand \bibitemNoStop [0]{.\EOS\space}%
\providecommand \EOS [0]{\spacefactor3000\relax}%
\providecommand \BibitemShut  [1]{\csname bibitem#1\endcsname}%
\let\auto@bib@innerbib\@empty
%</preamble>
\bibitem [{\citenamefont {Wilson}\ and\ \citenamefont
  {Fisher}(1972)}]{wilson1972critical}%
  \BibitemOpen
  \bibfield  {author} {\bibinfo {author} {\bibfnamefont {K.~G.}\ \bibnamefont
  {Wilson}}\ and\ \bibinfo {author} {\bibfnamefont {M.~E.}\ \bibnamefont
  {Fisher}},\ }\bibfield  {title} {\bibinfo {title} {{Critical Exponents in
  3.99 Dimensions}},\ }\href {https://doi.org/10.1103/PhysRevLett.28.240}
  {\bibfield  {journal} {\bibinfo  {journal} {Phys. Rev. Lett.}\ }\textbf
  {\bibinfo {volume} {28}},\ \bibinfo {pages} {240} (\bibinfo {year}
  {1972})}\BibitemShut {NoStop}%
\bibitem [{\citenamefont {Sachdev}(2001)}]{sachdev2001quantum}%
  \BibitemOpen
  \bibfield  {author} {\bibinfo {author} {\bibfnamefont {S.}~\bibnamefont
  {Sachdev}},\ }\href {https://doi.org/10.1017/CBO9780511613958} {\emph
  {\bibinfo {title} {{Quantum Phase Transitions}}}}\ (\bibinfo  {publisher}
  {Cambridge University Press},\ \bibinfo {year} {2001})\BibitemShut {NoStop}%
\bibitem [{\citenamefont {Labuhn}\ \emph {et~al.}(2016)\citenamefont {Labuhn},
  \citenamefont {Barredo}, \citenamefont {Ravets}, \citenamefont
  {de~L{\'e}s{\'e}leuc}, \citenamefont {Macr{\`i}}, \citenamefont {Lahaye},\
  and\ \citenamefont {Browaeys}}]{Labuhn2016}%
  \BibitemOpen
  \bibfield  {author} {\bibinfo {author} {\bibfnamefont {H.}~\bibnamefont
  {Labuhn}}, \bibinfo {author} {\bibfnamefont {D.}~\bibnamefont {Barredo}},
  \bibinfo {author} {\bibfnamefont {S.}~\bibnamefont {Ravets}}, \bibinfo
  {author} {\bibfnamefont {S.}~\bibnamefont {de~L{\'e}s{\'e}leuc}}, \bibinfo
  {author} {\bibfnamefont {T.}~\bibnamefont {Macr{\`i}}}, \bibinfo {author}
  {\bibfnamefont {T.}~\bibnamefont {Lahaye}},\ and\ \bibinfo {author}
  {\bibfnamefont {A.}~\bibnamefont {Browaeys}},\ }\bibfield  {title} {\bibinfo
  {title} {{Tunable two-dimensional arrays of single Rydberg atoms for
  realizing quantum Ising models}},\ }\href
  {https://doi.org/10.1038/nature18274} {\bibfield  {journal} {\bibinfo
  {journal} {Nature}\ }\textbf {\bibinfo {volume} {534}},\ \bibinfo {pages}
  {667} (\bibinfo {year} {2016})}\BibitemShut {NoStop}%
\bibitem [{\citenamefont {Bernien}\ \emph {et~al.}(2017)\citenamefont
  {Bernien}, \citenamefont {Schwartz}, \citenamefont {Keesling}, \citenamefont
  {Levine}, \citenamefont {Omran}, \citenamefont {Pichler}, \citenamefont
  {Choi}, \citenamefont {Zibrov}, \citenamefont {Endres}, \citenamefont
  {Greiner}, \citenamefont {Vuleti{\'{c}}},\ and\ \citenamefont
  {Lukin}}]{Bernien2017}%
  \BibitemOpen
  \bibfield  {author} {\bibinfo {author} {\bibfnamefont {H.}~\bibnamefont
  {Bernien}}, \bibinfo {author} {\bibfnamefont {S.}~\bibnamefont {Schwartz}},
  \bibinfo {author} {\bibfnamefont {A.}~\bibnamefont {Keesling}}, \bibinfo
  {author} {\bibfnamefont {H.}~\bibnamefont {Levine}}, \bibinfo {author}
  {\bibfnamefont {A.}~\bibnamefont {Omran}}, \bibinfo {author} {\bibfnamefont
  {H.}~\bibnamefont {Pichler}}, \bibinfo {author} {\bibfnamefont
  {S.}~\bibnamefont {Choi}}, \bibinfo {author} {\bibfnamefont {A.~S.}\
  \bibnamefont {Zibrov}}, \bibinfo {author} {\bibfnamefont {M.}~\bibnamefont
  {Endres}}, \bibinfo {author} {\bibfnamefont {M.}~\bibnamefont {Greiner}},
  \bibinfo {author} {\bibfnamefont {V.}~\bibnamefont {Vuleti{\'{c}}}},\ and\
  \bibinfo {author} {\bibfnamefont {M.~D.}\ \bibnamefont {Lukin}},\ }\bibfield
  {title} {\bibinfo {title} {Probing many-body dynamics on a 51-atom quantum
  simulator},\ }\href {https://doi.org/10.1038/nature24622} {\bibfield
  {journal} {\bibinfo  {journal} {Nature}\ }\textbf {\bibinfo {volume} {551}},\
  \bibinfo {pages} {579} (\bibinfo {year} {2017})}\BibitemShut {NoStop}%
\bibitem [{\citenamefont {Evered}\ \emph {et~al.}(2023)\citenamefont {Evered},
  \citenamefont {Bluvstein}, \citenamefont {Kalinowski}, \citenamefont {Ebadi},
  \citenamefont {Manovitz}, \citenamefont {Zhou}, \citenamefont {Li},
  \citenamefont {Geim}, \citenamefont {Wang}, \citenamefont {Maskara},
  \citenamefont {Levine}, \citenamefont {Semeghini}, \citenamefont {Greiner},
  \citenamefont {Vuleti{\'{c}}},\ and\ \citenamefont {Lukin}}]{Evered2023}%
  \BibitemOpen
  \bibfield  {author} {\bibinfo {author} {\bibfnamefont {S.~J.}\ \bibnamefont
  {Evered}}, \bibinfo {author} {\bibfnamefont {D.}~\bibnamefont {Bluvstein}},
  \bibinfo {author} {\bibfnamefont {M.}~\bibnamefont {Kalinowski}}, \bibinfo
  {author} {\bibfnamefont {S.}~\bibnamefont {Ebadi}}, \bibinfo {author}
  {\bibfnamefont {T.}~\bibnamefont {Manovitz}}, \bibinfo {author}
  {\bibfnamefont {H.}~\bibnamefont {Zhou}}, \bibinfo {author} {\bibfnamefont
  {S.~H.}\ \bibnamefont {Li}}, \bibinfo {author} {\bibfnamefont {A.~A.}\
  \bibnamefont {Geim}}, \bibinfo {author} {\bibfnamefont {T.~T.}\ \bibnamefont
  {Wang}}, \bibinfo {author} {\bibfnamefont {N.}~\bibnamefont {Maskara}},
  \bibinfo {author} {\bibfnamefont {H.}~\bibnamefont {Levine}}, \bibinfo
  {author} {\bibfnamefont {G.}~\bibnamefont {Semeghini}}, \bibinfo {author}
  {\bibfnamefont {M.}~\bibnamefont {Greiner}}, \bibinfo {author} {\bibfnamefont
  {V.}~\bibnamefont {Vuleti{\'{c}}}},\ and\ \bibinfo {author} {\bibfnamefont
  {M.~D.}\ \bibnamefont {Lukin}},\ }\bibfield  {title} {\bibinfo {title}
  {High-fidelity parallel entangling gates on a neutral-atom quantum
  computer},\ }\href {https://doi.org/10.1038/s41586-023-06481-y} {\bibfield
  {journal} {\bibinfo  {journal} {Nature}\ }\textbf {\bibinfo {volume} {622}},\
  \bibinfo {pages} {268} (\bibinfo {year} {2023})}\BibitemShut {NoStop}%
\bibitem [{\citenamefont {Bluvstein}\ \emph {et~al.}(2024)\citenamefont
  {Bluvstein}, \citenamefont {Evered}, \citenamefont {Geim}, \citenamefont
  {Li}, \citenamefont {Zhou}, \citenamefont {Manovitz}, \citenamefont {Ebadi},
  \citenamefont {Cain}, \citenamefont {Kalinowski}, \citenamefont {Hangleiter},
  \citenamefont {Bonilla~Ataides}, \citenamefont {Maskara}, \citenamefont
  {Cong}, \citenamefont {Gao}, \citenamefont {Sales~Rodriguez}, \citenamefont
  {Karolyshyn}, \citenamefont {Semeghini}, \citenamefont {Gullans},
  \citenamefont {Greiner}, \citenamefont {Vuleti{\'{c}}},\ and\ \citenamefont
  {Lukin}}]{Bluvstein2024}%
  \BibitemOpen
  \bibfield  {author} {\bibinfo {author} {\bibfnamefont {D.}~\bibnamefont
  {Bluvstein}}, \bibinfo {author} {\bibfnamefont {S.~J.}\ \bibnamefont
  {Evered}}, \bibinfo {author} {\bibfnamefont {A.~A.}\ \bibnamefont {Geim}},
  \bibinfo {author} {\bibfnamefont {S.~H.}\ \bibnamefont {Li}}, \bibinfo
  {author} {\bibfnamefont {H.}~\bibnamefont {Zhou}}, \bibinfo {author}
  {\bibfnamefont {T.}~\bibnamefont {Manovitz}}, \bibinfo {author}
  {\bibfnamefont {S.}~\bibnamefont {Ebadi}}, \bibinfo {author} {\bibfnamefont
  {M.}~\bibnamefont {Cain}}, \bibinfo {author} {\bibfnamefont {M.}~\bibnamefont
  {Kalinowski}}, \bibinfo {author} {\bibfnamefont {D.}~\bibnamefont
  {Hangleiter}}, \bibinfo {author} {\bibfnamefont {J.~P.}\ \bibnamefont
  {Bonilla~Ataides}}, \bibinfo {author} {\bibfnamefont {N.}~\bibnamefont
  {Maskara}}, \bibinfo {author} {\bibfnamefont {I.}~\bibnamefont {Cong}},
  \bibinfo {author} {\bibfnamefont {X.}~\bibnamefont {Gao}}, \bibinfo {author}
  {\bibfnamefont {P.}~\bibnamefont {Sales~Rodriguez}}, \bibinfo {author}
  {\bibfnamefont {T.}~\bibnamefont {Karolyshyn}}, \bibinfo {author}
  {\bibfnamefont {G.}~\bibnamefont {Semeghini}}, \bibinfo {author}
  {\bibfnamefont {M.~J.}\ \bibnamefont {Gullans}}, \bibinfo {author}
  {\bibfnamefont {M.}~\bibnamefont {Greiner}}, \bibinfo {author} {\bibfnamefont
  {V.}~\bibnamefont {Vuleti{\'{c}}}},\ and\ \bibinfo {author} {\bibfnamefont
  {M.~D.}\ \bibnamefont {Lukin}},\ }\bibfield  {title} {\bibinfo {title}
  {Logical quantum processor based on reconfigurable atom arrays},\ }\href
  {https://doi.org/10.1038/s41586-023-06927-3} {\bibfield  {journal} {\bibinfo
  {journal} {Nature}\ }\textbf {\bibinfo {volume} {626}},\ \bibinfo {pages}
  {58} (\bibinfo {year} {2024})}\BibitemShut {NoStop}%
\bibitem [{\citenamefont {de~Léséleuc}\ \emph {et~al.}(2019)\citenamefont
  {de~Léséleuc}, \citenamefont {Lienhard}, \citenamefont {Scholl},
  \citenamefont {Barredo}, \citenamefont {Weber}, \citenamefont {Lang},
  \citenamefont {Büchler}, \citenamefont {Lahaye},\ and\ \citenamefont
  {Browaeys}}]{deLeseleuc2019}%
  \BibitemOpen
  \bibfield  {author} {\bibinfo {author} {\bibfnamefont {S.}~\bibnamefont
  {de~Léséleuc}}, \bibinfo {author} {\bibfnamefont {V.}~\bibnamefont
  {Lienhard}}, \bibinfo {author} {\bibfnamefont {P.}~\bibnamefont {Scholl}},
  \bibinfo {author} {\bibfnamefont {D.}~\bibnamefont {Barredo}}, \bibinfo
  {author} {\bibfnamefont {S.}~\bibnamefont {Weber}}, \bibinfo {author}
  {\bibfnamefont {N.}~\bibnamefont {Lang}}, \bibinfo {author} {\bibfnamefont
  {H.~P.}\ \bibnamefont {Büchler}}, \bibinfo {author} {\bibfnamefont
  {T.}~\bibnamefont {Lahaye}},\ and\ \bibinfo {author} {\bibfnamefont
  {A.}~\bibnamefont {Browaeys}},\ }\bibfield  {title} {\bibinfo {title}
  {{Observation of a symmetry-protected topological phase of interacting bosons
  with Rydberg atoms}},\ }\href {https://doi.org/10.1126/science.aav9105}
  {\bibfield  {journal} {\bibinfo  {journal} {Science}\ }\textbf {\bibinfo
  {volume} {365}},\ \bibinfo {pages} {775} (\bibinfo {year}
  {2019})}\BibitemShut {NoStop}%
\bibitem [{\citenamefont {Samajdar}\ \emph {et~al.}(2021)\citenamefont
  {Samajdar}, \citenamefont {Ho}, \citenamefont {Pichler}, \citenamefont
  {Lukin},\ and\ \citenamefont {Sachdev}}]{Samajdar2021}%
  \BibitemOpen
  \bibfield  {author} {\bibinfo {author} {\bibfnamefont {R.}~\bibnamefont
  {Samajdar}}, \bibinfo {author} {\bibfnamefont {W.~W.}\ \bibnamefont {Ho}},
  \bibinfo {author} {\bibfnamefont {H.}~\bibnamefont {Pichler}}, \bibinfo
  {author} {\bibfnamefont {M.~D.}\ \bibnamefont {Lukin}},\ and\ \bibinfo
  {author} {\bibfnamefont {S.}~\bibnamefont {Sachdev}},\ }\bibfield  {title}
  {\bibinfo {title} {{Quantum phases of Rydberg atoms on a kagome lattice}},\
  }\href {https://doi.org/10.1073/pnas.2015785118} {\bibfield  {journal}
  {\bibinfo  {journal} {Proc. Natl. Acad. Sci. U.S.A}\ }\textbf {\bibinfo
  {volume} {118}},\ \bibinfo {pages} {e2015785118} (\bibinfo {year}
  {2021})}\BibitemShut {NoStop}%
\bibitem [{\citenamefont {Verresen}\ \emph {et~al.}(2021)\citenamefont
  {Verresen}, \citenamefont {Lukin},\ and\ \citenamefont
  {Vishwanath}}]{Verresen2021}%
  \BibitemOpen
  \bibfield  {author} {\bibinfo {author} {\bibfnamefont {R.}~\bibnamefont
  {Verresen}}, \bibinfo {author} {\bibfnamefont {M.~D.}\ \bibnamefont
  {Lukin}},\ and\ \bibinfo {author} {\bibfnamefont {A.}~\bibnamefont
  {Vishwanath}},\ }\bibfield  {title} {\bibinfo {title} {{Prediction of Toric
  Code Topological Order from Rydberg Blockade}},\ }\href
  {https://doi.org/10.1103/PhysRevX.11.031005} {\bibfield  {journal} {\bibinfo
  {journal} {Phys. Rev. X}\ }\textbf {\bibinfo {volume} {11}},\ \bibinfo
  {pages} {031005} (\bibinfo {year} {2021})}\BibitemShut {NoStop}%
\bibitem [{\citenamefont {Semeghini}\ \emph {et~al.}(2021)\citenamefont
  {Semeghini}, \citenamefont {Levine}, \citenamefont {Keesling}, \citenamefont
  {Ebadi}, \citenamefont {Wang}, \citenamefont {Bluvstein}, \citenamefont
  {Verresen}, \citenamefont {Pichler}, \citenamefont {Kalinowski},
  \citenamefont {Samajdar}, \citenamefont {Omran}, \citenamefont {Sachdev},
  \citenamefont {Vishwanath}, \citenamefont {Greiner}, \citenamefont
  {Vuletić},\ and\ \citenamefont {Lukin}}]{Semeghini2021}%
  \BibitemOpen
  \bibfield  {author} {\bibinfo {author} {\bibfnamefont {G.}~\bibnamefont
  {Semeghini}}, \bibinfo {author} {\bibfnamefont {H.}~\bibnamefont {Levine}},
  \bibinfo {author} {\bibfnamefont {A.}~\bibnamefont {Keesling}}, \bibinfo
  {author} {\bibfnamefont {S.}~\bibnamefont {Ebadi}}, \bibinfo {author}
  {\bibfnamefont {T.~T.}\ \bibnamefont {Wang}}, \bibinfo {author}
  {\bibfnamefont {D.}~\bibnamefont {Bluvstein}}, \bibinfo {author}
  {\bibfnamefont {R.}~\bibnamefont {Verresen}}, \bibinfo {author}
  {\bibfnamefont {H.}~\bibnamefont {Pichler}}, \bibinfo {author} {\bibfnamefont
  {M.}~\bibnamefont {Kalinowski}}, \bibinfo {author} {\bibfnamefont
  {R.}~\bibnamefont {Samajdar}}, \bibinfo {author} {\bibfnamefont
  {A.}~\bibnamefont {Omran}}, \bibinfo {author} {\bibfnamefont
  {S.}~\bibnamefont {Sachdev}}, \bibinfo {author} {\bibfnamefont
  {A.}~\bibnamefont {Vishwanath}}, \bibinfo {author} {\bibfnamefont
  {M.}~\bibnamefont {Greiner}}, \bibinfo {author} {\bibfnamefont
  {V.}~\bibnamefont {Vuletić}},\ and\ \bibinfo {author} {\bibfnamefont
  {M.~D.}\ \bibnamefont {Lukin}},\ }\bibfield  {title} {\bibinfo {title}
  {Probing topological spin liquids on a programmable quantum simulator},\
  }\href {https://doi.org/10.1126/science.abi8794} {\bibfield  {journal}
  {\bibinfo  {journal} {Science}\ }\textbf {\bibinfo {volume} {374}},\ \bibinfo
  {pages} {1242} (\bibinfo {year} {2021})}\BibitemShut {NoStop}%
\bibitem [{\citenamefont {Kornja{\v{c}}a}\ \emph {et~al.}(2023)\citenamefont
  {Kornja{\v{c}}a}, \citenamefont {Samajdar}, \citenamefont {Macr{\`i}},
  \citenamefont {Gemelke}, \citenamefont {Wang},\ and\ \citenamefont
  {Liu}}]{Kornjaca2023}%
  \BibitemOpen
  \bibfield  {author} {\bibinfo {author} {\bibfnamefont {M.}~\bibnamefont
  {Kornja{\v{c}}a}}, \bibinfo {author} {\bibfnamefont {R.}~\bibnamefont
  {Samajdar}}, \bibinfo {author} {\bibfnamefont {T.}~\bibnamefont {Macr{\`i}}},
  \bibinfo {author} {\bibfnamefont {N.}~\bibnamefont {Gemelke}}, \bibinfo
  {author} {\bibfnamefont {S.-T.}\ \bibnamefont {Wang}},\ and\ \bibinfo
  {author} {\bibfnamefont {F.}~\bibnamefont {Liu}},\ }\bibfield  {title}
  {\bibinfo {title} {{Trimer quantum spin liquid in a honeycomb array of
  Rydberg atoms}},\ }\href {https://doi.org/10.1038/s42005-023-01470-z}
  {\bibfield  {journal} {\bibinfo  {journal} {Commun. Phys.}\ }\textbf
  {\bibinfo {volume} {6}},\ \bibinfo {pages} {358} (\bibinfo {year}
  {2023})}\BibitemShut {NoStop}%
\bibitem [{\citenamefont {Bluvstein}\ \emph {et~al.}(2021)\citenamefont
  {Bluvstein}, \citenamefont {Omran}, \citenamefont {Levine}, \citenamefont
  {Keesling}, \citenamefont {Semeghini}, \citenamefont {Ebadi}, \citenamefont
  {Wang}, \citenamefont {Michailidis}, \citenamefont {Maskara}, \citenamefont
  {Ho}, \citenamefont {Choi}, \citenamefont {Serbyn}, \citenamefont {Greiner},
  \citenamefont {Vuletić},\ and\ \citenamefont {Lukin}}]{Bluvstein2021}%
  \BibitemOpen
  \bibfield  {author} {\bibinfo {author} {\bibfnamefont {D.}~\bibnamefont
  {Bluvstein}}, \bibinfo {author} {\bibfnamefont {A.}~\bibnamefont {Omran}},
  \bibinfo {author} {\bibfnamefont {H.}~\bibnamefont {Levine}}, \bibinfo
  {author} {\bibfnamefont {A.}~\bibnamefont {Keesling}}, \bibinfo {author}
  {\bibfnamefont {G.}~\bibnamefont {Semeghini}}, \bibinfo {author}
  {\bibfnamefont {S.}~\bibnamefont {Ebadi}}, \bibinfo {author} {\bibfnamefont
  {T.~T.}\ \bibnamefont {Wang}}, \bibinfo {author} {\bibfnamefont {A.~A.}\
  \bibnamefont {Michailidis}}, \bibinfo {author} {\bibfnamefont
  {N.}~\bibnamefont {Maskara}}, \bibinfo {author} {\bibfnamefont {W.~W.}\
  \bibnamefont {Ho}}, \bibinfo {author} {\bibfnamefont {S.}~\bibnamefont
  {Choi}}, \bibinfo {author} {\bibfnamefont {M.}~\bibnamefont {Serbyn}},
  \bibinfo {author} {\bibfnamefont {M.}~\bibnamefont {Greiner}}, \bibinfo
  {author} {\bibfnamefont {V.}~\bibnamefont {Vuletić}},\ and\ \bibinfo
  {author} {\bibfnamefont {M.~D.}\ \bibnamefont {Lukin}},\ }\bibfield  {title}
  {\bibinfo {title} {{Controlling quantum many-body dynamics in driven Rydberg
  atom arrays}},\ }\href {https://doi.org/10.1126/science.abg2530} {\bibfield
  {journal} {\bibinfo  {journal} {Science}\ }\textbf {\bibinfo {volume}
  {371}},\ \bibinfo {pages} {1355} (\bibinfo {year} {2021})}\BibitemShut
  {NoStop}%
\bibitem [{\citenamefont {Keesling}\ \emph {et~al.}(2019)\citenamefont
  {Keesling}, \citenamefont {Omran}, \citenamefont {Levine}, \citenamefont
  {Bernien}, \citenamefont {Pichler}, \citenamefont {Choi}, \citenamefont
  {Samajdar}, \citenamefont {Schwartz}, \citenamefont {Silvi}, \citenamefont
  {Sachdev}, \citenamefont {Zoller}, \citenamefont {Endres}, \citenamefont
  {Greiner}, \citenamefont {Vuleti{\'{c}}},\ and\ \citenamefont
  {Lukin}}]{Keesling2019}%
  \BibitemOpen
  \bibfield  {author} {\bibinfo {author} {\bibfnamefont {A.}~\bibnamefont
  {Keesling}}, \bibinfo {author} {\bibfnamefont {A.}~\bibnamefont {Omran}},
  \bibinfo {author} {\bibfnamefont {H.}~\bibnamefont {Levine}}, \bibinfo
  {author} {\bibfnamefont {H.}~\bibnamefont {Bernien}}, \bibinfo {author}
  {\bibfnamefont {H.}~\bibnamefont {Pichler}}, \bibinfo {author} {\bibfnamefont
  {S.}~\bibnamefont {Choi}}, \bibinfo {author} {\bibfnamefont {R.}~\bibnamefont
  {Samajdar}}, \bibinfo {author} {\bibfnamefont {S.}~\bibnamefont {Schwartz}},
  \bibinfo {author} {\bibfnamefont {P.}~\bibnamefont {Silvi}}, \bibinfo
  {author} {\bibfnamefont {S.}~\bibnamefont {Sachdev}}, \bibinfo {author}
  {\bibfnamefont {P.}~\bibnamefont {Zoller}}, \bibinfo {author} {\bibfnamefont
  {M.}~\bibnamefont {Endres}}, \bibinfo {author} {\bibfnamefont
  {M.}~\bibnamefont {Greiner}}, \bibinfo {author} {\bibfnamefont
  {V.}~\bibnamefont {Vuleti{\'{c}}}},\ and\ \bibinfo {author} {\bibfnamefont
  {M.~D.}\ \bibnamefont {Lukin}},\ }\bibfield  {title} {\bibinfo {title}
  {{Quantum Kibble--Zurek mechanism and critical dynamics on a programmable
  Rydberg simulator}},\ }\href {https://doi.org/10.1038/s41586-019-1070-1}
  {\bibfield  {journal} {\bibinfo  {journal} {Nature}\ }\textbf {\bibinfo
  {volume} {568}},\ \bibinfo {pages} {207} (\bibinfo {year}
  {2019})}\BibitemShut {NoStop}%
\bibitem [{\citenamefont {Ebadi}\ \emph {et~al.}(2021)\citenamefont {Ebadi},
  \citenamefont {Wang}, \citenamefont {Levine}, \citenamefont {Keesling},
  \citenamefont {Semeghini}, \citenamefont {Omran}, \citenamefont {Bluvstein},
  \citenamefont {Samajdar}, \citenamefont {Pichler}, \citenamefont {Ho},
  \citenamefont {Choi}, \citenamefont {Sachdev}, \citenamefont {Greiner},
  \citenamefont {Vuleti{\'{c}}},\ and\ \citenamefont {Lukin}}]{Ebadi2021}%
  \BibitemOpen
  \bibfield  {author} {\bibinfo {author} {\bibfnamefont {S.}~\bibnamefont
  {Ebadi}}, \bibinfo {author} {\bibfnamefont {T.~T.}\ \bibnamefont {Wang}},
  \bibinfo {author} {\bibfnamefont {H.}~\bibnamefont {Levine}}, \bibinfo
  {author} {\bibfnamefont {A.}~\bibnamefont {Keesling}}, \bibinfo {author}
  {\bibfnamefont {G.}~\bibnamefont {Semeghini}}, \bibinfo {author}
  {\bibfnamefont {A.}~\bibnamefont {Omran}}, \bibinfo {author} {\bibfnamefont
  {D.}~\bibnamefont {Bluvstein}}, \bibinfo {author} {\bibfnamefont
  {R.}~\bibnamefont {Samajdar}}, \bibinfo {author} {\bibfnamefont
  {H.}~\bibnamefont {Pichler}}, \bibinfo {author} {\bibfnamefont {W.~W.}\
  \bibnamefont {Ho}}, \bibinfo {author} {\bibfnamefont {S.}~\bibnamefont
  {Choi}}, \bibinfo {author} {\bibfnamefont {S.}~\bibnamefont {Sachdev}},
  \bibinfo {author} {\bibfnamefont {M.}~\bibnamefont {Greiner}}, \bibinfo
  {author} {\bibfnamefont {V.}~\bibnamefont {Vuleti{\'{c}}}},\ and\ \bibinfo
  {author} {\bibfnamefont {M.~D.}\ \bibnamefont {Lukin}},\ }\bibfield  {title}
  {\bibinfo {title} {Quantum phases of matter on a 256-atom programmable
  quantum simulator},\ }\href {https://doi.org/10.1038/s41586-021-03582-4}
  {\bibfield  {journal} {\bibinfo  {journal} {Nature}\ }\textbf {\bibinfo
  {volume} {595}},\ \bibinfo {pages} {227} (\bibinfo {year}
  {2021})}\BibitemShut {NoStop}%
\bibitem [{\citenamefont {Zhang}\ \emph {et~al.}(2025)\citenamefont {Zhang},
  \citenamefont {Cantú}, \citenamefont {Liu}, \citenamefont {Bylinskii},
  \citenamefont {Braverman}, \citenamefont {Huber}, \citenamefont
  {Amato-Grill}, \citenamefont {Lukin}, \citenamefont {Gemelke}, \citenamefont
  {Keesling}, \citenamefont {Wang}, \citenamefont {Meurice},\ and\
  \citenamefont {Tsai}}]{Zhang2024}%
  \BibitemOpen
  \bibfield  {author} {\bibinfo {author} {\bibfnamefont {J.}~\bibnamefont
  {Zhang}}, \bibinfo {author} {\bibfnamefont {S.~H.}\ \bibnamefont {Cantú}},
  \bibinfo {author} {\bibfnamefont {F.}~\bibnamefont {Liu}}, \bibinfo {author}
  {\bibfnamefont {A.}~\bibnamefont {Bylinskii}}, \bibinfo {author}
  {\bibfnamefont {B.}~\bibnamefont {Braverman}}, \bibinfo {author}
  {\bibfnamefont {F.}~\bibnamefont {Huber}}, \bibinfo {author} {\bibfnamefont
  {J.}~\bibnamefont {Amato-Grill}}, \bibinfo {author} {\bibfnamefont
  {A.}~\bibnamefont {Lukin}}, \bibinfo {author} {\bibfnamefont
  {N.}~\bibnamefont {Gemelke}}, \bibinfo {author} {\bibfnamefont
  {A.}~\bibnamefont {Keesling}}, \bibinfo {author} {\bibfnamefont {S.-T.}\
  \bibnamefont {Wang}}, \bibinfo {author} {\bibfnamefont {Y.}~\bibnamefont
  {Meurice}},\ and\ \bibinfo {author} {\bibfnamefont {S.-W.}\ \bibnamefont
  {Tsai}},\ }\bibfield  {title} {\bibinfo {title} {{Probing quantum floating
  phases in Rydberg atom arrays}},\ }\href
  {https://doi.org/https://doi.org/10.1038/s41467-025-55947-2} {\bibfield
  {journal} {\bibinfo  {journal} {Nat. Commun.}\ }\textbf {\bibinfo {volume}
  {16}},\ \bibinfo {pages} {712} (\bibinfo {year} {2025})}\BibitemShut
  {NoStop}%
\bibitem [{\citenamefont {Samajdar}\ and\ \citenamefont
  {Huse}(2024)}]{samajdar2024quantum}%
  \BibitemOpen
  \bibfield  {author} {\bibinfo {author} {\bibfnamefont {R.}~\bibnamefont
  {Samajdar}}\ and\ \bibinfo {author} {\bibfnamefont {D.~A.}\ \bibnamefont
  {Huse}},\ }\href {https://arxiv.org/abs/2401.15144} {\bibinfo {title}
  {Quantum and classical coarsening and their interplay with the kibble-zurek
  mechanism}} (\bibinfo {year} {2024}),\ \Eprint
  {https://arxiv.org/abs/2401.15144} {arXiv:2401.15144 [quant-ph]} \BibitemShut
  {NoStop}%
\bibitem [{\citenamefont {Manovitz}\ \emph {et~al.}(2025)\citenamefont
  {Manovitz}, \citenamefont {Li}, \citenamefont {Ebadi}, \citenamefont
  {Samajdar}, \citenamefont {Geim}, \citenamefont {Evered}, \citenamefont
  {Bluvstein}, \citenamefont {Zhou}, \citenamefont {Koyluoglu}, \citenamefont
  {Feldmeier}, \citenamefont {Dolgirev}, \citenamefont {Maskara}, \citenamefont
  {Kalinowski}, \citenamefont {Sachdev}, \citenamefont {Huse}, \citenamefont
  {Greiner}, \citenamefont {Vuletić},\ and\ \citenamefont
  {Lukin}}]{Manovitz_2025}%
  \BibitemOpen
  \bibfield  {author} {\bibinfo {author} {\bibfnamefont {T.}~\bibnamefont
  {Manovitz}}, \bibinfo {author} {\bibfnamefont {S.~H.}\ \bibnamefont {Li}},
  \bibinfo {author} {\bibfnamefont {S.}~\bibnamefont {Ebadi}}, \bibinfo
  {author} {\bibfnamefont {R.}~\bibnamefont {Samajdar}}, \bibinfo {author}
  {\bibfnamefont {A.~A.}\ \bibnamefont {Geim}}, \bibinfo {author}
  {\bibfnamefont {S.~J.}\ \bibnamefont {Evered}}, \bibinfo {author}
  {\bibfnamefont {D.}~\bibnamefont {Bluvstein}}, \bibinfo {author}
  {\bibfnamefont {H.}~\bibnamefont {Zhou}}, \bibinfo {author} {\bibfnamefont
  {N.~U.}\ \bibnamefont {Koyluoglu}}, \bibinfo {author} {\bibfnamefont
  {J.}~\bibnamefont {Feldmeier}}, \bibinfo {author} {\bibfnamefont {P.~E.}\
  \bibnamefont {Dolgirev}}, \bibinfo {author} {\bibfnamefont {N.}~\bibnamefont
  {Maskara}}, \bibinfo {author} {\bibfnamefont {M.}~\bibnamefont {Kalinowski}},
  \bibinfo {author} {\bibfnamefont {S.}~\bibnamefont {Sachdev}}, \bibinfo
  {author} {\bibfnamefont {D.~A.}\ \bibnamefont {Huse}}, \bibinfo {author}
  {\bibfnamefont {M.}~\bibnamefont {Greiner}}, \bibinfo {author} {\bibfnamefont
  {V.}~\bibnamefont {Vuletić}},\ and\ \bibinfo {author} {\bibfnamefont
  {M.~D.}\ \bibnamefont {Lukin}},\ }\bibfield  {title} {\bibinfo {title}
  {Quantum coarsening and collective dynamics on a programmable simulator},\
  }\href {https://doi.org/10.1038/s41586-024-08353-5} {\bibfield  {journal}
  {\bibinfo  {journal} {Nature}\ }\textbf {\bibinfo {volume} {638}},\ \bibinfo
  {pages} {86–92} (\bibinfo {year} {2025})}\BibitemShut {NoStop}%
\bibitem [{\citenamefont {Darbha}\ \emph
  {et~al.}(2024{\natexlab{a}})\citenamefont {Darbha}, \citenamefont
  {Kornja\ifmmode~\check{c}\else \v{c}\fi{}a}, \citenamefont {Liu},
  \citenamefont {Balewski}, \citenamefont {Hirsbrunner}, \citenamefont {Lopes},
  \citenamefont {Wang}, \citenamefont {Van~Beeumen}, \citenamefont {Camps},\
  and\ \citenamefont {Klymko}}]{darbha2024false}%
  \BibitemOpen
  \bibfield  {author} {\bibinfo {author} {\bibfnamefont {S.}~\bibnamefont
  {Darbha}}, \bibinfo {author} {\bibfnamefont {M.}~\bibnamefont
  {Kornja\ifmmode~\check{c}\else \v{c}\fi{}a}}, \bibinfo {author}
  {\bibfnamefont {F.}~\bibnamefont {Liu}}, \bibinfo {author} {\bibfnamefont
  {J.}~\bibnamefont {Balewski}}, \bibinfo {author} {\bibfnamefont {M.~R.}\
  \bibnamefont {Hirsbrunner}}, \bibinfo {author} {\bibfnamefont {P.~L.~S.}\
  \bibnamefont {Lopes}}, \bibinfo {author} {\bibfnamefont {S.-T.}\ \bibnamefont
  {Wang}}, \bibinfo {author} {\bibfnamefont {R.}~\bibnamefont {Van~Beeumen}},
  \bibinfo {author} {\bibfnamefont {D.}~\bibnamefont {Camps}},\ and\ \bibinfo
  {author} {\bibfnamefont {K.}~\bibnamefont {Klymko}},\ }\bibfield  {title}
  {\bibinfo {title} {False vacuum decay and nucleation dynamics in neutral atom
  systems},\ }\href {https://doi.org/10.1103/PhysRevB.110.155103} {\bibfield
  {journal} {\bibinfo  {journal} {Phys. Rev. B}\ }\textbf {\bibinfo {volume}
  {110}},\ \bibinfo {pages} {155103} (\bibinfo {year}
  {2024}{\natexlab{a}})}\BibitemShut {NoStop}%
\bibitem [{\citenamefont {Darbha}\ \emph
  {et~al.}(2024{\natexlab{b}})\citenamefont {Darbha}, \citenamefont
  {Kornja\ifmmode~\check{c}\else \v{c}\fi{}a}, \citenamefont {Liu},
  \citenamefont {Balewski}, \citenamefont {Hirsbrunner}, \citenamefont {Lopes},
  \citenamefont {Wang}, \citenamefont {Van~Beeumen}, \citenamefont {Klymko},\
  and\ \citenamefont {Camps}}]{darbha2024longlived}%
  \BibitemOpen
  \bibfield  {author} {\bibinfo {author} {\bibfnamefont {S.}~\bibnamefont
  {Darbha}}, \bibinfo {author} {\bibfnamefont {M.}~\bibnamefont
  {Kornja\ifmmode~\check{c}\else \v{c}\fi{}a}}, \bibinfo {author}
  {\bibfnamefont {F.}~\bibnamefont {Liu}}, \bibinfo {author} {\bibfnamefont
  {J.}~\bibnamefont {Balewski}}, \bibinfo {author} {\bibfnamefont {M.~R.}\
  \bibnamefont {Hirsbrunner}}, \bibinfo {author} {\bibfnamefont {P.~L.~S.}\
  \bibnamefont {Lopes}}, \bibinfo {author} {\bibfnamefont {S.-T.}\ \bibnamefont
  {Wang}}, \bibinfo {author} {\bibfnamefont {R.}~\bibnamefont {Van~Beeumen}},
  \bibinfo {author} {\bibfnamefont {K.}~\bibnamefont {Klymko}},\ and\ \bibinfo
  {author} {\bibfnamefont {D.}~\bibnamefont {Camps}},\ }\bibfield  {title}
  {\bibinfo {title} {Long-lived oscillations of metastable states in neutral
  atom systems},\ }\href {https://doi.org/10.1103/PhysRevB.110.155114}
  {\bibfield  {journal} {\bibinfo  {journal} {Phys. Rev. B}\ }\textbf {\bibinfo
  {volume} {110}},\ \bibinfo {pages} {155114} (\bibinfo {year}
  {2024}{\natexlab{b}})}\BibitemShut {NoStop}%
\bibitem [{\citenamefont {Vodeb}\ \emph {et~al.}(2025)\citenamefont {Vodeb},
  \citenamefont {Desaules}, \citenamefont {Hallam}, \citenamefont {Rava},
  \citenamefont {Humar}, \citenamefont {Willsch}, \citenamefont {Jin},
  \citenamefont {Willsch}, \citenamefont {Michielsen},\ and\ \citenamefont
  {Papi{\'c}}}]{Papic_nucleation}%
  \BibitemOpen
  \bibfield  {author} {\bibinfo {author} {\bibfnamefont {J.}~\bibnamefont
  {Vodeb}}, \bibinfo {author} {\bibfnamefont {J.-Y.}\ \bibnamefont {Desaules}},
  \bibinfo {author} {\bibfnamefont {A.}~\bibnamefont {Hallam}}, \bibinfo
  {author} {\bibfnamefont {A.}~\bibnamefont {Rava}}, \bibinfo {author}
  {\bibfnamefont {G.}~\bibnamefont {Humar}}, \bibinfo {author} {\bibfnamefont
  {D.}~\bibnamefont {Willsch}}, \bibinfo {author} {\bibfnamefont
  {F.}~\bibnamefont {Jin}}, \bibinfo {author} {\bibfnamefont {M.}~\bibnamefont
  {Willsch}}, \bibinfo {author} {\bibfnamefont {K.}~\bibnamefont
  {Michielsen}},\ and\ \bibinfo {author} {\bibfnamefont {Z.}~\bibnamefont
  {Papi{\'c}}},\ }\bibfield  {title} {\bibinfo {title} {Stirring the false
  vacuum via interacting quantized bubbles on a 5,564-qubit quantum annealer},\
  }\href {https://doi.org/10.1038/s41567-024-02765-w} {\bibfield  {journal}
  {\bibinfo  {journal} {Nat. Phys.}\ }\textbf {\bibinfo {volume} {21}},\
  \bibinfo {pages} {386} (\bibinfo {year} {2025})}\BibitemShut {NoStop}%
\bibitem [{\citenamefont {Zhu}\ \emph {et~al.}(2024)\citenamefont {Zhu},
  \citenamefont {Liu}, \citenamefont {Lagnese}, \citenamefont {Surace},
  \citenamefont {Zhang}, \citenamefont {He}, \citenamefont {Halimeh},
  \citenamefont {Dalmonte}, \citenamefont {Morampudi}, \citenamefont {Wilczek},
  \citenamefont {Yuan},\ and\ \citenamefont
  {Pan}}]{zhu2024probingfalsevacuumdecay}%
  \BibitemOpen
  \bibfield  {author} {\bibinfo {author} {\bibfnamefont {Z.-H.}\ \bibnamefont
  {Zhu}}, \bibinfo {author} {\bibfnamefont {Y.}~\bibnamefont {Liu}}, \bibinfo
  {author} {\bibfnamefont {G.}~\bibnamefont {Lagnese}}, \bibinfo {author}
  {\bibfnamefont {F.~M.}\ \bibnamefont {Surace}}, \bibinfo {author}
  {\bibfnamefont {W.-Y.}\ \bibnamefont {Zhang}}, \bibinfo {author}
  {\bibfnamefont {M.-G.}\ \bibnamefont {He}}, \bibinfo {author} {\bibfnamefont
  {J.~C.}\ \bibnamefont {Halimeh}}, \bibinfo {author} {\bibfnamefont
  {M.}~\bibnamefont {Dalmonte}}, \bibinfo {author} {\bibfnamefont {S.~C.}\
  \bibnamefont {Morampudi}}, \bibinfo {author} {\bibfnamefont {F.}~\bibnamefont
  {Wilczek}}, \bibinfo {author} {\bibfnamefont {Z.-S.}\ \bibnamefont {Yuan}},\
  and\ \bibinfo {author} {\bibfnamefont {J.-W.}\ \bibnamefont {Pan}},\ }\href
  {https://arxiv.org/abs/2411.12565} {\bibinfo {title} {Probing false vacuum
  decay on a cold-atom gauge-theory quantum simulator}} (\bibinfo {year}
  {2024}),\ \Eprint {https://arxiv.org/abs/2411.12565} {arXiv:2411.12565
  [cond-mat.quant-gas]} \BibitemShut {NoStop}%
\bibitem [{\citenamefont {Surace}\ \emph {et~al.}(2024)\citenamefont {Surace},
  \citenamefont {Lerose}, \citenamefont {Katz}, \citenamefont {Bennewitz},
  \citenamefont {Schuckert}, \citenamefont {Luo}, \citenamefont {De},
  \citenamefont {Ware}, \citenamefont {Morong}, \citenamefont {Collins},
  \citenamefont {Monroe}, \citenamefont {Davoudi},\ and\ \citenamefont
  {Gorshkov}}]{surace2024stringbreakingdynamicsquantumadiabatic}%
  \BibitemOpen
  \bibfield  {author} {\bibinfo {author} {\bibfnamefont {F.~M.}\ \bibnamefont
  {Surace}}, \bibinfo {author} {\bibfnamefont {A.}~\bibnamefont {Lerose}},
  \bibinfo {author} {\bibfnamefont {O.}~\bibnamefont {Katz}}, \bibinfo {author}
  {\bibfnamefont {E.~R.}\ \bibnamefont {Bennewitz}}, \bibinfo {author}
  {\bibfnamefont {A.}~\bibnamefont {Schuckert}}, \bibinfo {author}
  {\bibfnamefont {D.}~\bibnamefont {Luo}}, \bibinfo {author} {\bibfnamefont
  {A.}~\bibnamefont {De}}, \bibinfo {author} {\bibfnamefont {B.}~\bibnamefont
  {Ware}}, \bibinfo {author} {\bibfnamefont {W.}~\bibnamefont {Morong}},
  \bibinfo {author} {\bibfnamefont {K.}~\bibnamefont {Collins}}, \bibinfo
  {author} {\bibfnamefont {C.}~\bibnamefont {Monroe}}, \bibinfo {author}
  {\bibfnamefont {Z.}~\bibnamefont {Davoudi}},\ and\ \bibinfo {author}
  {\bibfnamefont {A.~V.}\ \bibnamefont {Gorshkov}},\ }\href
  {https://arxiv.org/abs/2411.10652} {\bibinfo {title} {String-breaking
  dynamics in quantum adiabatic and diabatic processes}} (\bibinfo {year}
  {2024}),\ \Eprint {https://arxiv.org/abs/2411.10652} {arXiv:2411.10652
  [quant-ph]} \BibitemShut {NoStop}%
\bibitem [{\citenamefont {Luo}\ \emph {et~al.}(2025)\citenamefont {Luo},
  \citenamefont {Surace}, \citenamefont {De}, \citenamefont {Lerose},
  \citenamefont {Bennewitz}, \citenamefont {Ware}, \citenamefont {Schuckert},
  \citenamefont {Davoudi}, \citenamefont {Gorshkov}, \citenamefont {Katz},\
  and\ \citenamefont {Monroe}}]{Luo2025}%
  \BibitemOpen
  \bibfield  {author} {\bibinfo {author} {\bibfnamefont {D.}~\bibnamefont
  {Luo}}, \bibinfo {author} {\bibfnamefont {F.~M.}\ \bibnamefont {Surace}},
  \bibinfo {author} {\bibfnamefont {A.}~\bibnamefont {De}}, \bibinfo {author}
  {\bibfnamefont {A.}~\bibnamefont {Lerose}}, \bibinfo {author} {\bibfnamefont
  {E.~R.}\ \bibnamefont {Bennewitz}}, \bibinfo {author} {\bibfnamefont
  {B.}~\bibnamefont {Ware}}, \bibinfo {author} {\bibfnamefont {A.}~\bibnamefont
  {Schuckert}}, \bibinfo {author} {\bibfnamefont {Z.}~\bibnamefont {Davoudi}},
  \bibinfo {author} {\bibfnamefont {A.~V.}\ \bibnamefont {Gorshkov}}, \bibinfo
  {author} {\bibfnamefont {O.}~\bibnamefont {Katz}},\ and\ \bibinfo {author}
  {\bibfnamefont {C.}~\bibnamefont {Monroe}},\ }\href
  {https://arxiv.org/abs/2505.09607} {\bibinfo {title} {Quantum simulation of
  bubble nucleation across a quantum phase transition}} (\bibinfo {year}
  {2025}),\ \Eprint {https://arxiv.org/abs/2505.09607} {arXiv:2505.09607
  [quant-ph]} \BibitemShut {NoStop}%
\bibitem [{\citenamefont {Michailidis}\ \emph {et~al.}(2020)\citenamefont
  {Michailidis}, \citenamefont {Turner}, \citenamefont
  {Papi\ifmmode~\acute{c}\else \'{c}\fi{}}, \citenamefont {Abanin},\ and\
  \citenamefont {Serbyn}}]{PhysRevX.10.011055}%
  \BibitemOpen
  \bibfield  {author} {\bibinfo {author} {\bibfnamefont {A.~A.}\ \bibnamefont
  {Michailidis}}, \bibinfo {author} {\bibfnamefont {C.~J.}\ \bibnamefont
  {Turner}}, \bibinfo {author} {\bibfnamefont {Z.}~\bibnamefont
  {Papi\ifmmode~\acute{c}\else \'{c}\fi{}}}, \bibinfo {author} {\bibfnamefont
  {D.~A.}\ \bibnamefont {Abanin}},\ and\ \bibinfo {author} {\bibfnamefont
  {M.}~\bibnamefont {Serbyn}},\ }\bibfield  {title} {\bibinfo {title} {Slow
  quantum thermalization and many-body revivals from mixed phase space},\
  }\href {https://doi.org/10.1103/PhysRevX.10.011055} {\bibfield  {journal}
  {\bibinfo  {journal} {Phys. Rev. X}\ }\textbf {\bibinfo {volume} {10}},\
  \bibinfo {pages} {011055} (\bibinfo {year} {2020})}\BibitemShut {NoStop}%
\bibitem [{\citenamefont {Fradkin}\ \emph {et~al.}(2015)\citenamefont
  {Fradkin}, \citenamefont {Kivelson},\ and\ \citenamefont
  {Tranquada}}]{RevModPhys.87.457}%
  \BibitemOpen
  \bibfield  {author} {\bibinfo {author} {\bibfnamefont {E.}~\bibnamefont
  {Fradkin}}, \bibinfo {author} {\bibfnamefont {S.~A.}\ \bibnamefont
  {Kivelson}},\ and\ \bibinfo {author} {\bibfnamefont {J.~M.}\ \bibnamefont
  {Tranquada}},\ }\bibfield  {title} {\bibinfo {title} {Colloquium: Theory of
  intertwined orders in high temperature superconductors},\ }\href
  {https://doi.org/10.1103/RevModPhys.87.457} {\bibfield  {journal} {\bibinfo
  {journal} {Rev. Mod. Phys.}\ }\textbf {\bibinfo {volume} {87}},\ \bibinfo
  {pages} {457} (\bibinfo {year} {2015})}\BibitemShut {NoStop}%
\bibitem [{\citenamefont {Maffi}\ \emph {et~al.}(2024)\citenamefont {Maffi},
  \citenamefont {Tausendpfund}, \citenamefont {Rizzi},\ and\ \citenamefont
  {Burrello}}]{PhysRevLett.132.226502}%
  \BibitemOpen
  \bibfield  {author} {\bibinfo {author} {\bibfnamefont {L.}~\bibnamefont
  {Maffi}}, \bibinfo {author} {\bibfnamefont {N.}~\bibnamefont {Tausendpfund}},
  \bibinfo {author} {\bibfnamefont {M.}~\bibnamefont {Rizzi}},\ and\ \bibinfo
  {author} {\bibfnamefont {M.}~\bibnamefont {Burrello}},\ }\bibfield  {title}
  {\bibinfo {title} {{Quantum Simulation of the Tricritical Ising Model in
  Tunable Josephson Junction Ladders}},\ }\href
  {https://doi.org/10.1103/PhysRevLett.132.226502} {\bibfield  {journal}
  {\bibinfo  {journal} {Phys. Rev. Lett.}\ }\textbf {\bibinfo {volume} {132}},\
  \bibinfo {pages} {226502} (\bibinfo {year} {2024})}\BibitemShut {NoStop}%
\bibitem [{\citenamefont {Wang}\ \emph {et~al.}(2025)\citenamefont {Wang},
  \citenamefont {Li},\ and\ \citenamefont
  {Li}}]{wang2025tricriticalkibblezurekscalingrydberg}%
  \BibitemOpen
  \bibfield  {author} {\bibinfo {author} {\bibfnamefont {H.}~\bibnamefont
  {Wang}}, \bibinfo {author} {\bibfnamefont {X.}~\bibnamefont {Li}},\ and\
  \bibinfo {author} {\bibfnamefont {C.}~\bibnamefont {Li}},\ }\bibfield
  {title} {\bibinfo {title} {{Tricritical Kibble-Zurek scaling in Rydberg atom
  ladders}},\ }\bibfield  {journal} {\bibinfo  {journal} {Nature
  Communications}\ }\textbf {\bibinfo {volume} {16}},\ \href
  {https://doi.org/10.1038/s41467-025-65652-9} {10.1038/s41467-025-65652-9}
  (\bibinfo {year} {2025})\BibitemShut {NoStop}%
\bibitem [{\citenamefont {Gould}\ \emph {et~al.}(2022)\citenamefont {Gould},
  \citenamefont {G\"uyer},\ and\ \citenamefont
  {Rummukainen}}]{PhysRevD.106.114507}%
  \BibitemOpen
  \bibfield  {author} {\bibinfo {author} {\bibfnamefont {O.}~\bibnamefont
  {Gould}}, \bibinfo {author} {\bibfnamefont {S.}~\bibnamefont {G\"uyer}},\
  and\ \bibinfo {author} {\bibfnamefont {K.}~\bibnamefont {Rummukainen}},\
  }\bibfield  {title} {\bibinfo {title} {First-order electroweak phase
  transitions: A nonperturbative update},\ }\href
  {https://doi.org/10.1103/PhysRevD.106.114507} {\bibfield  {journal} {\bibinfo
   {journal} {Phys. Rev. D}\ }\textbf {\bibinfo {volume} {106}},\ \bibinfo
  {pages} {114507} (\bibinfo {year} {2022})}\BibitemShut {NoStop}%
\bibitem [{\citenamefont {Samajdar}\ \emph {et~al.}(2020)\citenamefont
  {Samajdar}, \citenamefont {Ho}, \citenamefont {Pichler}, \citenamefont
  {Lukin},\ and\ \citenamefont {Sachdev}}]{Samajdar2020}%
  \BibitemOpen
  \bibfield  {author} {\bibinfo {author} {\bibfnamefont {R.}~\bibnamefont
  {Samajdar}}, \bibinfo {author} {\bibfnamefont {W.~W.}\ \bibnamefont {Ho}},
  \bibinfo {author} {\bibfnamefont {H.}~\bibnamefont {Pichler}}, \bibinfo
  {author} {\bibfnamefont {M.~D.}\ \bibnamefont {Lukin}},\ and\ \bibinfo
  {author} {\bibfnamefont {S.}~\bibnamefont {Sachdev}},\ }\bibfield  {title}
  {\bibinfo {title} {{Complex Density Wave Orders and Quantum Phase Transitions
  in a Model of Square-Lattice Rydberg Atom Arrays}},\ }\href
  {https://doi.org/10.1103/PhysRevLett.124.103601} {\bibfield  {journal}
  {\bibinfo  {journal} {Phys. Rev. Lett.}\ }\textbf {\bibinfo {volume} {124}},\
  \bibinfo {pages} {103601} (\bibinfo {year} {2020})}\BibitemShut {NoStop}%
\bibitem [{\citenamefont {Scholl}\ \emph {et~al.}(2021)\citenamefont {Scholl},
  \citenamefont {Schuler}, \citenamefont {Williams}, \citenamefont
  {Eberharter}, \citenamefont {Barredo}, \citenamefont {Schymik}, \citenamefont
  {Lienhard}, \citenamefont {Henry}, \citenamefont {Lang}, \citenamefont
  {Lahaye}, \citenamefont {La\"uchli},\ and\ \citenamefont
  {Browaeys}}]{Scholl2021}%
  \BibitemOpen
  \bibfield  {author} {\bibinfo {author} {\bibfnamefont {P.}~\bibnamefont
  {Scholl}}, \bibinfo {author} {\bibfnamefont {M.}~\bibnamefont {Schuler}},
  \bibinfo {author} {\bibfnamefont {H.~J.}\ \bibnamefont {Williams}}, \bibinfo
  {author} {\bibfnamefont {A.~A.}\ \bibnamefont {Eberharter}}, \bibinfo
  {author} {\bibfnamefont {D.}~\bibnamefont {Barredo}}, \bibinfo {author}
  {\bibfnamefont {K.-N.}\ \bibnamefont {Schymik}}, \bibinfo {author}
  {\bibfnamefont {V.}~\bibnamefont {Lienhard}}, \bibinfo {author}
  {\bibfnamefont {L.-P.}\ \bibnamefont {Henry}}, \bibinfo {author}
  {\bibfnamefont {T.~C.}\ \bibnamefont {Lang}}, \bibinfo {author}
  {\bibfnamefont {T.}~\bibnamefont {Lahaye}}, \bibinfo {author} {\bibfnamefont
  {A.~M.}\ \bibnamefont {La\"uchli}},\ and\ \bibinfo {author} {\bibfnamefont
  {A.}~\bibnamefont {Browaeys}},\ }\bibfield  {title} {\bibinfo {title}
  {{Quantum simulation of 2D antiferromagnets with hundreds of Rydberg
  atoms}},\ }\href {https://doi.org/https://doi.org/10.1038/s41586-021-03585-1}
  {\bibfield  {journal} {\bibinfo  {journal} {Nature}\ }\textbf {\bibinfo
  {volume} {595}},\ \bibinfo {pages} {233} (\bibinfo {year}
  {2021})}\BibitemShut {NoStop}%
\bibitem [{\citenamefont {Wurtz}\ \emph {et~al.}(2023)\citenamefont {Wurtz},
  \citenamefont {Bylinskii}, \citenamefont {Braverman}, \citenamefont
  {Amato-Grill}, \citenamefont {Cantu}, \citenamefont {Huber}, \citenamefont
  {Lukin}, \citenamefont {Liu}, \citenamefont {Weinberg}, \citenamefont {Long},
  \citenamefont {Wang}, \citenamefont {Gemelke},\ and\ \citenamefont
  {Keesling}}]{aquila2023quera}%
  \BibitemOpen
  \bibfield  {author} {\bibinfo {author} {\bibfnamefont {J.}~\bibnamefont
  {Wurtz}}, \bibinfo {author} {\bibfnamefont {A.}~\bibnamefont {Bylinskii}},
  \bibinfo {author} {\bibfnamefont {B.}~\bibnamefont {Braverman}}, \bibinfo
  {author} {\bibfnamefont {J.}~\bibnamefont {Amato-Grill}}, \bibinfo {author}
  {\bibfnamefont {S.~H.}\ \bibnamefont {Cantu}}, \bibinfo {author}
  {\bibfnamefont {F.}~\bibnamefont {Huber}}, \bibinfo {author} {\bibfnamefont
  {A.}~\bibnamefont {Lukin}}, \bibinfo {author} {\bibfnamefont
  {F.}~\bibnamefont {Liu}}, \bibinfo {author} {\bibfnamefont {P.}~\bibnamefont
  {Weinberg}}, \bibinfo {author} {\bibfnamefont {J.}~\bibnamefont {Long}},
  \bibinfo {author} {\bibfnamefont {S.-T.}\ \bibnamefont {Wang}}, \bibinfo
  {author} {\bibfnamefont {N.}~\bibnamefont {Gemelke}},\ and\ \bibinfo {author}
  {\bibfnamefont {A.}~\bibnamefont {Keesling}},\ }\href
  {https://arxiv.org/abs/2306.11727} {\bibinfo {title} {Aquila: Quera's
  256-qubit neutral-atom quantum computer}} (\bibinfo {year} {2023}),\ \Eprint
  {https://arxiv.org/abs/2306.11727} {arXiv:2306.11727 [quant-ph]} \BibitemShut
  {NoStop}%
\bibitem [{\citenamefont {White}(1992{\natexlab{a}})}]{White1992}%
  \BibitemOpen
  \bibfield  {author} {\bibinfo {author} {\bibfnamefont {S.~R.}\ \bibnamefont
  {White}},\ }\bibfield  {title} {\bibinfo {title} {Density matrix formulation
  for quantum renormalization groups},\ }\href
  {https://doi.org/10.1103/PhysRevLett.69.2863} {\bibfield  {journal} {\bibinfo
   {journal} {Phys. Rev. Lett.}\ }\textbf {\bibinfo {volume} {69}},\ \bibinfo
  {pages} {2863} (\bibinfo {year} {1992}{\natexlab{a}})}\BibitemShut {NoStop}%
\bibitem [{\citenamefont {Fishman}\ \emph
  {et~al.}(2022{\natexlab{a}})\citenamefont {Fishman}, \citenamefont {White},\
  and\ \citenamefont {Stoudenmire}}]{itensor-r0.3}%
  \BibitemOpen
  \bibfield  {author} {\bibinfo {author} {\bibfnamefont {M.}~\bibnamefont
  {Fishman}}, \bibinfo {author} {\bibfnamefont {S.~R.}\ \bibnamefont {White}},\
  and\ \bibinfo {author} {\bibfnamefont {E.~M.}\ \bibnamefont {Stoudenmire}},\
  }\bibfield  {title} {\bibinfo {title} {{Codebase release 0.3 for ITensor}},\
  }\href {https://doi.org/10.21468/SciPostPhysCodeb.4-r0.3} {\bibfield
  {journal} {\bibinfo  {journal} {SciPost Phys. Codebases}\ ,\ \bibinfo {pages}
  {4}} (\bibinfo {year} {2022}{\natexlab{a}})}\BibitemShut {NoStop}%
\bibitem [{\citenamefont {Kalinowski}\ \emph {et~al.}(2022)\citenamefont
  {Kalinowski}, \citenamefont {Samajdar}, \citenamefont {Melko}, \citenamefont
  {Lukin}, \citenamefont {Sachdev},\ and\ \citenamefont
  {Choi}}]{Kalinowski2022}%
  \BibitemOpen
  \bibfield  {author} {\bibinfo {author} {\bibfnamefont {M.}~\bibnamefont
  {Kalinowski}}, \bibinfo {author} {\bibfnamefont {R.}~\bibnamefont
  {Samajdar}}, \bibinfo {author} {\bibfnamefont {R.~G.}\ \bibnamefont {Melko}},
  \bibinfo {author} {\bibfnamefont {M.~D.}\ \bibnamefont {Lukin}}, \bibinfo
  {author} {\bibfnamefont {S.}~\bibnamefont {Sachdev}},\ and\ \bibinfo {author}
  {\bibfnamefont {S.}~\bibnamefont {Choi}},\ }\bibfield  {title} {\bibinfo
  {title} {{Bulk and boundary quantum phase transitions in a square Rydberg
  atom array}},\ }\href {https://doi.org/10.1103/PhysRevB.105.174417}
  {\bibfield  {journal} {\bibinfo  {journal} {Phys. Rev. B}\ }\textbf {\bibinfo
  {volume} {105}},\ \bibinfo {pages} {174417} (\bibinfo {year}
  {2022})}\BibitemShut {NoStop}%
\bibitem [{\citenamefont {Miles}\ \emph {et~al.}(2023)\citenamefont {Miles},
  \citenamefont {Samajdar}, \citenamefont {Ebadi}, \citenamefont {Wang},
  \citenamefont {Pichler}, \citenamefont {Sachdev}, \citenamefont {Lukin},
  \citenamefont {Greiner}, \citenamefont {Weinberger},\ and\ \citenamefont
  {Kim}}]{miles2023machine}%
  \BibitemOpen
  \bibfield  {author} {\bibinfo {author} {\bibfnamefont {C.}~\bibnamefont
  {Miles}}, \bibinfo {author} {\bibfnamefont {R.}~\bibnamefont {Samajdar}},
  \bibinfo {author} {\bibfnamefont {S.}~\bibnamefont {Ebadi}}, \bibinfo
  {author} {\bibfnamefont {T.~T.}\ \bibnamefont {Wang}}, \bibinfo {author}
  {\bibfnamefont {H.}~\bibnamefont {Pichler}}, \bibinfo {author} {\bibfnamefont
  {S.}~\bibnamefont {Sachdev}}, \bibinfo {author} {\bibfnamefont {M.~D.}\
  \bibnamefont {Lukin}}, \bibinfo {author} {\bibfnamefont {M.}~\bibnamefont
  {Greiner}}, \bibinfo {author} {\bibfnamefont {K.~Q.}\ \bibnamefont
  {Weinberger}},\ and\ \bibinfo {author} {\bibfnamefont {E.-A.}\ \bibnamefont
  {Kim}},\ }\bibfield  {title} {\bibinfo {title} {Machine learning discovery of
  new phases in programmable quantum simulator snapshots},\ }\href
  {https://doi.org/10.1103/PhysRevResearch.5.013026} {\bibfield  {journal}
  {\bibinfo  {journal} {Phys. Rev. Res.}\ }\textbf {\bibinfo {volume} {5}},\
  \bibinfo {pages} {013026} (\bibinfo {year} {2023})}\BibitemShut {NoStop}%
\bibitem [{\citenamefont {O’Rourke}\ and\ \citenamefont
  {Chan}(2023)}]{Garnet_chan_rydberg}%
  \BibitemOpen
  \bibfield  {author} {\bibinfo {author} {\bibfnamefont {M.~J.}\ \bibnamefont
  {O’Rourke}}\ and\ \bibinfo {author} {\bibfnamefont {G.~K.-L.}\ \bibnamefont
  {Chan}},\ }\bibfield  {title} {\bibinfo {title} {{Entanglement in the quantum
  phases of an unfrustrated Rydberg atom array}},\ }\href
  {https://doi.org/10.1038/s41467-023-41166-0} {\bibfield  {journal} {\bibinfo
  {journal} {Nat. Commun.}\ }\textbf {\bibinfo {volume} {14}},\ \bibinfo
  {pages} {5397} (\bibinfo {year} {2023})}\BibitemShut {NoStop}%
\bibitem [{\citenamefont {Yan}\ \emph {et~al.}(2023)\citenamefont {Yan},
  \citenamefont {Wang}, \citenamefont {Samajdar}, \citenamefont {Sachdev},\
  and\ \citenamefont {Meng}}]{Yan2023}%
  \BibitemOpen
  \bibfield  {author} {\bibinfo {author} {\bibfnamefont {Z.}~\bibnamefont
  {Yan}}, \bibinfo {author} {\bibfnamefont {Y.-C.}\ \bibnamefont {Wang}},
  \bibinfo {author} {\bibfnamefont {R.}~\bibnamefont {Samajdar}}, \bibinfo
  {author} {\bibfnamefont {S.}~\bibnamefont {Sachdev}},\ and\ \bibinfo {author}
  {\bibfnamefont {Z.~Y.}\ \bibnamefont {Meng}},\ }\bibfield  {title} {\bibinfo
  {title} {{Emergent Glassy Behavior in a Kagome Rydberg Atom Array}},\ }\href
  {https://doi.org/10.1103/PhysRevLett.130.206501} {\bibfield  {journal}
  {\bibinfo  {journal} {Phys. Rev. Lett.}\ }\textbf {\bibinfo {volume} {130}},\
  \bibinfo {pages} {206501} (\bibinfo {year} {2023})}\BibitemShut {NoStop}%
\bibitem [{\citenamefont {Jaksch}\ \emph {et~al.}(2000)\citenamefont {Jaksch},
  \citenamefont {Cirac}, \citenamefont {Zoller}, \citenamefont {Rolston},
  \citenamefont {C\^ot\'e},\ and\ \citenamefont {Lukin}}]{PhysRevLett.85.2208}%
  \BibitemOpen
  \bibfield  {author} {\bibinfo {author} {\bibfnamefont {D.}~\bibnamefont
  {Jaksch}}, \bibinfo {author} {\bibfnamefont {J.~I.}\ \bibnamefont {Cirac}},
  \bibinfo {author} {\bibfnamefont {P.}~\bibnamefont {Zoller}}, \bibinfo
  {author} {\bibfnamefont {S.~L.}\ \bibnamefont {Rolston}}, \bibinfo {author}
  {\bibfnamefont {R.}~\bibnamefont {C\^ot\'e}},\ and\ \bibinfo {author}
  {\bibfnamefont {M.~D.}\ \bibnamefont {Lukin}},\ }\bibfield  {title} {\bibinfo
  {title} {Fast quantum gates for neutral atoms},\ }\href
  {https://doi.org/10.1103/PhysRevLett.85.2208} {\bibfield  {journal} {\bibinfo
   {journal} {Phys. Rev. Lett.}\ }\textbf {\bibinfo {volume} {85}},\ \bibinfo
  {pages} {2208} (\bibinfo {year} {2000})}\BibitemShut {NoStop}%
\bibitem [{\citenamefont {Lukin}\ \emph {et~al.}(2001)\citenamefont {Lukin},
  \citenamefont {Fleischhauer}, \citenamefont {Cote}, \citenamefont {Duan},
  \citenamefont {Jaksch}, \citenamefont {Cirac},\ and\ \citenamefont
  {Zoller}}]{PhysRevLett.87.037901}%
  \BibitemOpen
  \bibfield  {author} {\bibinfo {author} {\bibfnamefont {M.~D.}\ \bibnamefont
  {Lukin}}, \bibinfo {author} {\bibfnamefont {M.}~\bibnamefont {Fleischhauer}},
  \bibinfo {author} {\bibfnamefont {R.}~\bibnamefont {Cote}}, \bibinfo {author}
  {\bibfnamefont {L.~M.}\ \bibnamefont {Duan}}, \bibinfo {author}
  {\bibfnamefont {D.}~\bibnamefont {Jaksch}}, \bibinfo {author} {\bibfnamefont
  {J.~I.}\ \bibnamefont {Cirac}},\ and\ \bibinfo {author} {\bibfnamefont
  {P.}~\bibnamefont {Zoller}},\ }\bibfield  {title} {\bibinfo {title} {Dipole
  blockade and quantum information processing in mesoscopic atomic ensembles},\
  }\href {https://doi.org/10.1103/PhysRevLett.87.037901} {\bibfield  {journal}
  {\bibinfo  {journal} {Phys. Rev. Lett.}\ }\textbf {\bibinfo {volume} {87}},\
  \bibinfo {pages} {037901} (\bibinfo {year} {2001})}\BibitemShut {NoStop}%
\bibitem [{\citenamefont {Landau}(1965)}]{Landau1937}%
  \BibitemOpen
  \bibfield  {author} {\bibinfo {author} {\bibfnamefont {L.}~\bibnamefont
  {Landau}},\ }\bibfield  {title} {\bibinfo {title} {On the theory of phase
  transitions},\ }in\ \href
  {https://doi.org/https://doi.org/10.1016/B978-0-08-010586-4.50034-1} {\emph
  {\bibinfo {booktitle} {Collected Papers of L.D. Landau}}},\ \bibinfo {editor}
  {edited by\ \bibinfo {editor} {\bibfnamefont {D.}~\bibnamefont {{Ter
  Haar}}}}\ (\bibinfo  {publisher} {Pergamon},\ \bibinfo {year} {1965})\ pp.\
  \bibinfo {pages} {193--216}\BibitemShut {NoStop}%
\bibitem [{\citenamefont {Park}\ and\ \citenamefont
  {Sachdev}(2001)}]{Park2001}%
  \BibitemOpen
  \bibfield  {author} {\bibinfo {author} {\bibfnamefont {K.}~\bibnamefont
  {Park}}\ and\ \bibinfo {author} {\bibfnamefont {S.}~\bibnamefont {Sachdev}},\
  }\bibfield  {title} {\bibinfo {title} {{Bond-operator theory of doped
  antiferromagnets: From Mott insulators with bond-centered charge order to
  superconductors with nodal fermions}},\ }\href
  {https://doi.org/10.1103/PhysRevB.64.184510} {\bibfield  {journal} {\bibinfo
  {journal} {Phys. Rev. B}\ }\textbf {\bibinfo {volume} {64}},\ \bibinfo
  {pages} {184510} (\bibinfo {year} {2001})}\BibitemShut {NoStop}%
\bibitem [{\citenamefont {Sachdev}\ and\ \citenamefont
  {Park}(2002)}]{sachdev2002ground}%
  \BibitemOpen
  \bibfield  {author} {\bibinfo {author} {\bibfnamefont {S.}~\bibnamefont
  {Sachdev}}\ and\ \bibinfo {author} {\bibfnamefont {K.}~\bibnamefont {Park}},\
  }\bibfield  {title} {\bibinfo {title} {Ground states of quantum
  antiferromagnets in two dimensions},\ }\href
  {https://doi.org/10.1006/aphy.2002.6232} {\bibfield  {journal} {\bibinfo
  {journal} {Ann. Phys.}\ }\textbf {\bibinfo {volume} {298}},\ \bibinfo {pages}
  {58} (\bibinfo {year} {2002})}\BibitemShut {NoStop}%
\bibitem [{\citenamefont {Shackleton}\ \emph {et~al.}(2021)\citenamefont
  {Shackleton}, \citenamefont {Thomson},\ and\ \citenamefont
  {Sachdev}}]{shackleton_deconfined_2021}%
  \BibitemOpen
  \bibfield  {author} {\bibinfo {author} {\bibfnamefont {H.}~\bibnamefont
  {Shackleton}}, \bibinfo {author} {\bibfnamefont {A.}~\bibnamefont
  {Thomson}},\ and\ \bibinfo {author} {\bibfnamefont {S.}~\bibnamefont
  {Sachdev}},\ }\bibfield  {title} {\bibinfo {title} {Deconfined criticality
  and a gapless {$\mathbb{Z}_2$} spin liquid in the square-lattice
  antiferromagnet},\ }\href {https://doi.org/10.1103/PhysRevB.104.045110}
  {\bibfield  {journal} {\bibinfo  {journal} {Phys. Rev. B}\ }\textbf {\bibinfo
  {volume} {104}},\ \bibinfo {pages} {045110} (\bibinfo {year}
  {2021})}\BibitemShut {NoStop}%
\bibitem [{\citenamefont {Hindmarsh}\ \emph {et~al.}(2021)\citenamefont
  {Hindmarsh}, \citenamefont {Lüben}, \citenamefont {Lumma},\ and\
  \citenamefont {Pauly}}]{Hindmarsh2021}%
  \BibitemOpen
  \bibfield  {author} {\bibinfo {author} {\bibfnamefont {M.}~\bibnamefont
  {Hindmarsh}}, \bibinfo {author} {\bibfnamefont {M.}~\bibnamefont {Lüben}},
  \bibinfo {author} {\bibfnamefont {J.}~\bibnamefont {Lumma}},\ and\ \bibinfo
  {author} {\bibfnamefont {M.}~\bibnamefont {Pauly}},\ }\bibfield  {title}
  {\bibinfo {title} {{Phase transitions in the early universe}},\ }\href
  {https://doi.org/10.21468/SciPostPhysLectNotes.24} {\bibfield  {journal}
  {\bibinfo  {journal} {SciPost Phys. Lect. Notes}\ ,\ \bibinfo {pages} {24}}
  (\bibinfo {year} {2021})}\BibitemShut {NoStop}%
\bibitem [{\citenamefont {Albash}\ and\ \citenamefont
  {Lidar}(2018)}]{RevModPhys.90.015002}%
  \BibitemOpen
  \bibfield  {author} {\bibinfo {author} {\bibfnamefont {T.}~\bibnamefont
  {Albash}}\ and\ \bibinfo {author} {\bibfnamefont {D.~A.}\ \bibnamefont
  {Lidar}},\ }\bibfield  {title} {\bibinfo {title} {Adiabatic quantum
  computation},\ }\href {https://doi.org/10.1103/RevModPhys.90.015002}
  {\bibfield  {journal} {\bibinfo  {journal} {Rev. Mod. Phys.}\ }\textbf
  {\bibinfo {volume} {90}},\ \bibinfo {pages} {015002} (\bibinfo {year}
  {2018})}\BibitemShut {NoStop}%
\bibitem [{\citenamefont {Hibat-Allah}\ \emph {et~al.}(2025)\citenamefont
  {Hibat-Allah}, \citenamefont {Merali}, \citenamefont {Torlai}, \citenamefont
  {Melko},\ and\ \citenamefont {Carrasquilla}}]{hibatallah2024}%
  \BibitemOpen
  \bibfield  {author} {\bibinfo {author} {\bibfnamefont {M.}~\bibnamefont
  {Hibat-Allah}}, \bibinfo {author} {\bibfnamefont {E.}~\bibnamefont {Merali}},
  \bibinfo {author} {\bibfnamefont {G.}~\bibnamefont {Torlai}}, \bibinfo
  {author} {\bibfnamefont {R.~G.}\ \bibnamefont {Melko}},\ and\ \bibinfo
  {author} {\bibfnamefont {J.}~\bibnamefont {Carrasquilla}},\ }\bibfield
  {title} {\bibinfo {title} {{Recurrent neural network wave functions for
  Rydberg atom arrays on kagome lattice}},\ }\bibfield  {journal} {\bibinfo
  {journal} {Communications Physics}\ }\textbf {\bibinfo {volume} {8}},\ \href
  {https://doi.org/10.1038/s42005-025-02226-7} {10.1038/s42005-025-02226-7}
  (\bibinfo {year} {2025})\BibitemShut {NoStop}%
\bibitem [{\citenamefont {Ritort}\ and\ \citenamefont
  {Sollich}(2003)}]{ritort2003glassy}%
  \BibitemOpen
  \bibfield  {author} {\bibinfo {author} {\bibfnamefont {F.}~\bibnamefont
  {Ritort}}\ and\ \bibinfo {author} {\bibfnamefont {P.}~\bibnamefont
  {Sollich}},\ }\bibfield  {title} {\bibinfo {title} {Glassy dynamics of
  kinetically constrained models},\ }\href
  {https://doi.org/10.1080/0001873031000093582} {\bibfield  {journal} {\bibinfo
   {journal} {Adv. Phys.}\ }\textbf {\bibinfo {volume} {52}},\ \bibinfo {pages}
  {219} (\bibinfo {year} {2003})}\BibitemShut {NoStop}%
\bibitem [{\citenamefont {Lan}\ \emph {et~al.}(2018)\citenamefont {Lan},
  \citenamefont {van Horssen}, \citenamefont {Powell},\ and\ \citenamefont
  {Garrahan}}]{PhysRevLett.121.040603}%
  \BibitemOpen
  \bibfield  {author} {\bibinfo {author} {\bibfnamefont {Z.}~\bibnamefont
  {Lan}}, \bibinfo {author} {\bibfnamefont {M.}~\bibnamefont {van Horssen}},
  \bibinfo {author} {\bibfnamefont {S.}~\bibnamefont {Powell}},\ and\ \bibinfo
  {author} {\bibfnamefont {J.~P.}\ \bibnamefont {Garrahan}},\ }\bibfield
  {title} {\bibinfo {title} {{Quantum Slow Relaxation and Metastability due to
  Dynamical Constraints}},\ }\href
  {https://doi.org/10.1103/PhysRevLett.121.040603} {\bibfield  {journal}
  {\bibinfo  {journal} {Phys. Rev. Lett.}\ }\textbf {\bibinfo {volume} {121}},\
  \bibinfo {pages} {040603} (\bibinfo {year} {2018})}\BibitemShut {NoStop}%
\bibitem [{\citenamefont {Angelone}\ \emph {et~al.}(2016)\citenamefont
  {Angelone}, \citenamefont {Mezzacapo},\ and\ \citenamefont
  {Pupillo}}]{angelone2016superglass}%
  \BibitemOpen
  \bibfield  {author} {\bibinfo {author} {\bibfnamefont {A.}~\bibnamefont
  {Angelone}}, \bibinfo {author} {\bibfnamefont {F.}~\bibnamefont
  {Mezzacapo}},\ and\ \bibinfo {author} {\bibfnamefont {G.}~\bibnamefont
  {Pupillo}},\ }\bibfield  {title} {\bibinfo {title} {{Superglass Phase of
  Interaction-Blockaded Gases on a Triangular Lattice}},\ }\href
  {https://doi.org/10.1103/PhysRevLett.116.135303} {\bibfield  {journal}
  {\bibinfo  {journal} {Phys. Rev. Lett.}\ }\textbf {\bibinfo {volume} {116}},\
  \bibinfo {pages} {135303} (\bibinfo {year} {2016})}\BibitemShut {NoStop}%
\bibitem [{\citenamefont {Feldmeier}\ \emph {et~al.}(2019)\citenamefont
  {Feldmeier}, \citenamefont {Pollmann},\ and\ \citenamefont
  {Knap}}]{feldmeier2019emergent}%
  \BibitemOpen
  \bibfield  {author} {\bibinfo {author} {\bibfnamefont {J.}~\bibnamefont
  {Feldmeier}}, \bibinfo {author} {\bibfnamefont {F.}~\bibnamefont
  {Pollmann}},\ and\ \bibinfo {author} {\bibfnamefont {M.}~\bibnamefont
  {Knap}},\ }\bibfield  {title} {\bibinfo {title} {{Emergent Glassy Dynamics in
  a Quantum Dimer Model}},\ }\href
  {https://doi.org/10.1103/PhysRevLett.123.040601} {\bibfield  {journal}
  {\bibinfo  {journal} {Phys. Rev. Lett.}\ }\textbf {\bibinfo {volume} {123}},\
  \bibinfo {pages} {040601} (\bibinfo {year} {2019})}\BibitemShut {NoStop}%
\bibitem [{\citenamefont {Mitra}(2018)}]{mitra2018quantum}%
  \BibitemOpen
  \bibfield  {author} {\bibinfo {author} {\bibfnamefont {A.}~\bibnamefont
  {Mitra}},\ }\bibfield  {title} {\bibinfo {title} {Quantum quench dynamics},\
  }\href {https://doi.org/10.1146/annurev-conmatphys-031016-025451} {\bibfield
  {journal} {\bibinfo  {journal} {Annu. Rev. Condens. Matter Phys.}\ }\textbf
  {\bibinfo {volume} {9}},\ \bibinfo {pages} {245} (\bibinfo {year}
  {2018})}\BibitemShut {NoStop}%
\bibitem [{\citenamefont {Friedemann}\ \emph {et~al.}(2018)\citenamefont
  {Friedemann}, \citenamefont {Duncan}, \citenamefont {Hirschberger},
  \citenamefont {Bauer}, \citenamefont {K{\"u}chler}, \citenamefont {Neubauer},
  \citenamefont {Brando}, \citenamefont {Pfleiderer},\ and\ \citenamefont
  {Grosche}}]{Friedemann2018}%
  \BibitemOpen
  \bibfield  {author} {\bibinfo {author} {\bibfnamefont {S.}~\bibnamefont
  {Friedemann}}, \bibinfo {author} {\bibfnamefont {W.~J.}\ \bibnamefont
  {Duncan}}, \bibinfo {author} {\bibfnamefont {M.}~\bibnamefont
  {Hirschberger}}, \bibinfo {author} {\bibfnamefont {T.~W.}\ \bibnamefont
  {Bauer}}, \bibinfo {author} {\bibfnamefont {R.}~\bibnamefont {K{\"u}chler}},
  \bibinfo {author} {\bibfnamefont {A.}~\bibnamefont {Neubauer}}, \bibinfo
  {author} {\bibfnamefont {M.}~\bibnamefont {Brando}}, \bibinfo {author}
  {\bibfnamefont {C.}~\bibnamefont {Pfleiderer}},\ and\ \bibinfo {author}
  {\bibfnamefont {F.~M.}\ \bibnamefont {Grosche}},\ }\bibfield  {title}
  {\bibinfo {title} {{Quantum tricritical points in NbFe$_2$}},\ }\href
  {https://doi.org/10.1038/nphys4242} {\bibfield  {journal} {\bibinfo
  {journal} {Nat. Phys.}\ }\textbf {\bibinfo {volume} {14}},\ \bibinfo {pages}
  {62} (\bibinfo {year} {2018})}\BibitemShut {NoStop}%
\bibitem [{\citenamefont {Wang}\ \emph {et~al.}(2021)\citenamefont {Wang},
  \citenamefont {He}, \citenamefont {Duan},\ and\ \citenamefont
  {Chen}}]{Wang2021}%
  \BibitemOpen
  \bibfield  {author} {\bibinfo {author} {\bibfnamefont {Y.-Z.}\ \bibnamefont
  {Wang}}, \bibinfo {author} {\bibfnamefont {S.}~\bibnamefont {He}}, \bibinfo
  {author} {\bibfnamefont {L.}~\bibnamefont {Duan}},\ and\ \bibinfo {author}
  {\bibfnamefont {Q.-H.}\ \bibnamefont {Chen}},\ }\bibfield  {title} {\bibinfo
  {title} {Quantum tricritical point emerging in the spin-boson model with two
  dissipative spins in staggered biases},\ }\href
  {https://doi.org/10.1103/PhysRevB.103.205106} {\bibfield  {journal} {\bibinfo
   {journal} {Phys. Rev. B}\ }\textbf {\bibinfo {volume} {103}},\ \bibinfo
  {pages} {205106} (\bibinfo {year} {2021})}\BibitemShut {NoStop}%
\bibitem [{\citenamefont {Papi{\'{c}}}(2022)}]{Papic2021}%
  \BibitemOpen
  \bibfield  {author} {\bibinfo {author} {\bibfnamefont {Z.}~\bibnamefont
  {Papi{\'{c}}}},\ }\bibinfo {title} {Weak ergodicity breaking through the lens
  of quantum entanglement},\ in\ \href
  {https://doi.org/10.1007/978-3-031-03998-0_13} {\emph {\bibinfo {booktitle}
  {Entanglement in Spin Chains: From Theory to Quantum Technology
  Applications}}},\ \bibinfo {editor} {edited by\ \bibinfo {editor}
  {\bibfnamefont {A.}~\bibnamefont {Bayat}}, \bibinfo {editor} {\bibfnamefont
  {S.}~\bibnamefont {Bose}},\ and\ \bibinfo {editor} {\bibfnamefont
  {H.}~\bibnamefont {Johannesson}}}\ (\bibinfo  {publisher} {Springer
  International Publishing},\ \bibinfo {address} {Cham},\ \bibinfo {year}
  {2022})\ pp.\ \bibinfo {pages} {341--395}\BibitemShut {NoStop}%
\bibitem [{\citenamefont {Moudgalya}\ \emph {et~al.}(2022)\citenamefont
  {Moudgalya}, \citenamefont {Bernevig},\ and\ \citenamefont
  {Regnault}}]{Moudgalya2022}%
  \BibitemOpen
  \bibfield  {author} {\bibinfo {author} {\bibfnamefont {S.}~\bibnamefont
  {Moudgalya}}, \bibinfo {author} {\bibfnamefont {B.~A.}\ \bibnamefont
  {Bernevig}},\ and\ \bibinfo {author} {\bibfnamefont {N.}~\bibnamefont
  {Regnault}},\ }\bibfield  {title} {\bibinfo {title} {{Quantum many-body scars
  and Hilbert space fragmentation: a review of exact results}},\ }\href
  {https://doi.org/10.1088/1361-6633/ac73a0} {\bibfield  {journal} {\bibinfo
  {journal} {Rep. Prog. Phys.}\ }\textbf {\bibinfo {volume} {85}},\ \bibinfo
  {pages} {086501} (\bibinfo {year} {2022})}\BibitemShut {NoStop}%
\bibitem [{\citenamefont {Schlosser}\ \emph {et~al.}(2023)\citenamefont
  {Schlosser}, \citenamefont {Tichelmann}, \citenamefont {Sch\"affner},
  \citenamefont {de~Mello}, \citenamefont {Hambach}, \citenamefont {Sch\"utz},\
  and\ \citenamefont {Birkl}}]{PhysRevLett.130.180601}%
  \BibitemOpen
  \bibfield  {author} {\bibinfo {author} {\bibfnamefont {M.}~\bibnamefont
  {Schlosser}}, \bibinfo {author} {\bibfnamefont {S.}~\bibnamefont
  {Tichelmann}}, \bibinfo {author} {\bibfnamefont {D.}~\bibnamefont
  {Sch\"affner}}, \bibinfo {author} {\bibfnamefont {D.~O.}\ \bibnamefont
  {de~Mello}}, \bibinfo {author} {\bibfnamefont {M.}~\bibnamefont {Hambach}},
  \bibinfo {author} {\bibfnamefont {J.}~\bibnamefont {Sch\"utz}},\ and\
  \bibinfo {author} {\bibfnamefont {G.}~\bibnamefont {Birkl}},\ }\bibfield
  {title} {\bibinfo {title} {Scalable multilayer architecture of assembled
  single-atom qubit arrays in a three-dimensional talbot tweezer lattice},\
  }\href {https://doi.org/10.1103/PhysRevLett.130.180601} {\bibfield  {journal}
  {\bibinfo  {journal} {Phys. Rev. Lett.}\ }\textbf {\bibinfo {volume} {130}},\
  \bibinfo {pages} {180601} (\bibinfo {year} {2023})}\BibitemShut {NoStop}%
\bibitem [{\citenamefont {Pause}\ \emph {et~al.}(2024)\citenamefont {Pause},
  \citenamefont {Sturm}, \citenamefont {Mittenb\"{u}hler}, \citenamefont
  {Amann}, \citenamefont {Preuschoff}, \citenamefont {Sch\"{a}ffner},
  \citenamefont {Schlosser},\ and\ \citenamefont {Birkl}}]{Pause:24}%
  \BibitemOpen
  \bibfield  {author} {\bibinfo {author} {\bibfnamefont {L.}~\bibnamefont
  {Pause}}, \bibinfo {author} {\bibfnamefont {L.}~\bibnamefont {Sturm}},
  \bibinfo {author} {\bibfnamefont {M.}~\bibnamefont {Mittenb\"{u}hler}},
  \bibinfo {author} {\bibfnamefont {S.}~\bibnamefont {Amann}}, \bibinfo
  {author} {\bibfnamefont {T.}~\bibnamefont {Preuschoff}}, \bibinfo {author}
  {\bibfnamefont {D.}~\bibnamefont {Sch\"{a}ffner}}, \bibinfo {author}
  {\bibfnamefont {M.}~\bibnamefont {Schlosser}},\ and\ \bibinfo {author}
  {\bibfnamefont {G.}~\bibnamefont {Birkl}},\ }\bibfield  {title} {\bibinfo
  {title} {Supercharged two-dimensional tweezer array with more than 1000
  atomic qubits},\ }\href {https://doi.org/10.1364/OPTICA.513551} {\bibfield
  {journal} {\bibinfo  {journal} {Optica}\ }\textbf {\bibinfo {volume} {11}},\
  \bibinfo {pages} {222} (\bibinfo {year} {2024})}\BibitemShut {NoStop}%
\bibitem [{\citenamefont {Manetsch}\ \emph {et~al.}(2025)\citenamefont
  {Manetsch}, \citenamefont {Nomura}, \citenamefont {Bataille}, \citenamefont
  {Lv}, \citenamefont {Leung},\ and\ \citenamefont {Endres}}]{Manetsch2024}%
  \BibitemOpen
  \bibfield  {author} {\bibinfo {author} {\bibfnamefont {H.~J.}\ \bibnamefont
  {Manetsch}}, \bibinfo {author} {\bibfnamefont {G.}~\bibnamefont {Nomura}},
  \bibinfo {author} {\bibfnamefont {E.}~\bibnamefont {Bataille}}, \bibinfo
  {author} {\bibfnamefont {X.}~\bibnamefont {Lv}}, \bibinfo {author}
  {\bibfnamefont {K.~H.}\ \bibnamefont {Leung}},\ and\ \bibinfo {author}
  {\bibfnamefont {M.}~\bibnamefont {Endres}},\ }\bibfield  {title} {\bibinfo
  {title} {A tweezer array with 6,100 highly coherent atomic qubits},\ }\href
  {https://doi.org/10.1038/s41586-025-09641-4} {\bibfield  {journal} {\bibinfo
  {journal} {Nature}\ }\textbf {\bibinfo {volume} {647}},\ \bibinfo {pages}
  {60–67} (\bibinfo {year} {2025})}\BibitemShut {NoStop}%
\bibitem [{\citenamefont {Schiffer}\ \emph {et~al.}(2024)\citenamefont
  {Schiffer}, \citenamefont {Wild}, \citenamefont {Maskara}, \citenamefont
  {Cain}, \citenamefont {Lukin},\ and\ \citenamefont
  {Samajdar}}]{PhysRevResearch.6.013271}%
  \BibitemOpen
  \bibfield  {author} {\bibinfo {author} {\bibfnamefont {B.~F.}\ \bibnamefont
  {Schiffer}}, \bibinfo {author} {\bibfnamefont {D.~S.}\ \bibnamefont {Wild}},
  \bibinfo {author} {\bibfnamefont {N.}~\bibnamefont {Maskara}}, \bibinfo
  {author} {\bibfnamefont {M.}~\bibnamefont {Cain}}, \bibinfo {author}
  {\bibfnamefont {M.~D.}\ \bibnamefont {Lukin}},\ and\ \bibinfo {author}
  {\bibfnamefont {R.}~\bibnamefont {Samajdar}},\ }\bibfield  {title} {\bibinfo
  {title} {Circumventing superexponential runtimes for hard instances of
  quantum adiabatic optimization},\ }\href
  {https://doi.org/10.1103/PhysRevResearch.6.013271} {\bibfield  {journal}
  {\bibinfo  {journal} {Phys. Rev. Res.}\ }\textbf {\bibinfo {volume} {6}},\
  \bibinfo {pages} {013271} (\bibinfo {year} {2024})}\BibitemShut {NoStop}%
\bibitem [{\citenamefont {Lukin}\ \emph {et~al.}(2024)\citenamefont {Lukin},
  \citenamefont {Schiffer}, \citenamefont {Braverman}, \citenamefont {Cantu},
  \citenamefont {Huber}, \citenamefont {Bylinskii}, \citenamefont
  {Amato-Grill}, \citenamefont {Maskara}, \citenamefont {Cain}, \citenamefont
  {Wild}, \citenamefont {Samajdar},\ and\ \citenamefont {Lukin}}]{lukin2024}%
  \BibitemOpen
  \bibfield  {author} {\bibinfo {author} {\bibfnamefont {A.}~\bibnamefont
  {Lukin}}, \bibinfo {author} {\bibfnamefont {B.~F.}\ \bibnamefont {Schiffer}},
  \bibinfo {author} {\bibfnamefont {B.}~\bibnamefont {Braverman}}, \bibinfo
  {author} {\bibfnamefont {S.~H.}\ \bibnamefont {Cantu}}, \bibinfo {author}
  {\bibfnamefont {F.}~\bibnamefont {Huber}}, \bibinfo {author} {\bibfnamefont
  {A.}~\bibnamefont {Bylinskii}}, \bibinfo {author} {\bibfnamefont
  {J.}~\bibnamefont {Amato-Grill}}, \bibinfo {author} {\bibfnamefont
  {N.}~\bibnamefont {Maskara}}, \bibinfo {author} {\bibfnamefont
  {M.}~\bibnamefont {Cain}}, \bibinfo {author} {\bibfnamefont {D.~S.}\
  \bibnamefont {Wild}}, \bibinfo {author} {\bibfnamefont {R.}~\bibnamefont
  {Samajdar}},\ and\ \bibinfo {author} {\bibfnamefont {M.~D.}\ \bibnamefont
  {Lukin}},\ }\href {https://arxiv.org/abs/2405.21019} {\bibinfo {title}
  {Quantum quench dynamics as a shortcut to adiabaticity}} (\bibinfo {year}
  {2024}),\ \Eprint {https://arxiv.org/abs/2405.21019} {arXiv:2405.21019
  [quant-ph]} \BibitemShut {NoStop}%
\bibitem [{\citenamefont {Bintz}\ \emph {et~al.}(2024)\citenamefont {Bintz},
  \citenamefont {Liu}, \citenamefont {Hauschild}, \citenamefont {Khalifa},
  \citenamefont {Chatterjee}, \citenamefont {Zaletel},\ and\ \citenamefont
  {Yao}}]{bintz2024}%
  \BibitemOpen
  \bibfield  {author} {\bibinfo {author} {\bibfnamefont {M.}~\bibnamefont
  {Bintz}}, \bibinfo {author} {\bibfnamefont {V.~S.}\ \bibnamefont {Liu}},
  \bibinfo {author} {\bibfnamefont {J.}~\bibnamefont {Hauschild}}, \bibinfo
  {author} {\bibfnamefont {A.}~\bibnamefont {Khalifa}}, \bibinfo {author}
  {\bibfnamefont {S.}~\bibnamefont {Chatterjee}}, \bibinfo {author}
  {\bibfnamefont {M.~P.}\ \bibnamefont {Zaletel}},\ and\ \bibinfo {author}
  {\bibfnamefont {N.~Y.}\ \bibnamefont {Yao}},\ }\href
  {https://arxiv.org/abs/2406.00098} {\bibinfo {title} {Dirac spin liquid in
  quantum dipole arrays}} (\bibinfo {year} {2024}),\ \Eprint
  {https://arxiv.org/abs/2406.00098} {arXiv:2406.00098 [cond-mat.str-el]}
  \BibitemShut {NoStop}%
\bibitem [{\citenamefont {Tian}\ \emph {et~al.}(2025)\citenamefont {Tian},
  \citenamefont {Samajdar},\ and\ \citenamefont {Gadway}}]{tian2025}%
  \BibitemOpen
  \bibfield  {author} {\bibinfo {author} {\bibfnamefont {M.}~\bibnamefont
  {Tian}}, \bibinfo {author} {\bibfnamefont {R.}~\bibnamefont {Samajdar}},\
  and\ \bibinfo {author} {\bibfnamefont {B.}~\bibnamefont {Gadway}},\
  }\bibfield  {title} {\bibinfo {title} {{Engineering Frustrated Rydberg Spin
  Models by Graphical Floquet Modulation}},\ }\href
  {https://doi.org/10.1103/yxbv-1dkk} {\bibfield  {journal} {\bibinfo
  {journal} {Phys. Rev. Lett.}\ }\textbf {\bibinfo {volume} {135}},\ \bibinfo
  {pages} {253001} (\bibinfo {year} {2025})}\BibitemShut {NoStop}%
\bibitem [{\citenamefont {Daley}\ \emph {et~al.}(2022)\citenamefont {Daley},
  \citenamefont {Bloch}, \citenamefont {Kokail}, \citenamefont {Flannigan},
  \citenamefont {Pearson}, \citenamefont {Troyer},\ and\ \citenamefont
  {Zoller}}]{Daley2022Practical}%
  \BibitemOpen
  \bibfield  {author} {\bibinfo {author} {\bibfnamefont {A.~J.}\ \bibnamefont
  {Daley}}, \bibinfo {author} {\bibfnamefont {I.}~\bibnamefont {Bloch}},
  \bibinfo {author} {\bibfnamefont {C.}~\bibnamefont {Kokail}}, \bibinfo
  {author} {\bibfnamefont {S.}~\bibnamefont {Flannigan}}, \bibinfo {author}
  {\bibfnamefont {N.}~\bibnamefont {Pearson}}, \bibinfo {author} {\bibfnamefont
  {M.}~\bibnamefont {Troyer}},\ and\ \bibinfo {author} {\bibfnamefont
  {P.}~\bibnamefont {Zoller}},\ }\bibfield  {title} {\bibinfo {title}
  {Practical quantum advantage in quantum simulation},\ }\href
  {https://doi.org/10.1038/s41586-022-04940-6} {\bibfield  {journal} {\bibinfo
  {journal} {Nature}\ }\textbf {\bibinfo {volume} {607}},\ \bibinfo {pages}
  {667} (\bibinfo {year} {2022})}\BibitemShut {NoStop}%
\bibitem [{\citenamefont {Kashyap}\ \emph {et~al.}(2025)\citenamefont
  {Kashyap}, \citenamefont {Styliaris}, \citenamefont {Mouradian},
  \citenamefont {Cirac},\ and\ \citenamefont {Trivedi}}]{Kashyap2025}%
  \BibitemOpen
  \bibfield  {author} {\bibinfo {author} {\bibfnamefont {V.}~\bibnamefont
  {Kashyap}}, \bibinfo {author} {\bibfnamefont {G.}~\bibnamefont {Styliaris}},
  \bibinfo {author} {\bibfnamefont {S.}~\bibnamefont {Mouradian}}, \bibinfo
  {author} {\bibfnamefont {J.~I.}\ \bibnamefont {Cirac}},\ and\ \bibinfo
  {author} {\bibfnamefont {R.}~\bibnamefont {Trivedi}},\ }\bibfield  {title}
  {\bibinfo {title} {Accuracy guarantees and quantum advantage in analog open
  quantum simulation with and without noise},\ }\href
  {https://doi.org/10.1103/PhysRevX.15.021017} {\bibfield  {journal} {\bibinfo
  {journal} {Phys. Rev. X}\ }\textbf {\bibinfo {volume} {15}},\ \bibinfo
  {pages} {021017} (\bibinfo {year} {2025})}\BibitemShut {NoStop}%
\bibitem [{\citenamefont {White}(1992{\natexlab{b}})}]{dmrg_1}%
  \BibitemOpen
  \bibfield  {author} {\bibinfo {author} {\bibfnamefont {S.~R.}\ \bibnamefont
  {White}},\ }\bibfield  {title} {\bibinfo {title} {Density matrix formulation
  for quantum renormalization groups},\ }\href
  {https://doi.org/10.1103/PhysRevLett.69.2863} {\bibfield  {journal} {\bibinfo
   {journal} {Phys. Rev. Lett.}\ }\textbf {\bibinfo {volume} {69}},\ \bibinfo
  {pages} {2863} (\bibinfo {year} {1992}{\natexlab{b}})}\BibitemShut {NoStop}%
\bibitem [{\citenamefont {White}(1993)}]{dmrg_2}%
  \BibitemOpen
  \bibfield  {author} {\bibinfo {author} {\bibfnamefont {S.~R.}\ \bibnamefont
  {White}},\ }\bibfield  {title} {\bibinfo {title} {Density-matrix algorithms
  for quantum renormalization groups},\ }\href
  {https://doi.org/10.1103/PhysRevB.48.10345} {\bibfield  {journal} {\bibinfo
  {journal} {Phys. Rev. B}\ }\textbf {\bibinfo {volume} {48}},\ \bibinfo
  {pages} {10345} (\bibinfo {year} {1993})}\BibitemShut {NoStop}%
\bibitem [{\citenamefont {Schollw\"ock}(2005)}]{dmrg_3}%
  \BibitemOpen
  \bibfield  {author} {\bibinfo {author} {\bibfnamefont {U.}~\bibnamefont
  {Schollw\"ock}},\ }\bibfield  {title} {\bibinfo {title} {The density-matrix
  renormalization group},\ }\href {https://doi.org/10.1103/RevModPhys.77.259}
  {\bibfield  {journal} {\bibinfo  {journal} {Rev. Mod. Phys.}\ }\textbf
  {\bibinfo {volume} {77}},\ \bibinfo {pages} {259} (\bibinfo {year}
  {2005})}\BibitemShut {NoStop}%
\bibitem [{\citenamefont {Fishman}\ \emph
  {et~al.}(2022{\natexlab{b}})\citenamefont {Fishman}, \citenamefont {White},\
  and\ \citenamefont {Stoudenmire}}]{itensor}%
  \BibitemOpen
  \bibfield  {author} {\bibinfo {author} {\bibfnamefont {M.}~\bibnamefont
  {Fishman}}, \bibinfo {author} {\bibfnamefont {S.~R.}\ \bibnamefont {White}},\
  and\ \bibinfo {author} {\bibfnamefont {E.~M.}\ \bibnamefont {Stoudenmire}},\
  }\bibfield  {title} {\bibinfo {title} {{The ITensor Software Library for
  Tensor Network Calculations}},\ }\href
  {https://doi.org/10.21468/SciPostPhysCodeb.4} {\bibfield  {journal} {\bibinfo
   {journal} {SciPost Phys. Codebases}\ ,\ \bibinfo {pages} {4}} (\bibinfo
  {year} {2022}{\natexlab{b}})}\BibitemShut {NoStop}%
\bibitem [{\citenamefont {White}\ and\ \citenamefont
  {Chernyshev}(2007)}]{PhysRevLett.99.127004}%
  \BibitemOpen
  \bibfield  {author} {\bibinfo {author} {\bibfnamefont {S.~R.}\ \bibnamefont
  {White}}\ and\ \bibinfo {author} {\bibfnamefont {A.~L.}\ \bibnamefont
  {Chernyshev}},\ }\bibfield  {title} {\bibinfo {title} {{Ne\'el Order in
  Square and Triangular Lattice Heisenberg Models}},\ }\href
  {https://doi.org/10.1103/PhysRevLett.99.127004} {\bibfield  {journal}
  {\bibinfo  {journal} {Phys. Rev. Lett.}\ }\textbf {\bibinfo {volume} {99}},\
  \bibinfo {pages} {127004} (\bibinfo {year} {2007})}\BibitemShut {NoStop}%
\bibitem [{\citenamefont {Balents}\ \emph {et~al.}(2005)\citenamefont
  {Balents}, \citenamefont {Bartosch}, \citenamefont {Burkov}, \citenamefont
  {Sachdev},\ and\ \citenamefont {Sengupta}}]{Balents2005}%
  \BibitemOpen
  \bibfield  {author} {\bibinfo {author} {\bibfnamefont {L.}~\bibnamefont
  {Balents}}, \bibinfo {author} {\bibfnamefont {L.}~\bibnamefont {Bartosch}},
  \bibinfo {author} {\bibfnamefont {A.}~\bibnamefont {Burkov}}, \bibinfo
  {author} {\bibfnamefont {S.}~\bibnamefont {Sachdev}},\ and\ \bibinfo {author}
  {\bibfnamefont {K.}~\bibnamefont {Sengupta}},\ }\bibfield  {title} {\bibinfo
  {title} {{Putting competing orders in their place near the Mott
  transition}},\ }\href {https://doi.org/10.1103/PhysRevB.71.144508} {\bibfield
   {journal} {\bibinfo  {journal} {Phys. Rev. B}\ }\textbf {\bibinfo {volume}
  {71}},\ \bibinfo {pages} {144508} (\bibinfo {year} {2005})}\BibitemShut
  {NoStop}%
\bibitem [{\citenamefont {Binder}(1981)}]{Binder1981}%
  \BibitemOpen
  \bibfield  {author} {\bibinfo {author} {\bibfnamefont {K.}~\bibnamefont
  {Binder}},\ }\bibfield  {title} {\bibinfo {title} {{Finite size scaling
  analysis of Ising model block distribution functions}},\ }\href
  {https://doi.org/10.1007/BF01293604} {\bibfield  {journal} {\bibinfo
  {journal} {Z. Phys. B}\ }\textbf {\bibinfo {volume} {43}},\ \bibinfo {pages}
  {119} (\bibinfo {year} {1981})}\BibitemShut {NoStop}%
\bibitem [{\citenamefont {Hasenbusch}\ \emph {et~al.}(1999)\citenamefont
  {Hasenbusch}, \citenamefont {Pinn},\ and\ \citenamefont
  {Vinti}}]{Hasenbusch1999}%
  \BibitemOpen
  \bibfield  {author} {\bibinfo {author} {\bibfnamefont {M.}~\bibnamefont
  {Hasenbusch}}, \bibinfo {author} {\bibfnamefont {K.}~\bibnamefont {Pinn}},\
  and\ \bibinfo {author} {\bibfnamefont {S.}~\bibnamefont {Vinti}},\ }\bibfield
   {title} {\bibinfo {title} {{Critical exponents of the three-dimensional
  Ising universality class from finite-size scaling with standard and improved
  actions}},\ }\href {https://doi.org/10.1103/PhysRevB.59.11471} {\bibfield
  {journal} {\bibinfo  {journal} {Phys. Rev. B}\ }\textbf {\bibinfo {volume}
  {59}},\ \bibinfo {pages} {11471} (\bibinfo {year} {1999})}\BibitemShut
  {NoStop}%
\bibitem [{\citenamefont {Goldenfeld}(1992)}]{Goldenfeld1992}%
  \BibitemOpen
  \bibfield  {author} {\bibinfo {author} {\bibfnamefont {N.}~\bibnamefont
  {Goldenfeld}},\ }\href {https://doi.org/10.1201/9780429493492} {\emph
  {\bibinfo {title} {{Lectures On Phase Transitions And The Renormalization
  Group}}}},\ Frontiers in Physics\ (\bibinfo  {publisher} {CRC Press},\
  \bibinfo {address} {Boca Raton},\ \bibinfo {year} {1992})\BibitemShut
  {NoStop}%
\bibitem [{\citenamefont {Pelissetto}\ and\ \citenamefont
  {Vicari}(2002)}]{Pelissetto2002}%
  \BibitemOpen
  \bibfield  {author} {\bibinfo {author} {\bibfnamefont {A.}~\bibnamefont
  {Pelissetto}}\ and\ \bibinfo {author} {\bibfnamefont {E.}~\bibnamefont
  {Vicari}},\ }\bibfield  {title} {\bibinfo {title} {Critical phenomena and
  renormalization-group theory},\ }\href
  {https://doi.org/https://doi.org/10.1016/S0370-1573(02)00219-3} {\bibfield
  {journal} {\bibinfo  {journal} {Phys. Rep.}\ }\textbf {\bibinfo {volume}
  {368}},\ \bibinfo {pages} {549} (\bibinfo {year} {2002})}\BibitemShut
  {NoStop}%
\bibitem [{\citenamefont {Silva}\ \emph {et~al.}(2018)\citenamefont {Silva},
  \citenamefont {Continentino},\ and\ \citenamefont {Barci}}]{Silva2018}%
  \BibitemOpen
  \bibfield  {author} {\bibinfo {author} {\bibfnamefont {N.~L.}\ \bibnamefont
  {Silva}}, \bibinfo {author} {\bibfnamefont {M.~A.}\ \bibnamefont
  {Continentino}},\ and\ \bibinfo {author} {\bibfnamefont {D.~G.}\ \bibnamefont
  {Barci}},\ }\bibfield  {title} {\bibinfo {title} {{Quantum corrections for
  the phase diagram of systems with competing order}},\ }\href
  {https://doi.org/10.1088/1361-648X/aac062} {\bibfield  {journal} {\bibinfo
  {journal} {J. Phys. Condens. Matter}\ }\textbf {\bibinfo {volume} {30}},\
  \bibinfo {pages} {225402} (\bibinfo {year} {2018})}\BibitemShut {NoStop}%
\bibitem [{\citenamefont {Kornjača}\ and\ \citenamefont
  {Flint}(2020)}]{kornjaca2020}%
  \BibitemOpen
  \bibfield  {author} {\bibinfo {author} {\bibfnamefont {M.}~\bibnamefont
  {Kornjača}}\ and\ \bibinfo {author} {\bibfnamefont {R.}~\bibnamefont
  {Flint}},\ }\href {https://arxiv.org/abs/2012.08511} {\bibinfo {title}
  {Signatures of spinorial order in {URu$_2$Si$_2$}: Landau-ginzburg theory of
  hastatic order}} (\bibinfo {year} {2020}),\ \Eprint
  {https://arxiv.org/abs/2012.08511} {arXiv:2012.08511 [cond-mat.str-el]}
  \BibitemShut {NoStop}%
\bibitem [{\citenamefont {Landau}\ and\ \citenamefont
  {Lifshitz}(1996)}]{Landau1996-tu}%
  \BibitemOpen
  \bibfield  {author} {\bibinfo {author} {\bibfnamefont {L.~D.}\ \bibnamefont
  {Landau}}\ and\ \bibinfo {author} {\bibfnamefont {E.~M.}\ \bibnamefont
  {Lifshitz}},\ }\href {https://doi.org/10.1016/C2009-0-24487-4} {\emph
  {\bibinfo {title} {Statistical Physics}}},\ \bibinfo {edition} {3rd}\ ed.,\
  Course Of Theoretical Physics\ (\bibinfo  {publisher}
  {Butterworth-Heinemann},\ \bibinfo {address} {Oxford, England},\ \bibinfo
  {year} {1996})\BibitemShut {NoStop}%
\bibitem [{\citenamefont {Kirkpatrick}\ \emph {et~al.}(1983)\citenamefont
  {Kirkpatrick}, \citenamefont {Gelatt},\ and\ \citenamefont
  {Vecchi}}]{Kirkpatrick1983}%
  \BibitemOpen
  \bibfield  {author} {\bibinfo {author} {\bibfnamefont {S.}~\bibnamefont
  {Kirkpatrick}}, \bibinfo {author} {\bibfnamefont {C.~D.}\ \bibnamefont
  {Gelatt}},\ and\ \bibinfo {author} {\bibfnamefont {M.~P.}\ \bibnamefont
  {Vecchi}},\ }\bibfield  {title} {\bibinfo {title} {Optimization by simulated
  annealing},\ }\href {https://doi.org/10.1126/science.220.4598.671} {\bibfield
   {journal} {\bibinfo  {journal} {Science}\ }\textbf {\bibinfo {volume}
  {220}},\ \bibinfo {pages} {671} (\bibinfo {year} {1983})}\BibitemShut
  {NoStop}%
\end{thebibliography}
\end{document}